\documentclass[10pt]{article}
\usepackage[T1]{fontenc}
\usepackage{amssymb}
\usepackage{slashed}
\usepackage{cancel}
\usepackage{mathrsfs} 
\usepackage{fullpage}
\usepackage{mathtools} 
\usepackage[colorlinks=true,linkcolor=blue,citecolor=magenta,linktocpage=true]{hyperref}
\usepackage[numbers,sort,compress]{natbib}
\usepackage{tocloft}
\usepackage{titlesec}
\usepackage{units}

\usepackage{empheq}
\usepackage[most]{tcolorbox}
\newtcbox{\othermathbox}[1][]{nobeforeafter,math upper,tcbox raise base,enhanced,rounded corners,colback=black!5,colframe=black}

\def\be#1\ee{\begin{align}#1\end{align}}
\def\bsub#1\esub{\begin{subequations}#1\end{subequations}}
\def\nn{\nonumber}
\def\q{\qquad}
\def\f{\frac}
\def\eps{\varepsilon}

\def\tf{\text{\tiny{TF}}}

\def\de{\mathrm{d}}

\def\D{\mathcal{D}}
\def\E{\mathcal{E}}
\def\F{\mathcal{F}}

\def\I{\mathcal{I}}
\def\J{\mathcal{J}}

\def\L{\mathcal{L}}
\def\M{\mathcal{M}}
\def\N{\mathcal{N}}
\def\O{\mathcal{O}}
\def\P{\mathcal{P}}

\def\S{\mathcal{S}}

\def\Y{\mathcal{Y}}

\def\ra{\rangle}
\def\la{\langle}
\def\pe{\phantom{\ =}}
\def\ethb{\overline{\eth}}
\def\Eth{\textup{\DH}}
\def\Ethb{\bar{\textup{\DH}}}

\numberwithin{equation}{section}
\begin{document}

\title{\Large{\textbf{\sffamily Twisting asymptotically-flat spacetimes}}}
\author{\sffamily Marc Geiller$^1$, Pujian Mao$^2$, Antoine Vincenti$^1$}
\date{}
\date{\small{\textit{
${}^1$ENS de Lyon, CNRS, LPENSL, UMR 5672, 69342 Lyon cedex 07, France\\
${}^2$Center for Joint Quantum Studies, Department of Physics,\\
School of Science, Tianjin University, 135 Yaguan Road, Tianjin 300350, China\\}}}

\maketitle

\begin{abstract}
We extend the Bondi formalism to describe asymptotically-flat spacetimes where the outgoing null geodesic congruence is not hypersurface-orthogonal, i.e. has a non-vanishing twist. In the Newman--Penrose formulation, the twist $\text{Im}(\rho)$ is sourced by a twist potential sitting in the transverse null dyad $(m,\bar{m})$, while in the metric formulation this potential arises from $g_{ra}\neq0$. We explain how to arrange and solve the Einstein equations for such generalized line elements, thereby providing an extension of the Bondi hierarchy to asymptotically-flat spacetimes with non-vanishing twist. We work out the twisting generalizations of all the well-known features pertaining to asymptotically-flat spacetimes in Bondi gauge, such as the solution space, the flux-balance laws, the asymptotic symmetries, and the transformation laws. The twist potential has a natural Carrollian interpretation as an Ehresmann connection, and gives rise to Carroll boosts as extra asymptotic symmetries. The supertranslation generators in this gauge with twist have no extension in the bulk and therefore only act at null infinity. This enables us to study finite supertranslations of any given solution. One of the advantages of the Bondi gauge with non-vanishing twist is also that it allows to write algebraically special solutions in a manifestly finite radial expansion, and with a repeated principal null direction such that $\Psi_0=\Psi_1=0$. This is in particular the case for the Kerr--Taub--NUT solution. The asymptotic symmetries of algebraically special solutions also have a finite radial expansion, which enables us to study, for example, the supertranslated Schwarzschild solution and its charges quite straightforwardly. We expect that these results will find applications in the development of flat holography for algebraically special solutions and in the study of their perturbations. We also study an analogue of the twist in three-dimensional spacetimes with non-vanishing cosmological constant, and find an 8-dimensional solution space which encompasses and generalizes the existing results in the literature.
\end{abstract}

\thispagestyle{empty}

\newpage
\setcounter{page}{1}
\hfill
\tableofcontents
\bigskip
\hrule

\newpage

\section{Introduction}

Asymptotically-flat spacetimes, which enable to describe idealized isolated self-gravitating systems, have played a central role in mathematical general relativity and the development of the theory of gravitational radiation \cite{Bondi:1960jsa,Bondi:1962px,Bondi:1962rkt,Sachs:1962zza,Sachs:1962wk,Penrose:1962ij,Penrose:1964ge,Penrose:1965am,Geroch1977}. When specializing to null infinity, asymptotically-flat spacetimes are also the framework where one can explore the interplay between asymptotic symmetries \cite{Bondi:1960jsa,Bondi:1962px,Bondi:1962rkt,Sachs:1962zza}, memory effects \cite{Braginsky:1986ia,1987Natur.327..123B,Christodoulou:1991cr,Blanchet:1992br,Thorne:1992sdb}, and soft theorems for massless fields \cite{Low:1954kd,Weinberg:1965nx}, as revealed in the seminal work \cite{Strominger:2013jfa,He:2014laa,Strominger:2014pwa}. These unforeseen connections between non-perturbative gravitational wave observables and infrared features of the classical and quantum gravitational scattering have been investigated extensively in recent years, and have led to the discovery of new memory effects \cite{Pasterski:2015tva,Compere:2016gwf,Compere:2019odm,Nichols:2017rqr,Nichols:2018qac,Flanagan:2018yzh,Flanagan:2019ezo,Grant:2021hga,Grant:2023ged,Compere:2018ylh,Seraj:2021rxd,Seraj:2022qyt}, new soft graviton theorems \cite{White:2014qia,Cachazo:2014fwa,Zlotnikov:2014sva,Kalousios:2014uva,Conde:2016rom,Campiglia:2016efb,Campiglia:2016jdj,Banerjee:2021cly,Freidel:2021dfs,Laddha:2018myi,Sahoo:2018lxl,Saha:2019tub}, and new asymptotic symmetries \cite{Barnich:2009se,Barnich:2010eb,Barnich:2011mi,Barnich:2011ct,Barnich:2013axa,Campiglia:2020qvc,Campiglia:2014yka,Campiglia:2015yka,Flanagan:2015pxa,Freidel:2021fxf,Geiller:2022vto,Geiller:2024amx}. This has also fostered the development of the celestial and Carrollian approaches to flat holography \cite{Pasterski:2016qvg,Pasterski:2017kqt,Donnay:2020guq,Pasterski:2021raf,Pasterski:2021rjz,Raclariu:2021zjz,Donnay:2022aba,Donnay:2022wvx,Mason:2023mti}.

Although asymptotically-flat spacetimes can be characterized in a coordinate-independent way through Penrose's conformal compactification \cite{Penrose:1964ge,Geroch1977}, it is very convenient for practical purposes to work with a metric written in the Bondi gauge \cite{Madler:2016xju}. Explicitly, this relies on a choice of coordinates $x^\mu=(u,r,x^a)$ adapted to the null geodesics of the spacetime, and chosen such that $\partial_\mu u$ is null while $\partial_\mu x^a$ is constant along the null rays. The resulting metric satisfies the three gauge conditions $g^{uu}=0$ and $g^{ua}=0$ on the inverse metric, or equivalently $g_{rr}=0$ and $g_{ra}=0$, which can conveniently be thought of as defining the partial Bondi gauge \cite{Geiller:2022vto}. A complete gauge fixing can then be obtained by specifying a fourth condition, which is usually achieved through a choice of radial coordinate $r$. In the Bondi--Sachs gauge \cite{Madler:2016xju} the coordinate $r$ is chosen to be an areal distance, while in the Newman--Unti gauge \cite{Newman:1962cia} it is chosen as the affine parameter for the null geodesics\footnote{Other choices can also naturally be seen as full gauge fixings of the partial Bondi gauge, namely the unimodular gauge (where one would set $\sqrt{-g}=1$) and the double null gauge \cite{PhysRevD.110.024018,PhysRevD.110.024020}. MG thanks Simone Speziale for pointing this out.}. By working with a metric in Bondi gauge, one can solve the Einstein equations near future null infinity in a radial expansion, and then uncover expressions for the Bondi--Trautman mass loss and the angular momentum loss \cite{Tamburino:1966zz,Geroch:1981ut,Ashtekar:1981bq,Dray:1984rfa,Wald:1999wa}, as well as for the BMS charges associated with the asymptotic symmetries. The Bondi gauge also has important applications in numerical relativity (see \cite{Madler:2016xju} and references therein).

The aim of the present work is to extend the Bondi formalism to the case of asymptotically-flat spacetimes with a non-vanishing twist of the outgoing null congruence. Before motivating this extension and explaining how it is achieved, let us first recall the most notable generalizations of the Bondi gauge which have already been introduced in recent years, motivated in part by the above-mentioned results.
\begin{itemize}
\item \textbf{Extended, generalized, and BMS--Weyl.} These enlargements of the Bondi framework are obtained by progressively relaxing conditions on the angular part of the asymptotic symmetry generator and on the leading sphere metric entering the expansion $g_{ab}=r^2q_{ab}+\O(r)$. This is motivated by the BMS/CFT correspondence \cite{Barnich:2010eb,Barnich:2011mi}, by the construction of an asymptotic symmetry counterpart to the subleading soft graviton theorem \cite{Campiglia:2014yka,Campiglia:2015yka,Campiglia:2020qvc}, and by the question of how much one can relax the boundary conditions and extend the asymptotic symmetry group \cite{Freidel:2021fxf,Geiller:2022vto,Geiller:2024amx}. Here we consider the framework of BMS--Weyl, where the sphere metric $q_{ab}$ is arbitrary, and where the asymptotic symmetries include diffeomorphisms of the sphere and local conformal rescalings.
\item \textbf{Time dependency in the celestial metric.} In usual treatments of asymptotically-flat spacetimes the leading metric $q_{ab}$ is taken as time-independent, implying that $\I^+$ has no intrinsic shear. This is sometimes known as the ``Bondi condition''. However, the extension to include time dependency in $q_{ab}$ or its conformal factor \cite{Barnich:2010eb,Geiller:2022vto} has two particularly important applications, which are the study of Robinson--Trautman solutions \cite{Ciambelli:2017wou,Compere:2018ylh,Adami:2024mtu,barnich-seraj-RT} and the description of gravitational radiation in asymptotically-(A)dS spacetimes \cite{Poole:2018koa,Compere:2019bua,MG-AdS,McNees:2025acf,Compere:2023ktn,Compere:2024ekl,Arenas-Henriquez:2025rpt}. In the present work, although we consider the case of a vanishing cosmological constant, we allow for time dependency in $q_{ab}$.
\item \textbf{Logarithmic terms.} The Bondi formalism can also be extended to describe so-called logarithmically-asymptotically-flat spacetimes, which do not satisfy the Penrose peeling property. This is done by including extra data in the expansion of the angular metric $g_{ab}$, which sources logarithmic terms $\ln r$ in the metric and leads to a so-called polyhomogeneous expansion \cite{1985FoPh...15..605W,Chrusciel:1993hx,Kroon:1998tu,ValienteKroon:2002gb,Godazgar:2020peu,Freidel:2024tpl,Geiller:2024ryw}. The question of whether physically-realistic sources of gravitational radiation produce spacetimes whose asymptotic structure exhibits such failures of peeling has been around for a long time \cite{1986mgm..conf..365D,2002nmgm.meet...44C,1985FoPh...15..605W,Friedrich:1983vx,PhysRevD.19.3483,PhysRevD.19.3495,doi:10.1098/rspa.1981.0101,Andersson:1993we,Valiente-Kroon:2002xys,Kroon:2004me,Kehrberger:2021uvf,Kehrberger:2021vhp,Kehrberger:2021azo,Kehrberger:2024clh,Kehrberger:2024aak,Gajic:2022pst,Kehrberger:2023btg,Bieri:2023cyn}, and has recently been revived thanks to the discovery of the logarithmic soft graviton theorem \cite{Sahoo:2018lxl,Laddha:2018myi,Laddha:2018vbn,Saha:2019tub,Agrawal:2023zea,Choi:2024ygx,Choi:2024ajz} and results from scattering amplitudes \cite{DeAngelis:2025vlf}. Here we do not consider such logarithmic terms, but a careful analysis of the solution space will nonetheless be required in order to enforce the so-called no-log condition.
\item \textbf{Free induced boundary metric.} While the induced boundary metric on $\I^+$ is parametrized in part by the celestial metric $q_{ab}$, it can also feature extra free functions of $(u,x^a)$ which arise from the Einstein equations as radial integration constants and parametrize the leading order terms in $g_{ur}=\O(1)$ and $g_{ua}=\O(r^2)$. These boundary conditions, which are weaker than what is typically used in the Bondi gauge, were studied in \cite{Poole:2018koa,Compere:2019bua,Geiller:2022vto,McNees:2024iyu}. The relaxation of $g_{ua}$ has applications to the description of radiation in asymptotically-(A)dS spacetimes \cite{Bonga:2023eml,MG-AdS,McNees:2025acf} and of certain exact solutions such as shockwaves \cite{He:2023qha} and the C-metric in Bondi gauge (see appendix \ref{app:C-metric}). Here we will only keep track of these integration constants in the three-dimensional case studied in section \ref{sec:3d}.
\item \textbf{Other types of asymptotics.} The Bondi gauge can also be used beyond asymptotic flatness to describe spacetimes which are for example asymptotically-(A)dS \cite{Poole:2018koa,Compere:2019bua,Mao:2019ahc,Compere:2020lrt,Geiller:2022vto,Bonga:2023eml,MG-AdS,McNees:2025acf} or asymptotically-FLRW \cite{Bonga:2020fhx,Enriquez-Rojo:2020miw,Enriquez-Rojo:2021blc,Enriquez-Rojo:2022onp}. In this work we will only consider a non-vanishing cosmological constant in the three-dimensional case.
\item \textbf{Other dimensions.} Many of the above aspects were also studied in three-dimensional gravity, where the Bondi gauge enables us to elegantly describe the asymptotic symmetries at null infinity \cite{Ashtekar:1996cd,Barnich:2006av} and to recover the seminal results of Brown--Henneaux in AdS$_3$ \cite{Brown:1986nw,Barnich:2012aw}. The role of the integration constants parametrizing a free boundary metric was studied in \cite{Ciambelli:2020eba,Ciambelli:2020ftk,Campoleoni:2022wmf,Ruzziconi:2020wrb,Geiller:2021vpg}. Here we extend all of these results in section \ref{sec:3d} by introducing an even larger solution space.
\end{itemize}

Despite its notable advantages and successes, the Bondi gauge has a built-in inconvenience which is not addressed by any of the above-mentioned generalizations, namely the fact that certain solutions cannot be included in a ``simple'' form and require an infinite radial expansion. This is illustrated for example by the Kerr solution. Indeed, the Kerr metric has $g_{rr}\neq0$ in Boyer--Lindquist coordinates and $g_{ra}\neq0$ in Eddington--Finkelstein coordinates\footnote{One can note that these were the original coordinates used in the derivation of the Kerr metric \cite{Kerr:1963ud} (see also \cite{Visser:2007fj} for an historical account).}, which is incompatible in either case with the Bondi gauge conditions $g_{rr}=g_{ra}=0$. Starting from the Eddington--Finkelstein coordinates, one can force the metric in a form with $g_{ra}=0$, but at the expense of introducing an infinite radial expansion. This was done by Fletcher and Lun in \cite{FletcherLun1996,2003CQGra..20.4153F}, and then extended to Kerr--(A)dS in \cite{Hoque:2021nti}. The resulting Kerr metric in Bondi gauge is reconstructed near null infinity in a radial expansion, and in particular has non-vanishing (although time-independent) shear. While this is not problematic per se, it leads to unwanted complications when studying properties of Kerr in Bondi gauge. Relatedly, when using the Newman--Penrose formalism (NP hereafter) to study the Kerr metric in Bondi gauge, one finds that the angular momentum is encoded in the leading order of $\Psi_1\neq0$, and also that $\Psi_0\neq0$. This can all be traced back to the fact that the principal null directions of Kerr have twist (or rotation) and do not generate null hypersurfaces. In Eddington--Finkelstein coordinates, the twist is sourced by the term responsible for $g_{ra}\neq0$. When using a tetrad aligned with the principal null directions and encoding this twist, one finds that Kerr has $\Psi_0=\Psi_1=0$, as expected for algebraically special solutions. Similar complications arise with the supertranslated Schwarzschild solution, which has an infinite radial expansion when written in Bondi gauge with $g_{ra}=0$ \cite{Hawking:2016msc,Hawking:2016sgy,Compere:2016hzt,Compere:2016jwb}, but can be written in finite form when allowing for $g_{ra}\neq0$. More generally, this happens for all twisting algebraically special solutions.

The twist, also known as rotation, is a geometric property of null geodesics alongside the expansion and the shear. Its role has been extensively studied in the case of algebraically special solutions \cite{Stephani:2003tm}, but algebraically general solutions with non-vanishing twist have received only little attention, especially in the context of asymptotically-flat spacetimes. Recent progress in this direction was however made in \cite{Mao:2024jpt} using the NP formalism. There, the quantities of interest are the spin coefficients $\sigma$ and $\rho$. The former is the shear of the null geodesics $\ell=\partial_r$, while $\rho$ has real and imaginary parts corresponding respectively to the expansion and the twist of $\ell$. More precisely, the twist is obtained as a derivative operator acting on a ``twist potential'' $Z$, which is the retarded time component $m^u=Z$ of the null vector forming the tetrad $(\ell,n,m,\bar{m})$. This twist potential sources a non-vanishing component $g_{ra}\neq0$ in the metric, which therefore cannot fit within the Bondi formalism. Incidentally, it turns out that relaxations of the Bondi gauge allowing for $g_{ra}\neq0$ have also been studied in recent years, although without reference to the twist and for different motivations altogether. Indeed, generalizations known as the Carroll covariant Newman--Unti and Bondi--Sachs gauges were introduced respectively in \cite{Ciambelli:2020eba,Ciambelli:2020ftk,Campoleoni:2022wmf,Mittal:2022ywl,Campoleoni:2023fug,Arenas-Henriquez:2025rpt,Fiorucci:2025twa} and \cite{Hartong:2025jpp}. The purpose of these works was to restore Carroll covariance on the boundary, with applications to the fluid/gravity correspondence and to the definition of a stress-energy tensor in the context of asymptotically-flat spacetimes \cite{Fiorucci:2025twa,Hartong:2025jpp} (see also \cite{Adami:2024rkr} for three-dimensional asymptotically-flat spacetimes). Our goal here is to merge these recent results and to present, in both the metric and NP formalisms, an extension of the Bondi formalism to account for a non-vanishing twist in the description of asymptotically-flat spacetimes. We note that, in the above references, the analysis of \cite{Fiorucci:2025twa} is performed in the first order formulation using tetrad variables, and is therefore very close to the spirit of the NP construction presented here.

The outline of this work and of the various results is as follows. We begin in section \ref{sec:tetrad} with a presentation of twisting asymptotically-flat spacetimes in the NP formalism, following and extending the results of \cite{Mao:2024jpt}. For this, we review the geometrical meaning of the twist, and then construct the solution space in terms of the spin coefficients, the Weyl scalars, and the tetrad vectors. We then study the flux-balance laws and the extension of the non-linear NP charges to the case of non-vanishing twist.

Section \ref{sec:metric} is then devoted to the study of the metric formalism. Although one could think that this is obtained straightforwardly from the NP formalism, there are numerous subtleties which need to be worked out in detail. We start by explaining in \ref{sec:metric with twist} how the twist potential sources $g_{ra}\neq0$ in the metric. In \ref{sec:gauge conditions} we then study the gauge conditions. In the tetrad formalism, we solve the NP equations using the gauge conditions $\kappa=\epsilon=\pi=0$. In the metric formalism, we show that this reduces in part to the conditions $g_{ur}=-1$ and $\partial_rg_{ra}=0$, which can equivalently be written as $\Gamma^\mu_{rr}=0$. These conditions are imposed on top of $g_{rr}=0$. The replacement of $g_{ra}=0$ in the standard Bondi gauge by $\partial_rg_{ra}=0$ is what allows to introduce a non-vanishing twist potential. This also explains why there is a residual symmetry associated with the twist potential. In subsection \ref{sec:no log} we study the condition for the absence of logarithmic terms in the metric, and \ref{sec:carroll} presents the Carrollian interpretation of the twist potential as an Ehresmann connection. In \ref{sec:hypersurface} we then explain how the so-called hypersurface equations can be arranged and solved in order to determine the radial expansion of the metric. Subsection \ref{sec:hierarchy} presents the extension of the Bondi hierarchy to the case of a non-vanishing twist potential. This crucial result explains how the Einstein equations can be disentangled and rearranged into hypersurface, evolution, and trivial equations. This Bondi hierarchy is what ultimately enables us to solve and control all of the Einstein equations in a radial expansion near future null infinity. Since certain solutions such as Kerr have a much simpler form in the Bondi gauge with twist, one could hope that this Bondi hierarchy will be useful for the study of black hole perturbation theory in the metric formulation. In \ref{sec:flux-balance} we then exploit the dictionary between the NP and metric formalisms to rewrite the flux-balance laws in tensorial form, thereby providing generalizations of the well-known mass loss and angular momentum loss formul\ae~to the case of non-vanishing twist.

In section \ref{sec:symmetries} we then study the asymptotic symmetries, the transformation laws, and take a brief look at the charges. This reveals in particular that the twist potential is associated with a new asymptotic symmetry parameter, which is the generator of the Carrollian boosts. We then compute the charge associated with this symmetry in the case of a round sphere metric, and show that it is integrable up to a corner term. The transformation laws are particularly interesting in order to understand the reduction to algebraically special solutions. Indeed, \eqref{transformation laws} shows that setting $\Psi_0=\Psi_1=0$ forces the Carroll boost to vanish. This then implies that the full bulk extension of the asymptotic Killing vector truncates and takes an exact form with no radial expansion. This mirrors the resummation of the metric in the algebraically special case.

Algebraically special solutions are studied in section \ref{sec: algebraically special}. Since the tetrad and the metric can now encode a non-vanishing twist potential, the vector $\ell$ can be chosen as a repeated principal null direction and one can set $\Psi_0=\Psi_1=0$ to obtain algebraically special solutions without discarding e.g. Kerr--Taub--NUT. In subsection \ref{sec:AS solution space} we present the construction of the algebraically special solution space, and the resummation of the metric and the tetrad thanks to the twist. In \ref{sec:AS symmetries} we present the truncated form of the BMS--Weyl asymptotic Killing vectors. Finally, subsection \ref{sec:AS type D} is devoted to the study of some examples of Petrov type D spacetimes. In particular, we show that the twist potential enables us to write the supertranslated Kerr--Taub--NUT metric in finite form. We compute the charges of this supertranslated solution, and recover in particular the result of \cite{Hawking:2016msc,Hawking:2016sgy,Compere:2016hzt,Compere:2016jwb} concerning the non-vanishing superrotation charge of supertranslated Schwarzschild. This result is obtained here in finite form without radial dependency. We then explain the relationship between the tetrad with twist potential used to build the algebraically special solutions and the Kinnersley tetrad which is adapted to type D solutions. Using the Kinnersley tetrad, we then finish with a discussion of the Killing tensors of the supertranslated Schwarzschild solution.

We close this work in section \ref{sec:3d} with a study of the three-dimensional case. We first explain how the Sachs optical scalars differ between three- and four-dimensional spacetimes, and in particular the meaning of the three-dimensional ``twist''. Using the NP formalism adapted to three-dimensional spacetimes, we then construct the solution space with non-vanishing cosmological constant. In particular, we keep track of all the integration constants which arise from the resolution of the Einstein equations. The upshot is a solution space parametrized by 8 functions of $(u,\phi)$, as summarized in table \ref{3d solution space table}. This is the largest known (A)dS$_3$ solution space in Bondi gauge.

Our notations and conventions, e.g. for indices and curvature tensors, are gathered in appendix \ref{app:notations}.

\section{Tetrad formalism}
\label{sec:tetrad}

We start by presenting the twisting asymptotically-flat solution space in tetrad variables using the Newman--Penrose (NP) formalism. This will allow to introduce important notations, and also to establish later on a dictionary between the metric and NP formalisms. In this first section we collect and streamline some of the results presented in \cite{Mao:2024jpt}, which we also extend to a more subleading order.

\subsection{Tetrad with a twist}

As the starting point, let us choose the signature to be mostly plus $(-,+,+,+)$ and pick a set of coordinates $x^\mu=(u,r,x^a)$. We then consider the internal metric
\be
\eta_{ij}=
\begin{pmatrix}
0&-1&0&0\\
-1&0&0&0\\
0&0&0&1\\
0&0&1&0
\end{pmatrix},
\ee
and the doubly-null tetrad $e_i=(\ell,n,m,\bar{m})$ formed by the vectors\footnote{At the difference with \cite{Barnich:2019vzx,Mao:2024jpt}, we denote the radial component $m^r$ by $\Omega$ instead of $\omega$ in order to keep the latter for the twist $\omega=\text{Im}(\rho)$ as in \cite{Chandrasekhar:1985kt}.}
\be\label{NP tetrad}
\ell=\partial_r,
\q\q
n=W\partial_u+U\partial_r+X^a\partial_a,
\q\q
m=Z\partial_u+\Omega\partial_r+m^a\partial_a.
\ee
The inverse spacetime metric is then given by
\be\label{inverse metric NP}
g^{\mu\nu}=e_i^\mu e_j^\nu\eta^{ij}=-2\ell^{(\mu}n^{\nu)}+2m^{(\mu}\bar{m}^{\nu)},
\ee
and the vectors satisfy $\ell^\mu n_\mu=-m^\mu\bar{m}_\mu=-1$, with all other contractions vanishing. The tetrad and the metric depend on four real $(W,U,X^a)$ and four complex $(Z,\Omega,m^a)$ functions of $x^\mu$. Using the compact notations introduced below in section \ref{sec:metric with twist}, the cotetrad corresponding to \eqref{NP tetrad} is given in \eqref{cotetrad}.

The novel ingredient of the twisting solution space is the complex scalar $Z$, which plays the role of a twist ``potential'' sourcing the twist $\text{Im}(\rho)\neq0$. In the metric formulation this twist potential becomes a real vector $Z^a=2\text{Re}(\bar{Z}m^a)$ sourcing in particular the $g^{uu}$, $g^{ua}$ and $g_{ra}$ components of the metric. As suggested by the name, the twist potential is more general than the twist since one can have $Z\neq0$ and $g_{ra}\neq0$ with nonetheless a vanishing twist. This is the case for example for the supertranslated Schwarzschild solution studied in section \ref{sec:supertranslated Schwarzschild}. In a slight abuse of language, we will often refer to the twist potential simply as the twist. Importantly, one should note that since $g^{uu}\neq0$ and $g^{ua}\neq0$ the coordinates $x^\mu$ are not Bondi coordinates. Instead, they should be thought of as a type of Eddington--Finkelstein coordinates. In another abuse of language, we will still refer to this gauge with $g_{ra}\neq0$ as a Bondi gauge with non-vanishing twist potential. This is because, as we show here, all the results of the standard Bondi gauge can smoothly be extended to include the twist potential. For the same reason, the gauge with $g_{ra}\neq0$ is called the Carroll covariant Newman--Unti gauge in \cite{Ciambelli:2020eba,Ciambelli:2020ftk,Campoleoni:2022wmf,Mittal:2022ywl,Campoleoni:2023fug,Fiorucci:2025twa} and the Carroll covariant Bondi--Sachs gauge in \cite{Hartong:2025jpp}.

The NP reformulation of the Einstein equations enables us to solve for the components of the tetrad in terms of the spin coefficients and the Weyl scalars, in terms of which one must then identify the free and initial data. Our conventions with mostly plus signature are given in appendix \ref{app:NP}, where we also give the explicit expression for the spin coefficients corresponding to the tetrad \eqref{NP tetrad}. Let us recall that a special role is played by the spin coefficients $(\kappa,\epsilon,\pi)$, which enable to rewrite the parallel transport of the tetrad vectors along $\ell$ as
\bsub
\be
\ell^\mu\nabla_\mu\ell&=(\epsilon+\bar{\epsilon})\ell-\bar{\kappa}m-\kappa\bar{m},\label{l nabla l}\\
\ell^\mu\nabla_\mu n&=-(\epsilon+\bar{\epsilon})n+\pi m+\bar{\pi}\bar{m},\\
\ell^\mu\nabla_\mu m&=(\epsilon-\bar{\epsilon})m+\bar{\pi}\ell-\kappa n.
\ee
\esub
Once the condition $\kappa=0$ is imposed, meaning that $\ell$ describes a congruence of null geodesics, one can always reach a gauge in which $\epsilon=0$ and $\pi=0$ by performing Lorentz transformations of type III and I respectively \cite{Chandrasekhar:1985kt}. In this case $\ell$ is affinely parametrized and the tetrad vectors are covariantly constant along $\ell$. The requirement that $\text{Re}(\epsilon)=0$ is known as the Newman--Unti gauge, and it implies that $r$ is the affine parameter for $\ell$. Importantly, one should note that in order to set\footnote{In section \ref{sec:Kinnersley} we relax this condition and transform to a Kinnersley tetrad with $\pi\neq0$ in order to describe type D solutions more conveniently.} $\pi=0$ it is necessary to have a radial component $\Omega\neq0$ in $m$. When $\kappa=0$ and $\epsilon=0$ we have furthermore that
\be\label{lDl}
\ell_{[\mu}\nabla_\nu\ell_{\rho]}=-2i\text{Im}(\rho)m_{[\mu}\bar{m}_\nu\ell_{\rho]},
\ee
where the anti-symmetrization runs over all three indices. This equation shows that the twist $\text{Im}(\rho)$ measures the failure of the null congruence to be hypersurface-orthogonal. The congruence which we consider in this work will therefore have expansion, shear, and twist. In the case of an affinely parametrized null congruence, we recall that these Sachs optical scalars can be expressed in a $d$-dimensional spacetime as
\bsub\label{optical scalars}
\be
&\text{(expansion)}&\theta&=\f{1}{2}\nabla_\mu\ell^\mu,\\
&\text{(shear)}&|\sigma|^2&=\f{1}{2}(\nabla^\mu\ell^\nu)\nabla_{(\mu}\ell_{\nu)}-\f{1}{2(d-2)}(\nabla_\mu\ell^\mu)^2,\label{optical shear}\\
&\text{(twist)}&\omega^2&=\f{1}{2}(\nabla^\mu\ell^\nu)\nabla_{[\mu}\ell_{\nu]}.
\ee
\esub
When $\kappa=\epsilon=0$, the relationship between these geometrical quantities and the NP spin coefficients given in appendix \ref{app:spin coeffs} is
\be
\theta=-\text{Re}(\rho),
\q\q
|\sigma|^2=\sigma\bar{\sigma},
\q\q
\omega^2=-\text{Im}(\rho)^2.
\ee
Let us now turn to the construction of the solution space.

\subsection{Solution space}
\label{sec:4d NP solution space}

The gauge choice $\kappa=\epsilon=\pi=0$ simplifies drastically the NP equations, which reduce to the expressions listed in appendix \ref{app:4d NP equations}. The advantage of this gauge is that it enables us to determine the radial expansion of the non-vanishing spin coefficients from equations which contain the directional derivative $D=\ell^\mu\partial_\mu=\partial_r$ alone, i.e. without coupling to the other directional derivatives $(\Delta,\delta,\bar{\delta})$. We now present the solution space arising from the resolution of all the NP equations. We start by listing all of the asymptotic expansions before giving their interpretation and explaining the structure of the solution space. We explain in appendix \ref{app:NP resolution} how the NP equations are solved in order to obtain the solution space.

First, the Weyl scalars satisfy peeling and are given by
\bsub\label{Psi}
\be
\Psi_0&=\f{\Psi_0^0}{r^5}+\f{\Psi_0^1}{r^6}+\O(r^{-7}),\label{NP Psi0}\\
\Psi_1&=\f{\Psi_1^0}{r^4}-\f{\Ethb\Psi_0^0+4i\Sigma\Psi_1^0}{r^5}+\O(r^{-6}),\\
\Psi_2&=\f{\Psi_2^0}{r^3}-\f{\Ethb\Psi_1^0+3i\Sigma\Psi_2^0}{r^4}+\O(r^{-5}),\\
\Psi_3&=\f{\Psi_3^0}{r^2}-\f{\Ethb\Psi_2^0+2i\Sigma\Psi_2^0}{r^3}+\O(r^{-4}),\\
\Psi_4&=\f{\Psi_4^0}{r}-\f{\Ethb\Psi_3^0+i\Sigma\Psi_4^0}{r^2}+\O(r^{-3}),
\ee
\esub
where the operator $\Ethb$ defined in \eqref{generalized eth} generalizes the standard spin-weighted derivative $\ethb$ in the presence of a non-vanishing twist potential. The spin weights of various quantities of interest are gathered in table \ref{helicity table} below. In addition to the gauge conditions $\kappa=\epsilon=\pi=0$, the spin coefficients are given by\footnote{We use a convenient book-keeping notation where the coefficients are labelled by their order in the radial expansion. For example the asymptotic shear, often denoted by $\sigma^0$, is here denoted by $\sigma_2$ since it appears at order $r^{-2}$.}
\bsub\label{radial spin coefficients}
\be
\rho&=-\f{1}{r}+\f{i\Sigma}{r^2}+\f{\Sigma^2-\sigma_2\bar{\sigma}_2}{r^3}-\f{i\Sigma\rho_3}{r^4}+\O(r^{-5}),\label{NP rho}\\
\sigma&=\f{\sigma_2}{r^2}-\f{2\sigma_2\rho_3+\Psi_0^0}{2r^4}-\f{\Psi_0^1}{3r^5}+\O(r^{-6}),\label{NP sigma}\\
\alpha&=\f{\alpha_1}{r}+\f{\bar{\alpha}_1\bar{\sigma}_2-i\Sigma\alpha_1}{r^2}-\f{\alpha_1\rho_3}{r^3}+\O(r^{-4}),\\
\beta&=-\f{\bar{\alpha}_1}{r}-\f{\alpha_1\sigma_2+i\Sigma\bar{\alpha}_1}{r^2}+\f{2\bar{\alpha}_1\rho_3-\Psi_1^0}{2r^3}+\O(r^{-4}),\\
\tau&=\f{\tau_1}{r}-\f{\sigma_2\bar{\tau}_1+i\Sigma\tau_1}{r^2}-\f{2\tau_1\rho_3+\Psi_1^0}{2r^3}+\O(r^{-4}),\\
\lambda&=\f{\lambda_1}{r}-\f{\bar{\sigma}_2\mu_1+i\Sigma\lambda_1}{r^2}+\O(r^{-3}),\\
\mu&=\f{\mu_1}{r}-\f{\sigma_2\lambda_1-i\Sigma\mu_1+\Psi_2^0}{r^2}+\O(r^{-3}),\label{NP mu}\\
\gamma&=\gamma_0+\f{\bar{\alpha}_1\bar{\tau}_1-\alpha_1\tau_1}{r}+\f{2\alpha_1\sigma_2\bar{\tau}_1-2\bar{\alpha}_1\bar{\sigma}_2\tau_1+2i\Sigma(\alpha_1\tau_1+\bar{\alpha}_1\bar{\tau}_1)-\Psi_2^0}{2r^2}+\O(r^{-3}),\\
\nu&=\nu_0-\f{\lambda_1\tau_1+\mu_1\bar{\tau}_1+\Psi_3^0}{r}+\O(r^{-2}),
\ee
\esub
where
\bsub\label{NP leading definitions}
\be
\Sigma&=\text{Im}(\Eth\bar{L}),\label{NP Sigma}\\
\rho_3&=\Sigma^2-\sigma_2\bar{\sigma}_2,\\
\alpha_1&=-\f{1}{2}D_a\bar{m}^a_1-\gamma_0\bar{L},\label{NP alpha1}\\
\tau_1&=-(\partial_u+2\gamma_0)L,\label{NP tau1}\\
\lambda_1&=(\partial_u+2\gamma_0)\bar{\sigma}_2-(\Ethb-\bar{\tau}_1)\bar{\tau}_1,\\
\mu_1&=2i\Sigma\gamma_0-\mu_R,\\
\mu_R&=\Eth\alpha_1+\Ethb\bar{\alpha}_1,\label{NP mu R}\\
\gamma_0&=\f{1}{4}\partial_u\ln\sqrt{q},\label{NP gamma0}\\
\nu_0&=2(\Ethb-\bar{\tau}_1)\gamma_0,\\
\Psi_3^0&=-\Eth\lambda_1+\Ethb\mu_1+2i\Sigma\nu_0,\label{Psi30}\\
\Psi_4^0&=(\Ethb-\bar{\tau}_1)\nu_0-(\partial_u+4\gamma_0)\lambda_1,\label{Psi40}\\
\Psi_2^0-\bar{\Psi}_2^0&=\bar{\lambda}_1\bar{\sigma}_2-\lambda_1\sigma_2+\Ethb\Omega_1-\Eth\bar{\Omega}_1+2i\Sigma(U_0-\mu_R).\label{NP Im Psi20}
\ee
\esub
Finally, the components of the tetrad are given by
\bsub\label{NP tetrad solution}
\be
W&=1-\f{2\text{Re}(L\bar{\tau}_1)}{r}+\f{2\text{Re}\big(L(\bar{\sigma}_2\tau_1-i\Sigma\bar{\tau}_1)\big)}{r^2}+\O(r^{-3}),\\
U&=-2\gamma_0r+\big(\Ethb-\bar{\tau}_1\big)\tau_1+\bar{\mu}_1-i(\partial_u+2\gamma_0)\Sigma-\f{\text{Re}\big(\Psi_2^0+2\tau_1\bar{\Omega}_1\big)}{r}+\O(r^{-2}),\label{U expansion}\\
X^a&=-\f{2\text{Re}(m^a_1\bar{\tau}_1)}{r}+\f{2\text{Re}\big(m^a_1(\bar{\sigma}_2\tau_1-i\Sigma\bar{\tau}_1)\big)}{r^2}+\f{\text{Re}\big(m^a_1(\bar{\Psi}_1^0+6\rho_3\bar{\tau}_1)\big)}{3r^3}+\O(r^{-4}),\label{NP solution for X}\\
Z&=\f{L}{r}+\f{i\Sigma L-\sigma_2\bar{L}}{r^2}-\f{\rho_3L}{r^3}+\O(r^{-4}),\label{NP solution for Z}\\
\Omega&=\f{\Ethb\sigma_2-i(\Eth-2\tau_1)\Sigma}{r}+\f{2i\Sigma\Omega_1-2\sigma_2\bar{\Omega}_1-\Psi_1^0}{2r^2}+\O(r^{-3}),\label{NP solution for Omega}\\
m^a&=\f{m^a_1}{r}+\f{i\Sigma m^a_1-\sigma_2\bar{m}^a_1}{r^2}-\f{\rho_3m^a_1}{r^3}+\f{\Psi_0^0\bar{m}^a_1-6\rho_3m^a_2}{6r^4}+\O(r^{-5}).\label{NP solution for ma}
\ee
\esub
The subleading orders of $(W,U,X^a)$ and $(Z,\Omega,m^a)$ can be found by solving respectively the components $(u,r,x^a)$ of \eqref{NP1j} and \eqref{NP1k}. Following standard terminology \cite{Stephani:2003tm}, we have denoted the leading term in the radial expansion of the twist potential $Z$ by $L(u,x^a)$ instead of $Z_1(u,x^a)$.

Let us now explain the various notations and the structure of this solution space. In the expansion \eqref{NP solution for ma} of the angular part of $m$, the leading term $m^a_1$ enables us to reconstruct the leading celestial metric as $q^{ab}=2m_1^{(a}\bar{m}_1^{b)}$. At leading order, the angular components of equation \eqref{NP2g} initially yield the constraints
\be\label{ma and gamma0 constraints}
(\partial_u+2\bar{\gamma}_0)m^a_1=0,
\q\q
(\partial_u+2\gamma_0)\bar{m}^a_1=0,
\q\q
\partial_uq_{ab}-2(\gamma_0+\bar{\gamma}_0)q_{ab}=0.
\ee
This implies that the time dependency of the frame must be contained in a conformal factor, which we can choose to be of the form $e^{\varphi+i\psi}$ with $(\varphi,\psi)$ real functions of $(u,x^a)$. We then have $2\gamma_0=i\psi-\varphi$, and in principle $\gamma_0$ and $\bar{\gamma}_0$ therefore parametrize two independent degrees of freedom. However, since the phase parametrized by $\psi$ will never appear in the metric and cannot play a role in the reconstruction of the solutions, we choose for simplicity the gauge in which\footnote{This choice fixes the imaginary part of the Weyl rescalings which can be revealed in the tetrad formulation \cite{Barnich:2016lyg,Barnich:2019vzx,Mao:2024jpt}, but does not affect the symmetry or charge content of the metric formulation.} $\psi=0$. We then obtain that $\gamma_0$ is real and given by \eqref{NP gamma0}. For practical calculations, especially when constructing the dictionary between NP and metric quantities, an explicit parametrization of $m^a_1$ can be chosen in the form
\be\label{frame m1}
m^a_1=\sqrt{\f{q_{\theta\theta}}{2q}}\left(\f{\sqrt{q}+iq_{\theta\phi}}{q_{\theta\theta}}\,\delta^a_\theta-i\delta^a_\phi\right).
\ee
We allow the metric $q_{ab}$ to be arbitrary in order to encompass the BMS--Weyl boundary conditions \cite{Barnich:2016lyg,Freidel:2021fxf}, and the associated covariant derivative $D_a$ has been used in order to write the leading coefficient $\alpha_1$ \eqref{NP alpha1}. With our choice $\psi=0$, the time dependency of this metric is constrained by $(\partial_u-4\gamma_0)q_{ab}=0$.

Then, it is important to note that the solution space contains two types of data in terms of which all the quantities of interest (i.e. the Weyl scalars, the spin coefficients and the tetrad) are expanded. The data has either unconstrained or constrained dependency on $u$.
\begin{itemize}
\item[$\bullet$] \textbf{Unconstrained data.} The functions $(\sigma_2,\gamma_0,L)$ of $(u,x^a)$ have an arbitrary dependency on time $u$. The complex scalar $\sigma_2$ encodes the two degrees of freedom of the asymptotic shear. The real scalar $\gamma_0$ contains the arbitrary time dependency of the leading transverse conformal factor. Finally, the complex scalar $L$ entering in the expansion \eqref{NP solution for Z} parametrizes the twist potential. It sources the twist $\Sigma=\text{Im}(\Eth\bar{L})$ or equivalently $\text{Im}(\rho)\neq0$ in \eqref{NP rho}.
\item[$\bullet$] \textbf{Constrained data.} The functions $\big(\text{Re}(\Psi_2^0),\Psi_1^0,\Psi_0^n\big)$ of $(u,x^a)$ have a dependency on $u$ constrained by evolution equations which we list below. The real part $\text{Re}(\Psi_2^0)$ represents the mass aspect and the complex scalar $\Psi_1^0$ contains the two degrees of freedom of the angular momentum aspect. Finally, for each $n\geq0$ the complex scalar $\Psi_0^n$ appearing in the expansion of $\Psi_0$ \eqref{NP Psi0} and $\sigma$ \eqref{NP sigma} contains the two degrees of freedom of a transverse-traceless tensor on the sphere. These are the two-dimensional tensors which parametrize the expansion of the transverse metric $\gamma_{ab}=2m_{(a}m_{b)}$ at order $r^{-1}$ and below. Note that one could actually add $m^a_1$ to the list of constrained data since its time dependency is constrained by \eqref{ma and gamma0 constraints}.
\end{itemize}

Let us now elaborate on the expansion of the transverse metric and on some hidden assumptions which went into the construction of the solution space. At order $r^2$, the metric $\gamma_{ab}=2m_{(a}m_{b)}$ is parametrized by $q_{ab}$. At order $r$ it is parametrized by the asymptotic shear $\sigma_2$, or equivalently the trace-free tensor $C_{ab}$ in the metric notation \eqref{gamma expansion}. At order $r^0$ it contains no independent degree of freedom, as can be seen from the absence of order $r^{-3}$ in $\sigma$ \eqref{NP sigma}. This condition, which is a choice that was enforced by hand in the resolution of the NP equations, ensures that there are no logarithmic branches appearing. We come back to this condition when studying the metric solution space in section \ref{sec:no log} below. At order $r^{-1}$ the metric $\gamma_{ab}$ is parametrized by $\Psi_0^0$, and similarly the order $r^{-n}$ involves $\Psi_0^{n-1}$. Note that in the expansion \eqref{NP Psi0} for $\Psi_0$ we have also chosen by hand to discard terms in powers of $\ln r$. An additional condition which was enforced by hand is $\text{Re}(\rho_2)=0$ in \eqref{NP rho}. This amounts to fixing the origin of the affine parameter \cite{Barnich:2011ty,Barnich:2012nkq,Geiller:2022vto}. In metric language, this means that $C_{ab}$ is indeed trace-free. Finally, let us note that the resolution of \eqref{NP1j} allows for solution of the form $W=W_0(u,x^a)+\O(r^{-1})$ and $X^a=X^a_0(u,x^b)+\O(r^{-1})$, and that have therefore discarded by hand the two free functions by setting $W_0=1$ and $X^a_0=0$. This choice simplifies drastically the structure of the solution space. We illustrate the role of $X^a_0$ with the C-metric in appendix \ref{app:C-metric}, and refer the reader to \cite{Geiller:2022vto} for a metric solution space where these additional functions are included.

\begin{table}[h]
\quad\begin{tabular}{|c|c|c|c|c|c|c|c|c|c|c|c|c|c|c|c|c|c|c|c|c|}
\hline
& $\Psi_0$ & $\Psi_1$ & $\Psi_2$ & $\Psi_3$ & $\Psi_4$ & $\partial_u$ & $\Eth$ & $\rho$ & $\sigma$ & $\alpha$ & $\beta$ & $\tau$ & $\lambda$ & $\mu$ & $\gamma$ & $\nu$ & $Z$ & $\Omega$ & $\Sigma$ & $m$ \\ \hline
$s$ & 2 & 1 & 0 & $-1$ & $-2$ & 0 & 1 & 0 & 2 & $-1$ & 1 & 1 & $-2$ & 0 & 0 & $-1$ & 1 & 1 & 0 & 1\\ \hline
\end{tabular}
\caption{Spin weights of various quantities appearing in the NP solution space.}
\label{helicity table}
\end{table}

\subsection{Evolution equations}
\label{sec:4d NP evolution equations}

On top of the expansions given above, there are evolution equations obeyed by some of the Weyl scalars and of the spin coefficients. First we have
\bsub
\be
(\partial_u+2\gamma_0)\alpha_1&=-(\Ethb-\bar{\tau}_1)\gamma_0,\label{alpha1 evolution}\\
(\partial_u+4\gamma_0)\mu_1&=(\Eth-\tau_1)\nu_0,\label{mu1 evolution}\\
(\partial_u+6\gamma_0)\Psi_3^0&=(\Eth-\tau_1)\Psi_4^0.\label{Psi30 evolution}
\ee
\esub
These are tautological relations which can be shown to hold automatically once the above expressions for the spin coefficients and the Weyl scalars are inserted. Then, by expanding the Bianchi identities \eqref{NP2l}-\eqref{NP2m}-\eqref{NP2n} at leading order we find
\bsub\label{NP evolution equations}
\be
(\partial_u+6\gamma_0)\Psi_0^0&=(\Eth-4\tau_1)\Psi_1^0+3\sigma_2\Psi_2^0,\label{EOM Psi00}\\
(\partial_u+6\gamma_0)\Psi_1^0&=(\Eth-3\tau_1)\Psi_2^0+2\sigma_2\Psi_3^0,\label{EOM Psi10}\\
(\partial_u+6\gamma_0)\Psi_2^0&=(\Eth-2\tau_1)\Psi_3^0+1\sigma_2\Psi_4^0.\label{EOM Psi20}
\ee
\esub
These are genuine evolution equations, apart from the imaginary part of \eqref{EOM Psi20} which follows automatically from the definition of the so-called dual mass $\text{Im}(\Psi_2^0)$ given in \eqref{NP Im Psi20}. In section \ref{sec:flux-balance} below we give the metric form of these evolution equations.

At this stage we have solved all of the NP equations of appendix \ref{app:4d NP equations} at leading order. It is then natural to ask what is left to be extracted from the subleading equations. There are three types of information appearing at subleading order:
\begin{itemize}
\item \textbf{Subleading radial expansions.} A first subset of the NP equations leads to the subleading expansions of the spin coefficients, the tetrad vectors, or the Weyl scalars. For example, solving \eqref{NP1a} at order $r^{-6}$ gives the next term in the expansion \eqref{NP rho}, which is $6\rho_5=\Psi_0^0\bar{\sigma}_2+\bar{\Psi}_0^0\sigma_2-6\rho_3^2$.
\item \textbf{Tautological evolution equations.} A second subset of the NP equations provides tautological evolution equations which are automatically satisfied when inserting the various coefficients in the expansions of the spin coefficients, the tetrad, or the Weyl scalars. For example, expanding \eqref{NP2d} at order $r^{-2}$ gives an evolution equation for $\mu_2$ which is automatically satisfied once the expression for $\mu_2$ given in \eqref{NP mu} is inserted and all the relations between the data are taken into account.
\item \textbf{Evolution equations for $\boldsymbol{\Psi_0^n}$.} Finally, the NP Bianchi identity \eqref{NP2l} leads to an infinite amount of independent evolution equations for the coefficients $\Psi_0^n$ entering the expansion of the Weyl scalar $\Psi_0$. For example, \eqref{EOM Psi00} is found by expanding \eqref{NP2l} at order $r^{-5}$, while the expansion at order $r^{-6}$ gives the evolution equation \eqref{EOM Psi01} for $\Psi_0^1$. Note that these subleading evolution equations may also be required in order to solve the above-mentioned tautological evolution equations.
\end{itemize}

This discussion of the three subsets of NP equations shows that all the field equations are under control, and that the subleading structure of the solution space only contains independent evolution equations arising from the Bianchi identity \eqref{NP2l}. However, an explicit direct proof of this structure is not available (to the best of our knowledge) in the NP formalism. This is to be contrasted with the metric formulation, where the so-called Bondi hierarchy \cite{Bondi:1962px,Tamburino:1966zz,Barnich:2010eb} actually guarantees that the Einstein equations can be split into hypersurface equations, evolution equations for the mass and angular momentum aspects, and an infinite tower of evolution equations for the subleading terms in the angular metric (i.e. the metric analogues of the $\Psi_0^n$'s). We demonstrate the existence of such a Bondi hierarchy of Einstein equations in the presence of a non-vanishing twist potential in section \ref{sec:hierarchy} below. It is the knowledge of the equivalence between the NP and metric formulations which enables us to conclude, as we did above, that all the equations are under control in the NP formulation even in the absence of a construction analogous to the metric Bondi hierarchy.

Finally, let us comment on the subleading evolution equations appearing when expanding \eqref{NP2l} further. The structure of these equations was investigated in the case of vanishing twist in \cite{Geiller:2024bgf} (see also \cite{Grant:2021hga,Compere:2022zdz}, in the metric formulation). For the sake of curiosity, we can write here the first subleading equation. It takes the form
\be\label{EOM Psi01}
(\partial_u+8\gamma_0)(\Psi_0^1+5i\Sigma\Psi_0^0)
&=-(\Ethb-\bar{\tau}_1)\Big((\Eth-5\tau_1)\Psi_0^0+4\sigma_2\Psi_1^0\Big)\cr
&=-(\ethb+2\gamma_0L)\Big((\Eth-5\tau_1)\Psi_0^0+4\sigma_2\Psi_1^0\Big)-\partial_u\Big(\bar{L}(\Eth-5\tau_1)\Psi_0^0+4\bar{L}\sigma_2\Psi_1^0\Big),\q\q
\ee
where the combination $\Ethb-\bar{\tau}_1$ on the first line is precisely such that a total $\partial_u$ appears on the second line. When $\gamma_0=0$, i.e. for a time-independent leading sphere metric, this result means that we can define a non-linear NP charge as
\be
Q_\text{NP}\coloneqq\Psi_0^1+5i\Sigma\Psi_0^0+\bar{L}\Big((\Eth-5\tau_1)\Psi_0^0+4\sigma_2\Psi_1^0\Big).
\ee
Since the evolution of this charge is a total $\ethb$, one can obtain 10 exactly conserved and real quantities by smearing $Q_\text{NP}$ with the appropriate spin-weighted spherical harmonics. This extends the result \cite{Newman:1965ik,Newman:1968uj} of Newman and Penrose to the twisting asymptotically-flat spacetimes under consideration. Presumably, one can also obtain from a similar construction for $\Psi_0^{n>1}$ an infinite amount of linearly conserved NP charges in the case of a non-vanishing twist. An interesting question is whether in the presence of twist the subleading evolution equations for $\Psi_0^n$ can be used to identify a $w_{1+\infty}$ algebra higher spin charges, as was done in the absence of twist in \cite{Freidel:2021ytz,Compere:2022zdz,Geiller:2024bgf,Cresto:2024mne,Cresto:2024fhd}.

The results above illustrate the power of using the NP formulation to solve the asymptotic Einstein equations. In section \ref{sec:3d} we apply the same procedure to build the solution space of three-dimensional gravity with $\Lambda\neq0$ and $g_{r\phi}\neq0$. Now that we have the NP solution space at hand, we turn to its construction in the metric formalism.

\section{Metric formalism}
\label{sec:metric}

The NP formalism presented above provides a deceptively simple way of building the twisting asymptotically-flat solution space and expanding it to arbitrary subleading order. However, it obscures by construction the role of the various metric Einstein field equations and the metric interpretation of the various gauge fixing conditions. In the absence of twist, the metric Einstein equations in Bondi gauge can be solved following the Bondi hierarchy \cite{Bondi:1962px,Tamburino:1966zz,Barnich:2010eb}. In this section we explain how this hierarchy extends to twisting asymptotically-flat spacetimes.

\subsection{Metric with a twist}
\label{sec:metric with twist}

In order to write the metric \eqref{inverse metric NP} in a compact form, let us first use the complex angular dyad $m^a$ to define
\be
\gamma^{ab}\coloneqq2m^{(a}\bar{m}^{b)},
\q\q
\eps^{ab}\coloneqq-2im^{[a}\bar{m}^{b]}=\f{1}{\sqrt{\gamma}}\epsilon^{ab},
\ee
where $\epsilon^{ab}$ is the Levi--Civita symbol satisfying $\epsilon^{\theta\phi}=+1$, and where $\gamma\coloneqq\det(\gamma_{ab})$ with $\gamma_{ab}$ the matrix inverse of $\gamma^{ab}$. We can then introduce
\be\label{definition a vectors}
\Omega^a\coloneqq2\text{Re}(\bar{\Omega}m^a),
\q\q
Z^a\coloneqq2\text{Re}(\bar{Z}m^a),
\q\q
\widetilde{\Omega}^a\coloneqq2\text{Im}(\bar{\Omega}m^a),
\q\q
\widetilde{Z}^a\coloneqq2\text{Im}(\bar{Z}m^a),
\ee
and use the angular metric to lower the indices and denote
\be
m_a\coloneqq\gamma_{ab}m^b,
\q\q
\Omega_a\coloneqq\gamma_{ab}\Omega^b,
\q\q
Z_a\coloneqq\gamma_{ab}Z^b,
\q\q
X_a\coloneqq\gamma_{ab}X^b.
\ee
Note that if we denote $\eps_{ab}\coloneqq\gamma_{ac}\gamma_{bd}\eps^{cd}$, then the matrix inverse of $\eps^{ab}$ is $(\eps^{ab})^{-1}=-\eps_{ab}$. With this notation the dual vectors are given by
\be
\widetilde{\Omega}^a=\eps^{ab}\Omega_b,
\q\q
\widetilde{Z}^a=\eps^{ab}Z_b,
\ee
and we have the contractions
\be
Z=Z_am^a,
\q\quad
\bar{Z}=Z_a\bar{m}^a
\q\quad
2Z\bar{Z}=Z_aZ^a,
\q\quad
2\text{Re}(\Omega\bar{Z})=\Omega_aZ^a,
\q\quad
2\text{Im}(\Omega\bar{Z})=\Omega_a\widetilde{Z}^a.
\ee
For convenience we can also turn the real-valued vector $X^a$ into a complex scalar by defining $X\coloneqq X_am^a$, so that it then satisfies $X^a=2\text{Re}(\bar{X}m^a)$ and $\widetilde{X}^a=2\text{Im}(\bar{X}m^a)$ as well. With these notations, the inverse spacetime metric \eqref{inverse metric NP} is then given by
\be
g^{\mu\nu}=
\begin{pmatrix}
Z_aZ^a&-W+Z_a\Omega^a&Z^a\\
-W+Z_a\Omega^a&-2U+\Omega_a\Omega^a&-X^a+\Omega^a\\
Z^b&-X^b+\Omega^b&\gamma^{ab}\\
\end{pmatrix}.
\ee
The metric takes the form
\be\label{metric}
g_{\mu\nu}=
\begin{pmatrix}
g_{ur}^2\Big(2U+X_a\big(X^a-2\Omega^a\big)\Big)&-\big(W-Z_aX^a\big)^{-1}&g_{ur}\big(X_a-\Omega_a\big)-g_{uu}Z_a\\
-\big(W-Z_aX^a\big)^{-1}&0&-g_{ur}Z_a\\
g_{ur}\big(X_b-\Omega_b\big)-g_{uu}Z_b&-g_{ur}Z_b&\gamma_{ab}+g_{uu}Z_aZ_b+2g_{ur}\big(\Omega_{(a}-X_{(a}\big)Z_{b)}\\
\end{pmatrix},
\ee
and its determinant is
\be
\sqrt{-g}=\sqrt{\gamma}\,\big(W-Z_aX^a\big)^{-1}=-g_{ur}\sqrt{\gamma}.
\ee
One should note in particular that $g_{ab}\neq\gamma_{ab}=2m_{(a}\bar{m}_{b)}$ when $Z\neq0$. Again, we recall that the spin coefficients corresponding to \eqref{NP tetrad} are given in \ref{app:spin coeffs} using the above notations. Using the NP results \eqref{NP tetrad solution} and the notations introduced above, one can immediately write down the form of the metric \eqref{metric} which solves the asymptotic Einstein equations. Here our goal however is to explain how this solution can be found starting from \eqref{metric} and solving the metric Einstein equations.

The metric \eqref{metric} differs from the ``standard'' Bondi form in two ways. First, the presence of the twist (potential) $Z_a$ implies that $g_{ra}\neq0$, although the gauge condition $g_{rr}=0$ still holds. Importantly, this relaxation of the Bondi gauge does not correspond to naively setting $g_{ra}\neq0$, since the twist $Z_a$ also enters crucially all the other components of the metric. This dependency on $Z_a$ is inherited directly from the form \eqref{NP tetrad} of the tetrad. Second, one should note that the metric also inherits from the tetrad a dependency on the vector $\Omega_a$, even in the case where $Z_a=0$. This is rather puzzling in view of the Bondi hierarchy. Indeed, the latter states that there are only three sets of hypersurface equations, namely the Einstein equations $G_{rr}=0$, $G_{ra}=0$, and $G_{ru}=0$, and that these equations determine the radial expansion of $W$, $X^a$, and $U$ respectively\footnote{This is known to hold when $Z_a=0$, but we show in section \ref{sec:hierarchy} that it also holds in the presence of a non-vanishing twist.}. There is therefore no remaining hypersurface equation which can determine the radial expansion of $\Omega_a$. There are two ways around this. One can either determine $\Omega_a$ by requiring that $\pi=0$ or, alternatively, if one wants to avoid referring to NP spin coefficients in the metric formalism, one should reabsorb $\Omega_a$ in a redefinition of $(W,U,X^a)$. This turns out to be the most convenient choice for our purposes, and we shall therefore redefine
\be\label{hat variables}
\hat{X}^a\coloneqq X^a-\Omega^a,
\q\q
\hat{W}\coloneqq W-Z_a\Omega^a,
\q\q
\hat{U}\coloneqq U-\f{1}{2}\Omega_a\Omega^a.
\ee
With these redefinitions, the metric becomes
\be\label{hat metric}
g_{\mu\nu}=
\begin{pmatrix}
g_{ur}^2\big(2\hat{U}+\hat{X}_a\hat{X}^a\big)&-\big(\hat{W}-Z_a\hat{X}^a\big)^{-1}&g_{ur}\hat{X}_a-g_{uu}Z_a\\
-\big(\hat{W}-Z_a\hat{X}^a\big)^{-1}&0&-g_{ur}Z_a\\
g_{ur}\hat{X}_b-g_{uu}Z_b&-g_{ur}Z_b&\gamma_{ab}+g_{uu}Z_aZ_b-2g_{ur}\hat{X}_{(a}Z_{b)}\\
\end{pmatrix},
\ee
and the corresponding spin coefficients in \ref{app:spin coeffs} can be found by setting $\Omega=0$ and $(W,U,X^a)=(\hat{W},\hat{U},\hat{X}^a)$. We can now in principle solve the Einstein equations to determine the radial expansion of $(\hat{W},\hat{U},\hat{X}^a)$. Before doing so however, we need to discuss the gauge conditions and the expansion of the transverse metric.

\subsection{Gauge conditions}
\label{sec:gauge conditions}

In the previous section we have solved the NP equations with the gauge conditions $\kappa=\epsilon=\pi=0$. This resolution has produced the tetrad with components \eqref{NP tetrad solution}, and this tetrad leads to the metric \eqref{metric}. We now ask the reverse question, namely, what are the gauge conditions to be imposed on the metric, in addition to $g_{rr}=0$, so that the resolution of the Einstein equations leads to a solution space equivalent to that obtained with the NP formalism? To answer this question, we can be inspired by the metric expression of the gauge conditions imposed on the spin coefficients. More precisely, we view the tetrad as fixed once and for all (and therefore its local Lorentz gauge freedom is fixed). This tetrad leads to spin coefficients whose explicit metric expressions are given in appendix \ref{app:spin coeffs}. Form these expressions, we can then read the conditions which must be satisfied by the metric in order to recover the gauge conditions which have been imposed on the spin coefficients in the previous section.

So far the metric \eqref{hat metric} only satisfies the gauge condition $g_{rr}=0$, and we have $g_{ra}\neq0$ because of the non-vanishing twist $Z_a$. Since the presence of the twist also implies a priori that $\kappa\neq0$ from \eqref{NP kappa}, it is natural to replace the standard Bondi gauge conditions $g_{ra}=0$ by the two real conditions $\kappa=0$. This then guarantees that $\ell$ forms a congruence of null geodesics. These conditions can be rewritten as $\partial_rZ_a=0$, and therefore have the advantage of fixing the radial expansion of $Z_a$. The solution is given by
\be\label{gauge Z=L}
Z_a(u,r,x^b)=L_a(u,x^b),
\ee
where $L_a$ is now the free vector which will parametrize the twist. It is related to $L$ in the expansion \eqref{NP solution for Z} of $Z$ by $L=L_am^a_1$. Importantly, in what follows the angular index of $L_a$ will be raised with the leading metric $q_{ab}$. More generally, from now on $q_{ab}$ will be used to lower and raise indices of all the quantities which are $r$-independent.

In addition to the three gauge conditions $g_{rr}=\partial_rZ_a=0$, we can now set a fourth condition by an appropriate choice of the radial coordinate $r$. In the absence of twist, this is usually achieved with the Newman--Unti (NU) or Bondi--Sachs (BS) gauge conditions \cite{Geiller:2022vto}. Here, since we want to establish a dictionary with the NP formalism presented above, it is natural to choose the NU gauge condition $\text{Re}(\epsilon)=0$. In metric notations, \eqref{NP epsilon} tells us that this condition is $\partial_r\hat{W}-Z_a\partial_r\hat{X}^a=0$. This is solved by
\be\label{gauge for W}
\hat{W}=\hat{W}_0+L_a\hat{X}^a=1+L_a\hat{X}^a,
\ee
where we have used \eqref{gauge Z=L} and set the integration constant to $\hat{W}_0(u,x^a)=1$. This choice amounts to fixing part of the induced boundary metric on $\I^+$ (it corresponds to $\beta_0=0$ in the notations of \cite{Compere:2019bua,Geiller:2022vto}). One can then see from \eqref{hat metric} that with this resolution of the NU gauge condition\footnote{In the literature the terminology ``Newman--Unti gauge'' is often used indistinguishably to mean $g_{ur}=-1$ and the fact that $\ell$ is affinely parametrized. However this last condition i.e. $\text{Re}(\epsilon)=0$, is clearly weaker than the first one. In the absence of twist it requires only that $\partial_rg_{ur}=0$, which therefore leaves room for the radial integration constant $W_0(u,x^a)$.} we have $g_{ur}=-1$. It is important to note however that this condition $g_{ur}=-1$ is the result of the NU condition $\text{Re}(\epsilon)=0$ and $\hat{W}_0=1$ combined with \eqref{gauge Z=L}. Unlike in the absence of twist, it does not follow from $\text{Re}(\epsilon)=0$ and $\hat{W}_0=1$ alone.

Geometrically, the three gauge conditions which we have imposed on top of $g_{rr}=0$ can be rewritten simply as $\ell^\mu\nabla_\mu\ell^\alpha=\Gamma^\alpha_{rr}=0$. Although this is a four-dimensional equation, \eqref{hat metric} and $\ell=\partial_r$ can be used to show that $(\ell^\mu\nabla_\mu\ell)_r$ actually vanishes identically because $\ell_\mu\Gamma^\mu_{rr}=0$, so that imposing $\ell^\mu\nabla_\mu\ell=0$ only fixes three non-trivial conditions. These three conditions are precisely solved by \eqref{gauge Z=L} and \eqref{gauge for W}. Explicitly, one can check that before imposing the gauge conditions we have
\be
\Gamma^u_{rr}=-g_{ur}Z^a\partial_rZ_a,
\q\q
\Gamma^r_{rr}=2\text{Re}(\epsilon),
\q\q
\Gamma^a_{rr}=-g_{ur}\gamma^{ab}\partial_rZ_b,
\ee
which shows that $\Gamma^r_{rr}=0$ is the NU gauge condition. One can also note in passing that
\be
\Gamma^A_{Ar}=\partial_r\ln\sqrt{\gamma}-g_{ur}\hat{X}^a\partial_rZ_a,
\ee
where $x^A=\{u,x^a\}$ are the coordinates on $\I^+$. When \eqref{gauge Z=L} is satisfied, this means that setting $\Gamma^A_{Ar}$ to $2/r$ implements the BS determinant condition. This is the choice which was made in \cite{Hartong:2025jpp}. Finally, let us note that when the two gauge conditions \eqref{gauge Z=L} and \eqref{gauge for W} are imposed, we find from \eqref{full rho} and \eqref{lDl} that
\be
\text{Im}(\rho)=\f{\sqrt{q}}{\sqrt{\gamma}}\Sigma,
\q\q
\epsilon^{\sigma\mu\nu\rho}\ell_\mu\nabla_\nu\ell_\rho\partial_\sigma=2\sqrt{q}\,\Sigma\,\partial_r,
\ee
where $\epsilon^{\sigma\mu\nu\rho}$ is the Levi--Civita symbol, and $\Sigma$ is given by \eqref{NP Sigma} in NP form and by \eqref{metric Sigma} in metric form. The radial expansion of the first expression reproduces consistently the imaginary part of \eqref{NP rho}. This also shows that in BS gauge $\text{Im}(\rho)$ has a finite radial expansion since the BS determinant condition implies that $\sqrt{\gamma}=r^2\sqrt{q}$.

Let us now discuss how the NU gauge condition affects the form of the angular metric. In order to solve the asymptotic Einstein equations with the metric \eqref{hat metric}, we need to prescribe an expansion for the transverse metric $\gamma_{ab}$. We choose
\be\label{gamma expansion}
\gamma_{ab}=r^2q_{ab}+rC_{ab}+\left(\f{1}{2}q_{ab}D+D_{ab}\right)+\sum_{n=1}^\infty\f{1}{r^n}\left(\f{1}{2}q_{ab}E^n+E^n_{ab}\right),
\ee
where the subleading terms after the shear $C_{ab}$ are written in terms of a trace and trace-free part with respect to the leading metric $q_{ab}$. This leading metric is constrained by the Einstein equation $G^\tf_{ab}|_{\O(r)}=0$ to satisfy $(\partial_u-4\gamma_0)q_{ab}=0$, implying that the time dependency of $q_{ab}$ is contained only in its conformal factor. As explained in detail in \cite{Geiller:2022vto,Geiller:2024amx}, the trace parts in \eqref{gamma expansion} are determined by the choice of radial coordinate. In BS gauge they are fixed by the determinant condition $\sqrt{\gamma}=r^2\sqrt{q}$, which at leading order implies in particular that $C_{ab}$ is trace-free. In NU gauge, which is the case of interest for us, the traces are determined instead by matching the gauge condition \eqref{gauge for W} with the solution for $\hat{W}$ obtained by solving the Einstein equation $G_{rr}=0$. Importantly, in NU gauge this leave the freedom of including a trace part in $C_{ab}$, even in the presence of a non-vanishing twist. In NP language this trace is encoded in $\text{Re}(\rho_2)$ \cite{Barnich:2012nkq,Barnich:2011ty}. Since in \eqref{NP rho} we have chosen to set this trace to zero, we also choose $C_{ab}$ in \eqref{gamma expansion} to be trace-free. We give below in \eqref{trace D} the expression for the trace $D$, which is obtained by matching $\hat{W}_2=L_a\hat{X}^a_2$ derived from \eqref{gauge for W} with the expression for $\hat{W}_2$ derived from $G_{rr}|_{\O(r^{-4})}=0$. Similarly, at order $r^{-n}$ the trace $E^n$ can be isolated by matching $\hat{W}_{n+2}=L_a\hat{X}^a_{n+2}$ with the expression for $\hat{W}_{n+2}$ obtained by solving $G_{rr}|_{\O(r^{-{(n+4)}})}=0$.

\subsection{No-log condition}
\label{sec:no log}

Let us now discuss the condition for the absence of logarithmic terms $\ln r$ in the solution space. Such logarithmic terms may arise from two sources \cite{Chrusciel:1993hx,Geiller:2024ryw}. First, they can be introduced by hand in the transverse metric, which we have excluded with our choice of expansion \eqref{gamma expansion}, or equivalently with the smooth expansion in \eqref{NP Psi0}. Second, they start to appear when solving $G_{ra}=0$ if the trace-free part $D_{ab}$ is chosen to contain independent degrees of freedom \cite{1985FoPh...15..605W,Geiller:2024ryw}. In the NP solution space given above we have excluded such degrees of freedom by hand by setting $\sigma_3=0$ in \eqref{NP sigma}. The spin coefficient $\sigma$ therefore jumps from $\sigma_2$ containing $C_{ab}$ to $\sigma_4$ containing $E^1_{ab}$ in $\Psi_0^0$. In order to translate this condition in metric language, we can study the transverse metric $\gamma_{ab}$ reconstructed from the frame $m^a$ given in \eqref{NP solution for ma}. We find that it is of the form \eqref{gamma expansion} with
\be\label{no-log Dab}
D_{ab}=-\Sigma\widetilde{C}_{ab}=\f{1}{2}{C_{(a}}^c\big(L_c\partial_uL_{b)}-L_{b)}\partial_uL_c+D_cL_{b)}-D_{b)}L_c\big).
\ee
We note in passing that in terms of NP quantities we have
\be
C_{ab}m^a_1m^b_1=2\sigma_2,
\q\q
D_{ab}m^a_1m^b_1=-2i\Sigma\sigma_2.
\ee
When the vector $\hat{X}^a$ is written in a radial expansion without logarithmic terms, there is no coefficient of the expansion appearing in $G_{ra}|_{\O(r^{-3})}=0$ and allowing a priori to solve this equation. Instead, this equation is solved once \eqref{no-log Dab} is inserted (in addition to the trace \eqref{trace D} coming from the NU gauge condition) which shows explicitly that this is indeed the generalization of the no-log condition to the case of twisting asymptotically-flat spacetimes.

Finally, let us point out that it is actually possible to solve the Einstein equations without imposing that $\ell^\mu\nabla_\mu\ell=0$ at all. This leads to a generalization of the so-called partial Bondi gauge \cite{Geiller:2022vto,Geiller:2024amx} (i.e. where no choice of radial coordinate has been made and where the traces in \eqref{gamma expansion} are free) to the case of a non-vanishing twist, where furthermore one can have an arbitrary radial expansion for the twist\footnote{For example, a metric in this twisting partial Bondi gauge can be obtained from the Robinson--Trautman solution by demanding that the transverse metric be a round sphere. Starting from
\be
\de s^2=\left(\f{2M}{r}-\f{R}{2}+2r\partial_u\ln P\right)\de u^2-2\de u\,\de r+\f{r^2}{P^2}\de\Omega^2,
\q\q
R=2\big(P^2+D^2\ln P\big),
\ee
where $P=P(u,x^a)$ and $\de\Omega^2$ is the metric on the two-sphere, the change of radial coordinate $r\mapsto rP$ leads to
\be
\de s^2=\left(\f{2M}{rP}-\f{R}{2}\right)\de u^2-2P\de u\,\de r-2r\partial_aP\de r\,\de x^a+r^2\de\Omega^2.
\ee
This form of the Robinson--Trautman solution has $g_{ra}\neq0$ and $g_{ur}\neq-1$, which fits in the family \eqref{hat metric} with a round sphere for $g_{ab}$. Of course, this rewriting does not spoil the fact that the Robinson--Trautman solution admits a twist-free and shear-free null geodesic congruence. It simply illustrates the fact that in order to obtain a round sphere metric one can turn on $g_{ra}\neq0$ instead of turning on $g_{ua}\neq0$ as done e.g. in \cite{Bonga:2023eml} (see section V.E and appendix B). The diffeomorphism which brings to a round sphere metric with $g_{ra}\neq0$ is much simpler than that leading to $g_{ua}\neq0$.}. Here we do not consider such a relaxed gauge since we want to reproduce the metric version of the NP solution space constructed in section \ref{sec:tetrad}. It is however interesting to note that the Einstein equations can be solved in such a generic context.

\subsection{Carrollian interpretation}
\label{sec:carroll}

The introduction of a non-vanishing twist potential $L_a$ in our solution space, while generalizing the standard Bondi gauge, does not affect the induced boundary geometry at future null infinity. The later is still given by
\be
\de s^2\big|_{\I^+}=\lim_{r\to\infty}\f{\de s^2}{r^2}=0\times\de u^2+q_{ab}\de x^a\de x^b\eqqcolon q_{AB}\de x^A\de x^B,
\ee
where $x^A=\{u,x^a\}$ are the coordinates on the boundary. The data defining this geometry consists in a nowhere-vanishing vector field $v=v^A\partial_A=\partial_u$ as well as a metric $q_{AB}$, together with the degeneracy condition $v^Aq_{AB}=0$, and with $(v,q)$ both defined up to conformal transformations. This is a weak conformal Carrollian structure \cite{Duval:2014uoa,Ciambelli:2019lap,Ciambelli:2025unn}. Now, just as in Lorentzian geometry where the metric can be described by a vielbein, one can describe the previous data in terms of a local Carroll--Cartan coframe \cite{Ciambelli:2025unn}. It consists in an Ehresmann connection (or clock form) $\tau$ such that $v^A\tau_A=1$, as well as a spatial coframe $\{e^i_A\}_{i=1,2}$ such that $q=\delta_{ij}e^i\otimes e^j$ and $v^Ae^i_A=0$. The triple $(\tau,v,q)$ forms a ruled Carrollian structure. As expected, switching to a description in terms of a coframe introduces new gauge redundancies. One can indeed manifestly rotate $e^i$ with a local spatial rotation, but also redefine the Ehresmann connection as $\tau\mapsto\tau+\lambda_ie^i$. This last internal symmetry is known as a local Carroll boost.

A natural question is then whether the data $(\tau,e^i)$ can be seen as induced from the bulk metric, just as it is the case for $(v,q)$. This turns out to indeed be the case, as explained in the construction of the Carroll covariant Newman--Unti and Bondi--Sachs gauges in \cite{Ciambelli:2020eba,Ciambelli:2020ftk,Campoleoni:2022wmf,Mittal:2022ywl,Campoleoni:2023fug,Hartong:2025jpp,Fiorucci:2025twa}. In these references, after translating the notations to the present work, the authors start from a gauge in which $g_{rr}=0$ and then impose $\Gamma^A_{rr}=0$, which in our case means that $\kappa=0$ or equivalently $\partial_rZ_a=0\Rightarrow Z_a=L_a$. Then they impose either the BS gauge by setting $\Gamma^A_{Ar}=0$ or the NU gauge by setting $\Gamma^r_{rr}=0$. The resulting parametrization of their bulk metric is
\be
\de s^2=-2\tau_A\de x^A\de r+g_{AB}\de x^A\de x^B,
\ee
with $g_{AB}=r^2\delta_{ij}e^i_Ae^j_B+\O(r)$ of degenerate signature $(0,+,+)$, $\tau_A=\mathcal{O}(1)$ and $\tau_Ae_i^A=0$. When pulled-back to null infinity, this leads exactly to the ruled Carrollian structure described above \cite{Ciambelli:2025mex}. Importantly, here the line element inferred from \eqref{metric} reduces to
\be
\de s^2=2\ell \de r+g_{AB}\de x^A\de x^B=\gamma_{ab}\de x^a\de x^b+\ell\big(\ell g_{uu}+2\de r+2(X_a-\Omega_a)\de x^a\big),
\ee
where the null 1-form is
\be
\ell=-\tau=-\de u+L_a\de x^a,
\ee
and where $g_{uu}=2U+X_a(X^a-2\Omega^a)$. This shows that the twist potential $L_a$ has a natural interpretation in terms of a Carrollian structure at the boundary. In the standard Bondi gauge with $g_{ra}=0$ the Ehresmann connection is fixed to $\tau=\de u$ and the local Carroll boost symmetry is reduced by the requirement that $\Upsilon_a=-\partial_af$. Accordingly, $\lambda_i$ above is forced to vanish and in our notations this translated into the condition $\Upsilon_a=-\partial_af$. On the contrary, when the twist potential is present in the solution space the boundary Carroll covariance is restored. From the point of view of working with a relaxed Ehresmann connection and thereby describing a twist potential in $g_{ra}\neq0$, the work \cite{Campoleoni:2023fug} is the closest to what we are presenting here. This reference has also the advantage of working with a non-vanishing cosmological constant, as the aim there was to study the flat limit.

\subsection{Hypersurface equations}
\label{sec:hypersurface}

We now present the resolution of the metric Einstein equations with the line element \eqref{hat metric} and the gauge conditions $\kappa=\text{Re}(\epsilon)=0$ (which when combined are equivalent to $Z_a=L_a$ and $g_{ur}=-1$). Just like in the absence of twist, the Einstein equations $(G_{ru},G_{ra},G_{rr})=0$ are hypersurface equations which determine respectively the radial expansion of $(\hat{U},\hat{X}^a,\hat{W})$. When $Z_a=0$, these equations can be solved one after the other. Solving $G_{rr}=0$ first determines $\hat{W}$, solving $G_{ra}=0$ then determines $\hat{X}^a$, and finally solving $G_{ur}=0$ determines $\hat{U}$. When $Z_a\neq0$ however, the equations are coupled and must be solved in a different order. More precisely, one must first solve the leading order of $(G_{ru},G_{ra},G_{rr})=0$, then the subleading order, and so on.

The explicit form of the hypersurface equations can be found by expressing the components of the Ricci tensor in terms of the spin coefficients using the results of appendix \ref{appendix:Ricci}. The most compact of these equations is $G_{rr}=R_{rr}=0$, where the first equality follows from the gauge condition $g_{rr}=0$. When $\kappa=\text{Re}(\epsilon)=0$ this equation takes the form
\be\label{rr EFE}
G_{rr}=R_{rr}=2(\partial_r\rho-\rho^2-\sigma\bar{\sigma})=2\partial_r\text{Re}(\rho)-2\text{Re}(\rho)^2+2\text{Im}(\rho)^2+\f{1}{8}\partial_r\gamma^{ab}\partial_r\gamma_{ab}+\f{1}{4}\f{\det(\partial_r\gamma_{ab})}{\det(\gamma_{ab})},
\ee
where we have used the expressions given in appendix \ref{app:spin coeffs} to show in particular that the result is real, as expected. The other two sets of hypersurface equations are very lengthy when the twist is non-vanishing, so we do not display them here.

Of course, the solution to the hypersurface equations has already been obtained above in NP form for the un-hatted quantities which enter the metric \eqref{metric}. Here our goal is to show that one can equivalently solve directly the metric equations. For the hatted variables, the solution corresponding to the fall-offs obtained in \eqref{NP tetrad solution} is an expansion of the form
\be\label{UXW fall-offs}
\hat{U}=r\hat{U}_{+1}+\sum_{n=0}^\infty\f{\hat{U}_n}{r^n},
\q\q
\hat{X}^a=\sum_{n=1}^\infty\f{\hat{X}^a_n}{r^n},
\q\q
\hat{W}=1+\sum_{n=1}^\infty\f{\hat{W}_n}{r^n}.
\ee
The first non-trivial equation is $G_{ru}|_{\O(r^{-1})}=0$, which is solved by
\be
\hat{U}_{+1}=-\f{1}{2}\partial_u\ln\sqrt{q}=-2\gamma_0.
\ee
The next equation is $G_{ra}|_{\O(r^{-1})}=0$, which is solved by
\be
\hat{X}^a_1=q^{ab}\partial_uL_b.
\ee
The next equation is $G_{rr}|_{\O(r^{-3})}=0$, which is solved by
\be
\hat{W}_1=L^a\partial_uL_a.
\ee
Note that this does indeed correspond to $\hat{W}_1=L_a\hat{X}^a_1$, as inferred from the gauge condition \eqref{gauge for W}. Consistently, one can check that these solutions agree with the results \eqref{NP tetrad solution} obtained in the NP formulation. From \eqref{definition a vectors} one can see that $\Omega^a=\O(r^{-2})$ and $\Omega_a=\O(1)$. The contributions of $\Omega$ in \eqref{hat variables} are therefore subleading, and one has $(\hat{U}_{+1},\hat{X}^a_1,\hat{W}_1)=(U_{+1},X^a_1,W_1)$. As part of the dictionary between metric and NP quantities, one can then check that $q^{ab}\partial_uL_b=-2\text{Re}(m^a_1\bar{\tau}_1)$ and $L^a\partial_uL_a=-2\text{Re}(L\bar{\tau}_1)$.

The hypersurface equations are now solved at leading order, and we can then go once again through the three sets of equations in order to solve them at subleading order. The equation $G_{ur}|_{\O(r^{-2})}=0$ is solved by
\be
\hat{U}_0=-\f{1}{4}R[q]-\f{1}{2}\partial_u\big(L^a\partial_uL_a\big)-\f{1}{2}D_a\partial_uL^a+\big(L_aL^a\partial_u-D_aL^a-L^a\partial_uL_a\big)\gamma_0.
\ee
The equation $G_{ra}|_{\O(r^{-2})}=0$ is solved by
\be
\hat{X}^a_2
&=-\f{1}{2}D_bC^{ab}-\f{1}{2}\partial_u\big(C^{ab}L_b\big)+\f{1}{2}D^bD^aL_b+\f{1}{2}\partial^a\big(L^b\partial_uL_b\big)\cr
&\pe+\f{1}{2}\Big(L_b\partial_u^2L^b+\partial_uL^b\partial_uL_b+\partial_u\big(D_bL^b\big)-L_b\partial_uL^b\partial_u-D_bL^b\partial_u-2L^bD_b\partial_u-L^bL_b\partial_u^2-D^2\Big)L^a\q\cr
&\pe-2\Big(2C^{ab}L_b+2\partial_uL^aL^bL_b-L^aL_b\partial_uL^b+2\partial^a\big(L^bL_b\big)+2L^aL^b\partial_b+L^bD_bL^a-3L^bD^aL_b\Big)\gamma_0.
\ee
Finally, the equation $G_{rr}|_{\O(r^{-4})}=0$ is solved by
\be\label{Grr solution for W2}
\hat{W}_2=L_a\hat{X}^a_2+\f{1}{4}D-\f{1}{16}C_{ab}C^{ab}-\f{1}{2}\Sigma^2,
\q\q
\Sigma^2=-\f{1}{2}\big(L^a\partial_uL^b+D^aL^b\big)\big(L_{[a}\partial_uL_{b]}+D_{[a}L_{b]}\big).
\ee
As announced, one can see that $\hat{W}_n$ is fixed both by $G_{rr}|_{\O(r^{-{n+2}})}=0$ and by the gauge condition \eqref{gauge for W}, which implies that $\hat{W}_n=L_a\hat{X}^a_n$. Matching the two expressions leads to a condition on the traces. For example, the matching of \eqref{Grr solution for W2} with $\hat{W}_2=L_a\hat{X}^a_2$ determines
\be\label{trace D}
D=2(\Sigma^2+\sigma_2\bar{\sigma}_2)=2\Sigma^2+\f{1}{4}C_{ab}C^{ab}.
\ee
In addition one can check that $\hat{U}_0=U_0$ and $\hat{X}^a_2=X^a_2-\Omega^a_2$, meaning that the metric and NP results agree as expected. The hypersurface equations are now solved at subleading order, and one can go on with the algorithm.

The sub-subleading equations are important because they involve the mass and angular momentum aspects as integration constants. One can check that $G_{ur}|_{\O(r^{-3})}=0$ is automatically satisfied, and that $G_{ur}|_{\O(r^{-4})}=0$ determines $\hat{U}_2$. This means that when solving $G_{ur}=0$ there is a jump from $\hat{U}_0$ to $\hat{U}_2$, and therefore that $\hat{U}_1$ is a radial integration constant. This is the mass aspect, which we will denote by $\hat{U}_1=M(u,x^a)$. Similarly, $G_{ra}|_{\O(r^{-3})}=0$ is automatically satisfied and $G_{ra}|_{\O(r^{-4})}=0$ determines $\hat{X}^a_4$. The radial integration constant is the angular momentum aspect $\hat{X}^a_3=P^a(u,x^b)$. We give the explicit relationship between $M$ and $\text{Re}(\Psi_2^0)$ on the one hand, and $P^a$ and $\Psi_1^0$ on the other hand in \eqref{leading Psi's} and \eqref{covariant functionals}. Finally, computing $\hat{W}_3$ from $G_{rr}|_{\O(r^{-5})}$ and matching it with the NU gauge condition $\hat{W}_3=L_a\hat{X}^a_3$ implies the vanishing of the trace $E^1=0$ in \eqref{gamma expansion}.

To conclude this section, let us remark that the hypersurface equations $G_{rr}=0$ and $G_{ra}=0$ involve respectively $\partial_r\hat{W}$ and $\partial_r^2\hat{X}^a$, and that we have therefore removed by hand two integration constants which are in principle allowed in the solution space. More precisely, with our choice of fall-offs \eqref{UXW fall-offs} we have set $\hat{W}_0=1$ and $\hat{X}^a_0=0$. The same choice was implicitly hidden in the NP solution space \eqref{NP tetrad solution}, where the same integration constants arise from \eqref{NP1j} and have been fixed by hand. This corresponds to fixing part of the induced boundary metric on $\I^+$, whose only remaining degrees of freedom are then in $q_{ab}$. The solution space in the absence of twist but including these integration constants can be found e.g. in \cite{Compere:2019bua,Geiller:2022vto}. In particular, a non-vanishing $\hat{X}^a_0$ (which is denoted $U^a_0$ in these references) can be used in order to encode radiation in asymptotically-(A)dS spacetimes \cite{Bonga:2023eml,MG-AdS,McNees:2025acf}. It also appears in the Bondi form of the C-metric, as explained in appendix \ref{app:C-metric}. For illustrative purposes, we have kept all these integration constants in the construction of the twisting solution space of three-dimensional gravity presented in section \ref{sec:3d}.

\subsection{Bondi hierarchy}
\label{sec:hierarchy}

We have seen that the four hypersurface equations $G_{\mu r}=0$ determine the radial expansion of the four free functions $(\hat{U},\hat{X}^a,\hat{W})$ entering the metric \eqref{hat metric}. We will now demonstrate that even in the presence of a non-vanishing twist potential, i.e. when $g_{ra}\neq0$, there is a way of reorganizing the remaining Einstein equations into trivial equations and evolution equations for $(M,P^a,E^n_{ab})$. This is the so-called Bondi hierarchy.

The existence of trivial equations can best be shown using the differential Bianchi identities \cite{Barnich:2010eb}. First, note that they can be rewritten as
\be
2\sqrt{-g}\,\nabla_\mu G^\mu_\alpha=2\partial_\mu\big(\sqrt{-g}\,G^\mu_\alpha\big)+\sqrt{-g}\,G_{\mu\nu}\partial_\alpha g^{\mu\nu}\stackrel{\text{B}}{=}0,
\ee
Where $\stackrel{\text{B}}{=}$ denoted an equality which holds using the Bianchi identity. When the hypersurface Einstein equations $G_{\mu r}=G^\mu_r=0$ are satisfied, we therefore get from the various components of this equation that
\bsub\label{Bianchi}
\be
(\alpha=r)\quad\Rightarrow\quad&G_{\mu\nu}\partial_r g^{\mu\nu}=G_{uu}\partial_rg^{uu}+2G_{ua}\partial_rg^{ua}+G_{ab}\partial_rg^{ab}\stackrel{\text{B}}{=}0,\label{Bianchi r}\\
(\alpha=a)\quad\Rightarrow\quad&2\partial_\mu\Big(\sqrt{-g}\,\big(G_{ua}g^{\mu u}+G_{ab}g^{\mu b}\big)\Big)+\sqrt{-g}\,\Big(G_{uu}\partial_ag^{uu}+2G_{ub}\partial_ag^{ub}+G_{bc}\partial_ag^{bc}\Big)\stackrel{\text{B}}{=}0,\label{Bianchi a}\\
(\alpha=u)\quad\Rightarrow\quad&2\partial_\mu\Big(\sqrt{-g}\,\big(G_{uu}g^{\mu u}+G_{ub}g^{\mu b}\big)\Big)+\sqrt{-g}\,\Big(G_{uu}\partial_ug^{uu}+2G_{ub}\partial_ug^{ub}+G_{bc}\partial_ug^{bc}\Big)\stackrel{\text{B}}{=}0.\label{Bianchi u}
\ee
\esub
Using the condition $\partial_rZ_a=0\Rightarrow Z_a=L_a$ and the fact that $g^{uu}=Z_aZ_bg^{ab}$ and $g^{ua}=Z_bg^{ab}$, we can write \eqref{Bianchi r} as
\be\label{Bianchi r rewritten}
\check{G}_{ab}\partial_rg^{ab}\stackrel{\text{B}}{=}0,
\q\q
\check{G}_{ab}\coloneqq G_{ab}+2G_{u(a}L_{b)}+G_{uu}L_aL_b.
\ee
One can note that $\check{G}_{ab}m^a\bar{m}^b=G_{\mu\nu}\hat{m}^\mu\bar{\hat{m}}^\nu$, where $\hat{m}$ is the tetrad vector corresponding to the metric \eqref{hat metric}. If we then split $\check{G}_{ab}$ into a trace-free and a trace part as
\be\label{Gab check TF}
\check{G}^\tf_{ab}\coloneqq\check{G}_{ab}-\f{1}{2}g_{ab}\la\check{G}\ra=0,
\q\q
\la\check{G}\ra\coloneqq g^{ab}\check{G}_{ab}=0,
\ee
the Bianchi identity rewritten as \eqref{Bianchi r rewritten} leads to
\be
\check{G}^\tf_{ab}\partial_rg^{ab}\stackrel{\text{B}}{=}\partial_r\ln\sqrt{\det(g_{ab})}\,\la\check{G}\ra.
\ee
This means that solving the trace-free part automatically implies the trace part. The latter can therefore be considered as a trivial equation. Since $G_{uu}$ and $G_{ua}$ both start at order $r^{-2}$, and since the gauge condition \eqref{gauge Z=L} gives $L_a=\O(1)$, we have that $\check{G}^\tf_{ab}|_{\O(r)}=G^\tf_{ab}|_{\O(r)}$. This is the equation which leads to the constraint $(\partial_u-4\gamma_0)q_{ab}=0$.

Let us now assume that we have solved the equations $\check{G}_{ab}=0$. Using the fact that we work with the hatted variables, and that in NU gauge when $\partial_rZ_a=0$ we have $g^{ur}=-\hat{W}=-1-L_a\hat{X}^a=-1+L_ag^{ra}$, we can expand \eqref{Bianchi a} and \eqref{Bianchi u} and rewrite them in the form
\bsub
\be
0&\stackrel{\text{B}}{=}\partial_u\Big(\sqrt{-g}\,L_a\big(G_{uu}g^{uu}+G_{ub}g^{ub}\big)\Big)+\partial_r\Big(\sqrt{-g}\,\big(G_{ua}+L_aG_{uu}+L_aG_{uu}g^{ur}+L_aG_{ub}g^{rb}\big)\Big)\nn\\
&\pe+L_a\partial_c\Big(\sqrt{-g}\,\big(G_{uu}g^{uc}+G_{ub}g^{bc}\big)\Big)+\sqrt{-g}\,\big(G_{uu}g^{uc}+G_{ub}g^{bc}\big)\big(\partial_cL_a-\partial_aL_c\big),\label{Bianchi a rewritten}\\
0&\stackrel{\text{B}}{=}\partial_u\Big(\sqrt{-g}\,\big(G_{uu}g^{uu}+G_{ub}g^{ub}\big)\Big)+\partial_r\Big(\sqrt{-g}\,\big(G_{uu}g^{ru}+G_{ub}g^{rb}\big)\Big)+\partial_c\Big(\sqrt{-g}\,\big(G_{uu}g^{cu}+G_{ub}g^{bc}\big)\Big)\q\nn\\
&\pe+\sqrt{-g}\,\big(G_{uu}g^{uc}+G_{ub}g^{bc}\big)\partial_uL_c.\label{Bianchi u rewritten}
\ee
\esub
Computing $\eqref{Bianchi a rewritten}-L_a\eqref{Bianchi u rewritten}$ and defining $\check{G}_{ua}\coloneqq G_{ua}+L_aG_{uu}$ then leads to
\be
\partial_r\big(\sqrt{-g}\,\check{G}_{ua}\big)+\sqrt{-g}\,\check{G}_{ub}g^{bc}\Big(\big(L_c\partial_u+\partial_c\big)L_a-\big(L_a\partial_u+\partial_a\big)L_c\Big)\stackrel{\text{B}}{=}0.
\ee
Importantly, this identity implies that the subleading terms in $\check{G}_{ua}$ are all proportional to its leading order $\check{G}_{ua}|_{\O(r^{-2})}$. One can therefore conclude that $\check{G}_{ua}$ contains a single non-trivial equation. This equation turns out to encode the time evolution of the angular momentum. Finally, assuming that $\check{G}_{ua}=0$, we get from \eqref{Bianchi u rewritten} that
\be
\partial_r\big(\sqrt{-g}\,G_{uu}\big)\stackrel{\text{B}}{=}0.
\ee
Once again, this implies that the subleading terms in $G_{uu}$ are all proportional to its leading order $G_{uu}|_{\O(r^{-2})}$, so that $G_{uu}$ contains only the information about the time evolution of the mass. With this computation, the Bondi hierarchy for twisting asymptotically-flat spacetimes reaches an end, and we are guaranteed that there are no more equations to be solved.

In summary, after the hypersurface equations $G_{\mu r}=0$ have been solved (to a sufficiently low order), there is a single equation sitting at order $r^{-2}$ in $\check{G}_{ua}$ and determining the time evolution of the angular momentum, a single equation sitting at order $r^{-2}$ in $\check{G}_{uu}$ and determining the time evolution of the mass, and an infinite amount of evolution equations for the tensors $E^n_{ab}$ in \eqref{gamma expansion} contained in $\check{G}^\tf_{ab}$. This Bondi hierarchy is essentially the same as the in absence of twist, but with the modifications $G\to\check{G}$ in order to account for the twist. We note that this Bondi hierarchy in the presence of twist would also hold in the same form if we included a polyhomogeneous expansion with $\log(r)$ terms. The only difference would appear in the exact way in which the various equations are solved, and the orders of the various expansions, but the organisation of the equations would remain unchanged. The reason for this is that the construction of the Bondi hierarchy presented above does not depend on what type of radial expansion is chosen for the metric. It only uses the Bianchi identities and the information about which metric components are non-vanishing.

\subsection{Flux-balance laws}
\label{sec:flux-balance}

As explained in the previous section, the Bondi hierarchy guarantees that once the hypersurface equations have been solved we are left only with evolution equations, or flux-balance laws. Since these equations are quite lengthy when written directly in metric form, it is useful to exploit the dictionary with the NP formalism to obtain more compact expressions. The evolution equations in NP form are given by the Bianchi identities \eqref{NP evolution equations}. In order to rewrite these equations, let us first note that we have
\bsub
\be
\Sigma&=\f{1}{2}(D^a+L^a\partial_u)\widetilde{L}_a,\label{metric Sigma}\\
\lambda_1&=N_{ab}\bar{m}^a_1\bar{m}^b_1,\\
\tau_1&=-\partial_uL_am^a_1,\\
\Omega_1&=\Omega_a^1m^a_1,\label{metric Omega1}\\
\mu_R&=\f{R}{4}-\big(D_aL^a+2L^a\partial_a\partial_u+L^aL_a\partial_u+L^a\partial_uL_a\big)\gamma_0\label{metric mu R},\\
\partial_um^a_1&=-2\gamma_0m^a_1,
\ee
\esub
with the definitions
\bsub
\be
N_{ab}&\coloneqq\f{1}{2}(\partial_u-2\gamma_0)C_{ab}+(\partial_u-4\gamma_0)L_{(a}\partial_uL_{b)}+D_{(a}\partial_uL_{b)},\\
\Omega_a^1&\coloneqq\f{1}{2}(D^b+L^b\partial_u)C_{ab}+\big(\widetilde{L}_a\partial_u-\widetilde{\partial}_a-2\partial_u\widetilde{L}_a\big)\Sigma.
\ee
\esub
Recall that since here we are manipulating quantities which are $r$-independent, the angular indices are lowered and raised with $q_{ab}$. Similarly, the dual notation $\widetilde{L}_a=\eps^{(q)}_{ab}L^b$ is defined with respect to $q_{ab}$. Using these notations, we can then write the leading terms in the Weyl scalars \eqref{Weyl scalars} as
\bsub\label{leading Psi's}
\be
\Psi_0^0&=-\E_{ab}^1m^a_1m^b_1,\\
\Psi_1^0&=-\P_am^a_1,\\
\Psi_2^0&=-\big(\M+i\widetilde{\M}\big),\\
\Psi_3^0&=-\J_a\bar{m}^a_1,\\
\Psi_4^0&=-\N_{ab}\bar{m}^a_1\bar{m}^b_1,
\ee
\esub
where
\bsub\label{covariant functionals}
\be
\E^1_{ab}&=-3E^1_{ab},\label{curly E}\\
\P_a&=-\f{3}{2}\big(P_a-C_{ab}\Omega^b_1+\rho_3\partial_uL_a\big),\label{curly P}\\
\M&=M-\Omega^a_1\partial_uL_a,\label{curly M}\\
\widetilde{\M}&=-\f{1}{4}N_{ab}\widetilde{C}^{ab}+\f{1}{2}(D^a+L^a\partial_u)\widetilde{\Omega}_a^1+\Sigma(\mu_R-U_0),\label{curly M tilde}\\
\J_a&=D^bN_{\la ab\ra}+L^b\partial_uN_{\la ab\ra}+2\big(\widetilde{\partial}_a+\widetilde{L}_a\partial_u\big)(\Sigma\gamma_0)+4\Sigma\big(\widetilde{\partial}_a+\partial_u\widetilde{L}_a+\widetilde{L}_a\partial_u\big)\gamma_0+(\partial_a+L_a\partial_u)\mu_R,\\
\N_{ab}
&=\partial_uN_{ab}-2\big(3\partial_u\gamma_0-4\gamma_0^2+\gamma_0\partial_u\big)\partial_uL_{(a}L_{b)}-2\big(\partial_u^2\gamma_0-4\gamma_0\partial_u\gamma_0\big)L_{(a}L_{b)}-2\partial_u\big(\gamma_0D_{(a}L_{b)}\big)\cr
&\pe-4\partial_u\big(L_{(a}\partial_{b)}\partial_u\gamma_0\big)-2D_{(a}\partial_{b)}\partial_u\gamma_0.
\ee
\esub
For this rewriting we have used in particular \eqref{Eth formulas} as well as the formulas gathered in appendix B of \cite{Geiller:2024bgf}. Note that when $\gamma_0=0$ and $L_a=0$ these expressions do not reduce to e.g. (3.12) of \cite{Geiller:2024bgf}, since here we are using the NU gauge while this reference is in BS gauge. We recover however consistently the results of \cite{Geiller:2022vto}, which were obtained in the partial Bondi gauge and can therefore be reduced to the NU gauge. The only difference is an overall minus sign in the definition of the Weyl scalars. In order to find \eqref{curly E}, we have used the vanishing of the trace $E^1=0$ and matched the expansion \eqref{gamma expansion} with the expression for $\gamma_{ab}=2m_{(a}\bar{m}_{b)}$ inferred from \eqref{NP solution for ma}.

In equations \eqref{curly P} and \eqref{curly M} we have given the relation between the NP quantities appearing in the Weyl scalars and the usual metric parametrization $M$ and $P^a$ of the mass and angular momentum aspects appearing at order $r^{-1}$ in $g_{uu}$ and $g_{ua}$. To obtain these relations, we have used \eqref{hat variables} and \eqref{NP tetrad solution} to write
\bsub
\be
M&\coloneqq\hat{U}_1=U_1-\f{1}{2}\Omega_a\Omega^a\big|_{\O(r^{-1})}=\M-2\tau_1\bar{\Omega}_1-\f{1}{2}\Omega_a\Omega^a\big|_{\O(r^{-1})},\\
P^a&\coloneqq\hat{X}^a_3=X^a_3-\Omega^a_3=-\f{1}{3}\P^a+2\text{Re}(\rho_3\bar{\tau}_1)-\Omega^a_3,
\ee
\esub
and then used the definition $\Omega^a=2\text{Re}(\bar{\Omega}m^a)$ to compute $\Omega^a_3$ on the one hand and show that $\Omega_a\Omega^a\big|_{\O(r^{-1})}=0$ on the other hand.

We can now rewrite the metric equivalent of the NP evolution equations \eqref{NP evolution equations} as well as the tautological relation \eqref{Psi30 evolution}. Using the notations above, separating $\Psi_2^0$ into its real and imaginary parts, and introducing $\M_{ab}\coloneqq\M q_{ab}+\widetilde{\M}\eps_{ab}$, we find
\bsub
\be
(\partial_u+2\gamma_0)\E^1_{ab}&=\big(D_{\la a}-4\gamma_0L_{\la a}+L_{\la a}\partial_u+4\partial_uL_{\la a}\big)\P_{b\ra}+\f{3}{2}\M_{ac}{C^c}_b,\\
(\partial_u+4\gamma_0)\P_a&=\big(\partial_a+L_a\partial_u+3\partial_uL_a\big)\M+\big(\widetilde{\partial}_a+\widetilde{L}_a\partial_u+3\partial_u\widetilde{L}_a\big)\widetilde{\M}+C_{ab}\J^b,\\
(\partial_u+6\gamma_0)\M&=\f{1}{2}\big(D_a+4\gamma_0L_a+L_a\partial_u+2\partial_uL_a\big)\J^a+\f{1}{4}C_{ab}\N^{ab},\label{EOM curly M}\\
(\partial_u+6\gamma_0)\widetilde{\M}&=\f{1}{2}\big(D_a+4\gamma_0L_a+L_a\partial_u+2\partial_uL_a\big)\widetilde{\J}^a+\f{1}{4}C_{ab}\widetilde{\N}^{ab},\label{EOM curly M dual}\\
(\partial_u+4\gamma_0)\J_a&=\big(D^a+4\gamma_0L^a+\partial_uL^a+L^a\partial_u\big)\N_{\la ab\ra}.
\ee
\esub
One should recall that \eqref{EOM curly M dual} is automatically satisfied and follows simply from the definition of the dual mass $\widetilde{\M}=-\text{Im}(\Psi_2^0)$ given in \eqref{NP Im Psi20} or equivalently in \eqref{curly M tilde}.

To conclude this section, we can study the mass loss when the leading boundary metric is time-independent, i.e. when $\gamma_0=0$. In this case, we find that \eqref{EOM curly M} can be rearranged as
\be
\partial_u\left(\M-\f{1}{2}L_a\J^a-\f{1}{4}C_{ab}N^{ab}\right)&=-\f{1}{4}N_{ab}N^{ab}+\f{1}{2}D_a\J^a+\f{1}{2}\partial_uL^a\big(D^bN_{\la ab\ra}+L^b\partial_uN_{\la ab\ra}\big).
\ee
Unfortunately, since the last term on the right-hand side cannot be written as a total $\partial_u$ or $D^a$ derivative, one cannot find from this expression a candidate energy whose smeared flux on the celestial sphere would be negative. This is however possible in the particular case $\partial_uL_a=0$. This condition can always be met with the help of the residual transformation (the Carroll boost) parametrized by $\Upsilon_a$ and acting on $L_a$ as in \eqref{transformation of L}. When $\partial_uL_a=0$ in addition to $\gamma_0=0$, we get from \eqref{NP tau1} that $\tau_1=0$ as well. Using the relation
\be
n^\mu\nabla_\mu\ell=(\gamma+\bar{\gamma})\ell-\tau\bar{m}-\bar{\tau}m,
\ee
this means that $\ell$ becomes parallelly transported with respect to $n$ on $\I^+$. It is interesting to note that this asymptotic condition enables us to recover a ``standard'' mass loss even in the presence of (then time-independent) twist potential.

\section{Asymptotic symmetries}
\label{sec:symmetries}

Now that we have a complete characterization of the solution space both in tetrad and metric variables, we turn to the analysis of the asymptotic symmetries. We first find the residual asymptotic Killing vector (AKV hereafter) and its action on the solution space, and then briefly discuss the asymptotic charges.

\subsection{Asymptotic Killing vectors}

Let us consider the vector field $\xi=\xi^u\partial_u+\xi^r\partial_r+\xi^a\partial_a$, and impose that it preserves the gauge choices. First, we have
\be
\pounds_\xi g_{rr}=2\partial_r\big(L_a\xi^a-\xi^u\big),
\ee
so preserving the condition $g_{rr}=0$ implies that
\be\label{AKV xi u}
\xi^u=f+L_a(\xi^a-Y^a),
\ee
where $f(u,x^a)$ and $Y^a(u,x^b)$ are radial integration constants. We have already singled out a particular field-dependent shift of $f$ in order to obtain later on an algebra of AKVs instead of an algebroid. Next, we find
\be
\pounds_\xi g_{ur}&=-\partial_u\big(f-L_aY^a\big)-\partial_r\xi^r-\big(\hat{X}_a\partial_r+\partial_uL_a\big)\xi^a.
\ee
Since the NU gauge condition implies $g_{ur}=-1$, this Lie derivative must be vanishing. This condition determines $\partial_r\xi^r$, which can then be used to find
\be
\pounds_\xi g_{ra}=\gamma_{ab}\partial_r\xi^b+\big(\partial_uL_a-\partial_a-L_a\partial_u\big)\big(f-L_bY^b\big)+\Sigma_{ab}\xi^b,
\q\q
\Sigma_{ab}\coloneqq-2\big(L_{[a}\partial_u+\partial_{[a}\big)L_{b]}.
\ee
We note in passing that $\Sigma_{ab}m^a_1\bar{m}^b_1=2i\Sigma$. Since our gauge choice in the presence of a non-vanishing twist is $g_{ra}=L_a$, this Lie derivative must be of order $\O(1)$. We can meet this requirement by fixing the expansion of $\xi^a$, which ends up containing two free functions $Y^a(u,x^b)$ and $\Upsilon^a(u,x^b)$. The result can then be plugged in $\pounds_\xi g_{ur}=0$ in order to find $\xi^r$. The latter contains a radial integration constant $\xi^r_0$, which is in turn constrained by the requirement that $\delta_\xi\big(q^{ab}C_{ab}\big)=0$. Finally, preserving the fall-off condition $g_{ua}=\O(r)$ imposes that $Y^a(x^b)$ be time-independent, while the conditions $g_{uu}=\O(r)$ and $g_{ab}=\O(r^2)$ are automatically satisfied.

At the end of the day, we find that the AKVs have integration constants $f(u,x^a)$, $Y^a(x^b)$, and $\Upsilon^a(u,x^b)$, and components given by the radial expansion
\bsub\label{AKVs}
\be
\xi^u&=f(u,x^a)+L_a\big(\xi^a-Y^a(x^b)\big),\\
\xi^r&=rW(u,x^a)+\sum_{n=0}^\infty\f{\xi^r_n}{r^n},\\
\xi^a&=Y^a(x^b)+\f{\Upsilon^a(u,x^b)}{r}+\sum_{n=2}^\infty\f{\xi^a_n}{r^n},
\ee
\esub
with
\bsub\label{subleading AKV}
\be
W&=-\partial_uf,\\
\xi^r_0&=-\f{1}{2}\big((\partial_u+4\gamma_0)L_a\xi^a_1+D_a\xi^a_1\big),\\
\xi^r_1&=-\f{1}{1}\Big(2q_{ab}\hat{X}^a_1\xi^b_2+q_{ab}\hat{X}^a_2\xi^b_1+C_{ab}\hat{X}^a_1\xi^b_1-\partial_uL_a\xi^a_2\Big),\\
\xi^r_2&=-\f{1}{2}\Big(3q_{ab}\hat{X}^a_1\xi^b_3+2q_{ab}\hat{X}^a_2\xi^b_2+q_{ab}\hat{X}^a_3\xi^b_1+2C_{ab}\hat{X}^a_1\xi^b_2+C_{ab}\hat{X}^a_2\xi^b_1+D_{ab}\hat{X}^a_1\xi^b_1-\partial_uL_a\xi^a_3\Big),\\
\xi^r_n&=-\f{1}{n}\Big((n+1)q_{ab}\hat{X}^a_1\xi^b_{n+1}+\dots+q_{ab}\hat{X}^a_{n+1}\xi^b_1+nC_{ab}\hat{X}^a_1\xi^b_n+\dots+C_{ab}\hat{X}^a_n\xi^b_1\cr
&\hspace{8.86cm}+\dots+E^{n-2}_{ab}\hat{X}^a_1\xi^b_1-\partial_uL_a\xi^a_{n+1}\Big),\\
\xi^a_1&=\Upsilon^a,\\
\xi^a_2&=\f{1}{2}\big(\Sigma^{ab}-C^{ab}\big)\xi^1_b,\\
\xi^a_3&=\f{1}{3}\big(\Sigma^{ab}-2C^{ab}\big)\xi^2_b-\f{1}{3}D^{ab}\xi^1_b,\\
\xi^a_4&=\f{1}{4}\big(\Sigma^{ab}-3C^{ab}\big)\xi^3_b-\f{2}{4}D^{ab}\xi^2_b-\f{1}{4}E^{ab}_1\xi^1_b,\\
\xi^a_n&=\f{1}{n}\big(\Sigma^{ab}-(n-1)C^{ab}\big)\xi^{n-1}_b-\f{n-2}{n}D^{ab}\xi^{n-2}_b+\dots-\f{1}{b}E^{ab}_{n-3}\xi^1_b.
\ee
\esub
We recall that the expression for $\xi^r_0$ is found by the requirement that $\delta_\xi\big(q^{ab}C_{ab}\big)=0$. Indeed, at first this condition is not automatically satisfied when working in NU gauge, but we must enforce that it holds since we have chosen $C_{ab}$ in \eqref{gamma expansion} to be trace-free.

\subsection{Transformation laws}

With the above expressions for the AKVs, we can compute the action of the asymptotic symmetries on the solution space. Denoting $\Upsilon\coloneqq\Upsilon_am^a_1$ and $\Y\coloneqq Y_am^a_1$, and recalling that $L=L_am^a_1$, we find
\bsub\label{transformation laws}
\be
\delta_\xi\ln\sqrt{q}&=f\partial_u\ln\sqrt{q}+D_aY^a+2W,\label{delta root q}\\
\delta_\xi q_{ab}&=\big(f\partial_u+\pounds_Y+2W\big)q_{ab},\label{delta qab}\\
\delta_\xi C_{ab}&=\big(f\partial_u+\pounds_Y+W\big)C_{ab}+2D_{\la a}\Upsilon_{b\ra}+2(\partial_u-4\gamma_0)\big(L_{\la a}\Upsilon_{b\ra}\big),\\
\delta_\xi\sigma_2&=\big(f\partial_u+\L_\Y-W\big)\sigma_2+\big(\Eth+(\partial_u+2\gamma_0)L\big)\Upsilon,\label{delta xi sigma}\\
\delta_\xi\M&=\big(f\partial_u+\pounds_Y-3W\big)\M-\Upsilon^a\J_a,\\
\delta_\xi\Psi_0^0&=\big(f\partial_u+\L_\Y-3W\big)\Psi_0^0-4\Upsilon\Psi_1^0,\label{delta xi Psi00}\\
\delta_\xi\Psi_1^0&=\big(f\partial_u+\L_\Y-3W\big)\Psi_1^0-3\Upsilon\Psi_2^0,\label{delta xi Psi10}\\
\delta_\xi\Psi_2^0&=\big(f\partial_u+\L_\Y-3W\big)\Psi_2^0-2\Upsilon\Psi_3^0,\\
\delta_\xi\Psi_3^0&=\big(f\partial_u+\L_\Y-3W\big)\Psi_3^0-1\Upsilon\Psi_4^0,\\
\delta_\xi\Psi_4^0&=\big(f\partial_u+\L_\Y-3W\big)\Psi_4^0,\\
\delta_\xi m^a_1&=\left(\partial_u+\f{1}{2}\big(\ethb\bar{\Y}-\eth\Y-\ethb\Y-\eth\bar{\Y}\big)-W\right)m^a_1-\eth\Y\bar{m}^a_1,\label{delta m1}\\
\delta_\xi L_a&=\big(f\partial_u+\pounds_Y+W\big)L_a-\Upsilon_a-\partial_af,\label{transformation of L}\\
\delta_\xi L&=\big(f\partial_u+\L_Y-W\big)L-\Upsilon-\Eth f,\\
\delta_\xi\Sigma
&=\big(f\partial_u+\L_Y-W\big)\Sigma-\f{1}{2}\big(D_a-\partial_uL_a+L_a\partial_u+4\gamma_0L_a\big)\widetilde{\Upsilon}^a\cr
&=\big(f\partial_u+\L_Y-W\big)\Sigma-\f{i}{2}\big(\Ethb\Upsilon-\Eth\bar{\Upsilon}+\bar{\tau}_1\Upsilon-\tau_1\bar{\Upsilon}\big).
\ee
\esub
In these expressions $\pounds_Y$ denotes the Lie derivative along the vector $Y^a$ while $\L_\Y$ denotes the spin-weighted Lie derivative acting on NP quantities of spin $s$ as
\be
\L_\Y\F_s=\left(\Y\ethb+\bar{\Y}\eth-\f{s}{2}\big(\eth+\ethb\big)\big(\Y-\bar{\Y}\big)\right)\F_s.
\ee
The transformation law \eqref{delta m1} follows from \eqref{delta qab} and the definition \eqref{frame m1}. These results will be particularly important when studying the reduction to algebraically special solutions in section \ref{sec: algebraically special} below.

Since the AKVs are manifestly field-dependent, one should compute their Lie bracket using the above transformation laws and the adjusted bracket \cite{Barnich:2011mi}. We find
\be
\big[\xi_1(f_1,Y_1,\Upsilon_1),\xi_2(f_2,Y_2,\Upsilon_2)\big]_\star
&=\big[\xi_1(f_1,Y_1,\Upsilon_1),\xi_2(f_2,Y_2,\Upsilon_2)\big]-\big(\delta_{\xi_1}\xi_2-\delta_{\xi_2}\xi_1\big)\cr
&=\xi_{12}(f_{12},Y_{12},\Upsilon_{12}),
\ee
with
\bsub
\be
f_{12}&=f_1\partial_uf_2+Y_1^a\partial_af_2-\delta_{\xi_1}f_2-(1\leftrightarrow2),\\
Y^a_{12}&=Y_1^b\partial_bY_2^a-\delta_{\xi_1}Y_2^a-(1\leftrightarrow2),\\
\Upsilon_{12}^a&=f_1\partial_u\Upsilon_2+\Upsilon_1^b\partial_bY_2^a+Y_1^b\partial_b\Upsilon_2^a-\Upsilon^a_1\partial_uf_2-\delta_{\xi_1}\Upsilon_2^a-(1\leftrightarrow2).
\ee
\esub
This is in agreement with the results found in \cite{Mao:2024jpt}. Notably, when $(f,Y,\Upsilon)$ are field-independent this algebra has field-independent structure functions as well. This property can be traced back to the field-dependent shift of $f$ which we made in \eqref{AKV xi u} when defining $\xi^u$. This is a typical example of how the field-dependent parametrization of the AKV can be used to land back on an algebra of vector fields instead of an algebroid. Interestingly, one can see that these subtleties already appear in the case of four-dimensional asymptotically flat spacetimes.

\subsection{Charges}

The charges can be computed using standard covariant phase space methods. In particular, in the metric formulation one can consider the Iyer--Wald charges \cite{Iyer:1994ys} defined as the $(ur)$ component of the variational formula
\be\label{IW charge}
\slashed{\delta}Q^{\mu\nu}(\xi)
&=\delta K^{\mu\nu}_\xi-K^{\mu\nu}_{\delta\xi}+2\xi^{[\mu}\theta^{\nu]}\cr
&=2\sqrt{-g}\left[\xi^{[\mu}\big(\nabla^{\nu]}\delta g-\nabla_\alpha\delta g^{\nu]\alpha}\big)+\xi_\alpha\nabla^{[\mu}\delta g^{\nu]\alpha}+\left(\f{1}{2}\delta gg^{[\mu\alpha}-\delta g^{[\mu\alpha}\right)\nabla_\alpha\xi^{\nu]}\right],
\ee
where $\delta g\coloneqq g_{\mu\nu}\delta g^{\mu\nu}$, where $K_\xi^{\mu\nu}=-2\nabla^{[\mu}\xi^{\nu]}$ is the Komar charge, and $\theta^\mu=2\sqrt{-g}\,g^{\alpha[\beta}\delta\Gamma^{\mu]}_{\alpha\beta}$ is the symplectic potential. Note that the second term on the first line is here in order to subtract variations acting on the field-dependency contained in the vector field $\xi$ and produced by the first term\footnote{Although the residual vector fields are field-dependent, the charge can never contain contributions arising from the variation of this field-dependency in the vector field. This can be traced back to the very definition of the charges \cite{Iyer:1994ys,Compere:2018aar}, which is the contraction of the symplectic structure with the variational vector field $\pounds_\xi$. Because of this operation, an arbitrary variation in the symplectic structure is replaced by a Lie derivative, and there is never a variation acting on the Lie derivative (and therefore on the vector field) itself.}. Alternatively, and closer to the spirit of the NP formulation, one can construct the charges from the first order tetrad formulation as was done in \cite{Barnich:2019vzx,Mao:2024jpt}. In the presence of a non-vanishing twist, and due to the fact that we are working with an arbitrary metric $q_{ab}$ with time dependency in $\gamma_0\propto\partial_u\ln\sqrt{q}$, the generic expression for these charges is extremely lengthy and not particularly illuminating. The charge also generically contains divergences in $r^2$ and $r$, which can in principle be renormalized using the standard techniques given e.g. in \cite{Freidel:2019ohg,McNees:2023tus,Riello:2024uvs,Delfante:2024npo}.

In order to illustrate the role of the twist and of the associated Carroll boost symmetry generator $\Upsilon$, let us focus on a sub-sector of the full result. For this, we pick a fixed round sphere metric $q_{ab}=\text{diag}(1,\sin^2\theta)$ so that $\delta q_{ab}=0$ and $\gamma_0=0$. In the formula for the Iyer--Wald charge, let us then consider only the contribution from $\Upsilon$. The result is
\be\label{charge with Upsilon}
\slashed{\delta}Q(\Upsilon)
&=\sqrt{q}\,\Upsilon^a\delta\Big(D^bC_{ab}+2L_a+D^2L_a-D^bD_aL_b+\partial_a(D_bL^b)+\partial_u\big(L^bD_bL_a-L^bD_aL_b+L_aD_bL^b\big)\Big)\nn\\
&\pe+\sqrt{q}\,D_a\left(\f{1}{2}\delta L^a\big(\partial_u(L_b\Upsilon^b)+D_b\Upsilon^b\big)-\delta\partial_uL^a(L_b\Upsilon^b)-\Upsilon^aD_b\delta L^b\right)\nn\\
&\pe+\sqrt{q}\,\partial_u\Big(\big(\partial_u(L_a\Upsilon^a)+D_a\Upsilon^a-L_a\Upsilon^a\partial_u\big)(L^b\delta L_b)+D_a\delta L^a\Big)+\O(r^{-1}).
\ee
This expression is valid even in the case of a field-dependent $\Upsilon$. It is moreover finite when taking the limit $r\to\infty$ to obtain the charges on a cut of $\I^+$. We have decomposed the result into contributions from an integrable part, a total divergence, and a corner term. Since the divergence is irrelevant and the corner term can be treated as an ambiguity, we obtain that the Carroll boost charge is integrable. However, since the twist $L$ has an arbitrary time dependency and does not satisfy a flux-balance law, nothing can be said about the (non)-conservation of the charge. The twist $L$ also contributes heavily to the $(f,Y,W)$ components of the charge, but we do not display them here. A detailed analysis of the total charge is postponed to future work. It would be particularly interesting to study the Wald--Zoupas prescription and the requirement of covariance in the present context \cite{Wald:1999wa,Grant:2021sxk,Odak:2022ndm,Rignon-Bret:2024gcx}. Note that in \eqref{charge with Upsilon} we have only written the contribution to the charge arising from $\Upsilon^a$. There are also contributions from $f$ and $Y^a$. We have checked, consistently, that by reducing to $L_a=0$ and $\Upsilon_a=-\partial_af$ the total charge reproduces the well-known result for BMS$_4$ \cite{Barnich:2011mi}.

As a closing remark, one can note that this charge is at odds with the result found in \cite{Mao:2024jpt} using the Hilbert--Palatini formulation. Indeed, when choosing a fixed round sphere metric the charge formula $Q_\Upsilon$ given by (58) in \cite{Mao:2024jpt} vanishes identically, while here is it not the case. The explanation for this difference is most likely related to the mismatch of charges computed from the tetrad and metric formulations \cite{Oliveri:2019gvm,Oliveri:2020xls,Freidel:2020xyx,Freidel:2020svx}, and of the so-called corner ambiguities which appear when comparing the symplectic structures arising from different Lagrangians for the same theory. This deserves a more detailed analysis in the presence of twist.

\section{Algebraically special solutions}
\label{sec: algebraically special}

We now turn to one of the most interesting aspects of the solution space with non-vanishing twist, which is the possibility of rewriting a large class of algebraically special solutions in finite form \cite{Stephani:2003tm}. As mentioned in the introduction, this property is neatly illustrated with the Kerr solution, which has an infinite radial expansion when forced into the ``standard'' Bondi gauge with $g_{ra}=0$ \cite{FletcherLun1996,2003CQGra..20.4153F,Venter:2005cs}, but evidently a finite form when allowing for $g_{ra}\neq0$ and working with outgoing Eddington--Finkelstein coordinates. In view of the solution space constructed here, this is because the outgoing Eddington--Finkelstein coordinates for Kerr are simply Bondi coordinates with non-vanishing twist potential\footnote{Recall however, as mentioned below \eqref{inverse metric NP}, that in the presence of a non-vanishing twist potential sourcing $g^{uu}\neq0$ and $g^{ua}\neq0$ the $u$ coordinate is not null, and is therefore strictly speaking not a Bondi coordinate.}. More generally, and because of their importance for the study of exact solutions of general relativity, one may wonder about the properties of algebraically special solutions when working in Bondi gauge.

Since algebraic speciality is a property of a solution itself, general covariance implies that this notion is independent of the choice of coordinates. When expressed in NP language, algebraic speciality is furthermore invariant under Lorentz transformations of the tetrad (although it may manifest itself differently if we compute non-invariant quantities like individual Weyl scalars or spin coefficients). In particular, given any representative tetrad one may compute the so-called speciality index $\S$ defined as \cite{Stephani:2003tm}
\bsub\label{S index}
\be
\S&\coloneqq I^3-27J^2,\\
I&\coloneqq\f{1}{32}\widetilde{W}_{\mu\nu\rho\sigma}\widetilde{W}^{\mu\nu\rho\sigma}=\Psi_0\Psi_4-4\Psi_1\Psi_3+3(\Psi_2)^2,\\
J&\coloneqq\f{1}{384}\widetilde{W}_{\mu\nu\rho\sigma}\widetilde{W}^{\rho\sigma}\vphantom{W}_{\alpha\beta}\widetilde{W}^{\alpha\beta\mu\nu}=\Psi_0\Psi_2\Psi_4-(\Psi_1)^2\Psi_4-\Psi_0(\Psi_3)^2+2\Psi_1\Psi_2\Psi_3-(\Psi_2)^3,
\ee
\esub
where $\widetilde{W}_{\mu\nu\rho\sigma}\coloneqq(1-i*)W_{\mu\nu\rho\sigma}$. This Lorentz-invariant scalar can be used to differentiate Petrov type I (or algebraically general solutions), when $\S\neq0$, from algebraically special solutions for which $\S=0$. In the latter case, the behavior of three further scalar invariants can then be used to differentiate the various algebraically special Petrov types. For practical calculations however, one works with a given choice of tetrad and algebraic speciality may not be manifest with this choice. For example, the ``standard'' non-twisting Bondi tetrad given in e.g. \cite{Newman:1961qr,Newman:2009,Barnich:2019vzx,Geiller:2024bgf} leads to $\Psi_0\neq0$ and $\Psi_1\neq0$ for Kerr in Bondi gauge. Additionally, in this case the speciality index $\S$ becomes somehow useless because one would have to check its vanishing to all orders in the (infinite) radial expansion.

Since the tetrad \eqref{NP tetrad} can encode a non-vanishing twist potential, one may choose $\ell$ as the repeated principal null direction and impose the algebraically special conditions $\Psi_0=\Psi_1=0$ without discarding ``too many'' interesting solutions. For example, while these conditions exclude Kerr when written in the standard non-twisting Bondi gauge, they still include Kerr when working with a non-vanishing twist\footnote{Although the tetrad \eqref{NP tetrad} enables us to make the algebraic speciality of e.g. Kerr--Taub--NUT manifest with $\Psi_0=\Psi_1=0$, it does not make the type D property manifest since it still leads to $\Psi_3\neq0$ and $\Psi_4\neq0$. However, one can check that $3\Psi_2\Psi_4-2(\Psi_3)^2=0$ vanishes as it should. To make the type D property more manifest, we introduce the map from \eqref{NP tetrad} to the Kinnersley tetrad in section \ref{sec:Kinnersley}.}. In short, although the Bondi gauge with non-vanishing twist constructed from \eqref{NP tetrad} does not reveal new solutions, it enables us to rewrite certain solutions in a much more elegant form. Let us now show explicitly how the twist enables us to resum the algebraically special solutions. This recovers the results of \cite{Stephani:2003tm} (see also \cite{Timofeev:1996vg} with a cosmological constant), which, in the same spirit as \cite{Mao:2024jpt}, can now be seen as arising as a sub-sector of the algebraically general solution space with twist. We also recall that the resummation of the algebraically special solutions in the context of the Bondi gauge with non-vanishing twist potential was studied in \cite{Petropoulos:2015fba,Ciambelli:2018wre,Mittal:2022ywl} with a strong emphasis on the boundary Carrollian geometry and the role of the boundary Cotton tensor. These references also study the case of asymptotically-(A)dS spacetimes, which gives a further motivation for extending the present work in the algebraically general case to a non-vanishing cosmological constant. The asymptotic structure of algebraically special solutions was also studied in \cite{LJMason_1998}, with a special emphasis on the (absence of) asymptotic simplicity \cite{Penrose:1962ij,Penrose:1964ge,Penrose:1965am}.

\subsection{Solution space}
\label{sec:AS solution space}

We now impose the conditions for algebraic speciality, namely $\Psi_0=\Psi_1=0$. From \eqref{S index} one can check that this indeed implies $\S=0$. By virtue of the Goldberg--Sachs theorem \cite{1962AcPPS..22...13G,Krasinski:2009vz}, this is equivalent to the existence of a shear-free congruence, which here is $\ell$ with $\sigma=0$. With these conditions, one can then solve exactly the NP equations of appendix \ref{app:4d NP equations} to obtain the Weyl scalars, the spin coefficients and the components of the tetrad. We find
\bsub\label{Psi ASS}
\be
\Psi_0&=0,\\
\Psi_1&=0,\\
\Psi_2&=\f{\Psi_2^0}{(r+i\Sigma)^3},\\
\Psi_3&=\f{\Psi_3^0}{(r+i\Sigma)^2}+\f{\Psi_3^1}{(r+i\Sigma)^3}+\f{\Psi_3^2}{(r+i\Sigma)^4},\\
\Psi_4&=\f{\Psi_4^0}{(r+i\Sigma)^1}+\f{\Psi_4^1}{(r+i\Sigma)^2}+\f{\Psi_4^2}{(r+i\Sigma)^3}+\f{\Psi_4^3}{(r+i\Sigma)^4}+\f{\Psi_4^4}{(r+i\Sigma)^5},
\ee
\esub
with
\bsub
\be
\Psi_3^1&=-\Ethb\Psi_2^0,\\
\Psi_3^2&=\f{3}{2}\Psi_2^0\big(i\Ethb\Sigma+\bar{\Omega}_1\big),\\
\Psi_4^1&=-\Ethb\Psi_3^0,\\
\Psi_4^2&=\Psi_3^0\big(i\Ethb\Sigma+\bar{\Omega}_1\big)-\f{1}{2}\Ethb\Psi_3^1+\f{3}{2}\Psi_2^0\lambda_1,\\
\Psi_4^3&=\Psi_3^1\big(i\Ethb\Sigma+\bar{\Omega}_1\big)-\f{1}{2}\Ethb\Psi_3^2,\\
\Psi_4^4&=\Psi_3^2\big(i\Ethb\Sigma+\bar{\Omega}_1\big).
\ee
\esub
The expressions for $(\Psi_3^0,\Psi_4^0,\lambda_1)$ are those given in \eqref{NP leading definitions}, but should be evaluated with $\sigma_2=0$ because of the algebraic speciality. Similarly, $\Omega_1$ can be found in \eqref{NP solution for Omega} or \eqref{metric Omega1} with $\sigma_2=0$ and $C_{ab}=0$. For the spin coefficients we find
\bsub
\be
\rho&=-\f{1}{r+i\Sigma},\\
\sigma&=0,\\
\alpha&=\f{\alpha_1}{r+i\Sigma},\\
\beta&=-\bar{\alpha},\\
\tau&=\f{\tau_1}{r+i\Sigma},\\
\lambda&=\f{\lambda_1}{r+i\Sigma},\\
\mu&=\f{\mu_1}{r-i\Sigma}-\f{r\Psi_2^0}{(r-i\Sigma)(r+i\Sigma)^2},\\
\gamma&=\gamma_0-\alpha_1\tau+\bar{\alpha}_1\bar{\tau}-\f{\Psi_2^0}{2(r+i\Sigma)^2},\\
\nu&=\nu_0-\lambda_1\tau+\bar{\tau}\left(\f{\Psi_2^0}{2(r+i\Sigma)}-\mu_1\right)-\f{\Psi_3^0}{r+i\Sigma}-\f{\Psi_3^1}{2(r+i\Sigma)^2}-\f{\Psi_3^2}{3(r+i\Sigma)^3},
\ee
\esub
where the leading coefficients $(\alpha_1,\tau_1,\lambda_1,\mu_1,\gamma_0,\nu_0)$ are given in \eqref{NP leading definitions} with $\sigma_2=0$. Finally, the components of the tetrad are given by
\bsub\label{AS tetrad}
\be
W&=1-2\text{Re}(L\bar{\tau}),\\
U&=-2\gamma_0r+\big(\Ethb-\bar{\tau}_1\big)\tau_1+\bar{\mu}_1-i(\partial_u+2\gamma_0)\Sigma-\text{Re}\left(\f{\Psi_2^0+2\tau_1\bar{\Omega}_1}{r+i\Sigma}\right),\\
X^a&=-2\text{Re}(m^a_1\bar{\tau}),\label{AS Xa}\\
Z&=\f{L}{r-i\Sigma},\\
\Omega&=\f{\Omega_1}{r-i\Sigma},\\
m^a&=\f{m^a_1}{r-i\Sigma}.\label{AS ma}
\ee
\esub
At this stage we have solved all of the radial NP equations \eqref{4d NP radial}. As announced, the solution space is in explicitly closed form with finite expansions in $-\rho^{-1}=r+i\Sigma$ and its complex conjugate. The line element reconstructed from this tetrad is
\be
\de s^2=(r^2+\Sigma^2)q_{ab}\de x^a\de x^b+\ell\big(\ell g_{uu}+2\de r+2(X_a-\Omega_a)\de x^a\big),
\ee
where $\ell=-(\de u-L_a\de x^a)$ and $g_{uu}=2U+X_a(X^a-2\Omega^a)$. Although this corresponds to the same solution space, this metric is slightly different from that of theorem 27.1 in \cite{Stephani:2003tm}. This is because in this reference the tetrad has $X^a=0$, and the data contained in $X^a$ is encoded instead in $\Omega^a$ (which is the complex scalar $W$ there).

As shown in \cite{Mao:2024jpt}, one can then go through the supplementary NP equations \eqref{4d NP supplementary} and check that they are exactly satisfied to all orders once the (finite set of) algebraic and differential constraints given in sections \ref{sec:4d NP solution space} and \ref{sec:4d NP evolution equations} are implemented. One should simply recall that this must be done with the conditions $\Psi_0=\Psi_1=0$ and $\sigma_2=0$.

At the end of the day, the algebraically special solution space constructed above is parametrized by the data $\big(m^a_1,L,\text{Re}(\Psi_2^0)\big)$. The twist potential $L$ has an arbitrary time dependency and gives rise to the twist $\Sigma=\text{Im}(\Eth\bar{L})$. The frame $m^a_1$ defines the arbitrary metric $q^{ab}=2m^{(a}_1m^{b)}_1$, and this relation can be inverted up to a $\text{U}(1)$ freedom as in \eqref{frame m1}. The time dependency of the frame is constrained by \eqref{NP2g} to satisfy $(\partial_u+2\gamma_0)m^a_1=0$, or equivalently for the metric we have $(\partial_u-4\gamma_0)q_{ab}=0$ with $4\gamma_0=\partial_u\ln\sqrt{q}$. The frame $m^a_1$ and the twist enable to define $m^a$ as in \eqref{AS ma}, and this frame then reconstructs the angular metric\footnote{Note that there is a subtle $\text{U}(1)$ ambiguity lurking around. Indeed, one may also reconstruct a frame for $\gamma^{ab}$ by using the analogue of \eqref{frame m1} and writing
\be
\underline{m}^a=\sqrt{\f{\gamma_{\theta\theta}}{2\gamma}}\left(\f{\sqrt{\gamma}+i\gamma_{\theta\phi}}{\gamma_{\theta\theta}}\,\delta^a_\theta-i\delta^a_\phi\right).
\ee
By construction this frame also satisfies $\gamma^{ab}=2\underline{m}^{(a}\bar{\underline{m}}^{b)}$. The frames $m^a$ defined in \eqref{AS ma} and $\underline{m}^a$ agree at leading order, so $m^a=\underline{m}^a+\O(r^{-2})$, but differ in general by a phase as $m^a=e^{i\varphi}\underline{m}^a$ with $\varphi=\arctan(\Sigma/r)$. One can see that this subtlety is due to the presence of a non-vanishing twist $\Sigma$.} $g^{ab}=\gamma^{ab}=2m^{(a}\bar{m}^{b)}=q^{ab}/(r^2+\Sigma^2)$. Finally, $\text{Re}(\Psi_2^0)\eqqcolon-\M$ is the mass aspect.

We must recall however that $\Psi_2^0$ is not completely arbitrary. From \eqref{EOM Psi10} and \eqref{EOM Psi20}, we find that in the algebraically special case it must satisfy the constraints
\bsub
\be
(\Eth-3\tau_1)\Psi_2^0&=0,\label{AS Psi20 constraint}\\
(\partial_u+6\gamma_0)\Psi_2^0&=(\Eth-2\tau_1)\Psi_3^0,\label{AS Psi20 evolution}
\ee
\esub
where specializing \eqref{Psi30} to the algebraically special case gives
\be
\Psi_3^0=\Eth(\Ethb-\bar{\tau}_1)\bar{\tau}_1+2i\big(\Ethb\Sigma+3\Sigma\Ethb-2\Sigma\bar{\tau}_1\big)\gamma_0-\Ethb\mu_R.
\ee
Now, one should recall that $\text{Im}(\Psi_2^0)$ is not an independent data, but is fully determined by \eqref{NP Im Psi20} (properly reduced to the algebraically special case). In particular, as in the algebraically general case the imaginary part of \eqref{AS Psi20 evolution}, or equivalently equation \eqref{EOM curly M dual}, follows automatically from the definition of the dual mass $\text{Im}(\Psi_2^0)$. In conclusion, we therefore find that the data $\big(m^a_1,L,\text{Re}(\Psi_2^0)\big)$ must satisfy the two equations contained in \eqref{AS Psi20 constraint} (since this is a spin 1 equation), and the single equation given by the real part of \eqref{AS Psi20 evolution}. In metric terms, using all the results and notations from section \ref{sec:flux-balance} and reducing to the algebraically special case, we find that these equations are
\bsub
\be
\big(\partial_a+L_a\partial_u+3\partial_uL_a\big)\M+\big(\widetilde{\partial}_a+\widetilde{L}_a\partial_u+3\partial_u\widetilde{L}_a\big)\widetilde{\M}&=0,\\
(\partial_u+6\gamma_0)\M-\f{1}{2}\big(D_a+4\gamma_0L_a+L_a\partial_u+2\partial_uL_a\big)\J^a&=0,
\ee
\esub
where $\Psi_2^0=-\big(\M+i\widetilde{\M}\big)$. These are the constraints to be solved in order to obtain an explicit algebraically special solution.

To conclude, let us note that when solving the radial equations we have discarded integration constants in $X^a=X^a_0-2\text{Re}(m^a_1\bar{\tau})$ and $W=W_0-2\text{Re}(L\bar{\tau})$ by setting them to $X^a_0(u,x^b)=0$ and $W_0(u,x^a)=1$, as we did in the algebraically general case. This makes the study of some algebraically special solutions slightly more convoluted. We illustrate this in appendix \ref{app:C-metric} with the C-metric, and keep the investigation of the solution space with integration constants and cosmological constant for future work.

\subsection{Asymptotic symmetries}
\label{sec:AS symmetries}

To study the asymptotic symmetries of the algebraically special solution space, one can simply start from the algebraically general case studied in section \ref{sec:symmetries} and perform the reduction to the algebraically special case. When setting $\Psi_0=\Psi_1=\sigma=0$, the transformation law \eqref{delta xi Psi00} implies consistently that the vanishing of $\Psi_0^0$ is preserved, but \eqref{delta xi sigma} and \eqref{delta xi Psi10} require to impose the vanishing of the Carrollian boost $\Upsilon=0$. Since in \eqref{AKVs} and \eqref{subleading AKV} we have shown that the subleading terms in the asymptotic Killing vector are all proportional to $\Upsilon^a$, this immediately implies that the AKV $\xi^\mu$ truncates and becomes
\be\label{AS AKV}
\xi^u=f(u,x^a),
\q\q
\xi^r=rW(u,x^a),
\q\q
\xi^a=Y^a(x^b),
\ee
with $W=-\partial_uf$. The transformation laws for the data $\big(m^a_1,q_{ab},L,\text{Re}(\Psi_2^0)\big)$ can then be read directly in \eqref{transformation laws} after setting $\Upsilon=0$.

The algebraically special solution space parametrized by the tetrad \eqref{AS tetrad} is sufficiently tractable to allow for the computation of general expressions for the asymptotic charges and for the symplectic structure. We postpone this detailed analysis of the charges, especially in the case of type D solutions, to future work, but in the next section we compute the charges in the case of explicit examples such as the supertranslated Schwarzschild and Kerr--Taub--NUT solutions.

\subsection{Type D solutions}
\label{sec:AS type D}

In this section we exploit the finite form of algebraically special solutions with twist potential in order to study the impact of supertranslations on the Kerr--Taub--NUT family and to compute the associated charges. We also discuss, for spacetimes of Petrov type D, the relationship between the tetrad constructed above and the Kinnersley tetrad which makes the type D nature of the solutions manifest. We then review how this tetrad can be used to construct a Killing--Yano and a Killing--St\"ackel tensor, with applications to the supertranslated Kerr--Taub--NUT solution.

\subsubsection{Supertranslated and superrotated Schwarzschild}
\label{sec:supertranslated Schwarzschild}

Let us start with the Schwarzschild solution in Bondi coordinates, which is
\be\label{Schwarzschild}
\de s^2=-\left(1-\f{2M}{r}\right)\de u^2-2\de u\,\de r+r^2\de\Omega^2.
\ee
Changing to stereographic coordinates using \eqref{stereographic}, and then replacing the angular variables by arbitrary functions as $z\mapsto f(z)$ and $\bar{z}\mapsto\bar{f}(\bar{z})$, we obtain
\be
\de s^2=-\left(1-\f{2M}{r}\right)\de u^2-2\de u\,\de r+\f{4f'\bar{f}'r^2}{\big(1+f\bar{f}\big)^2}\de z\,\de\bar{z}.
\ee
This is the superrotated Schwarzschild solution.

More interestingly, we can also study the impact of a finite supertranslation in the spirit of \cite{Hawking:2016msc,Hawking:2016sgy,Compere:2016hzt,Compere:2016jwb}. To do so, we simply shift the retarded time coordinate by an arbitrary function of the angles as $u\mapsto u-C(x^a)$, where $C$ is the supertranslation field \cite{Lu:2025fzm}. Then \eqref{Schwarzschild} becomes
\be\label{supertranslated schwarzschild}
\de s^2
&=-\left(1-\f{2M}{r}\right)(\de u-\de C)^2-2(\de u-\de C)\de r+r^2\de\Omega^2\cr
&=-\left(1-\f{2M}{r}\right)\de u^2-2\de u\,\de r+r^2\de\Omega^2+\left[\left(1-\f{2M}{r}\right)\big(2\de u-\partial_bC\de x^b\big)+2\de r\right]\big(\partial_aC\de x^a\big),
\ee
which is the finitely supertranslated Schwarzschild solution. This is indeed of the form \eqref{metric} with
\be
W=-1,
\q\!\!
U=-\f{1}{2}\left(1-\f{2M}{r}\right),
\q\!\!
X_a=\Omega_a=0,
\q\!\!
Z_a=L_a=\partial_aC,
\q\!\!
\gamma_{ab}=r^2
\begin{pmatrix}
1&0\\
0&\sin^2\theta
\end{pmatrix},
\ee
which also shows that the gauge conditions $g_{ur}=-1$ and $\partial_rZ_a=0$ are satisfied. Note that although the twist potential $L_a$ is non-vanishing, the actual twist is vanishing, i.e. $\Sigma=0$. This simple example illustrates a property analogous to that of the Kerr spacetime in the Bondi gauge with non-vanishing twist potential, namely the fact that the later enables us to write the metric in a finite radial expansion. Indeed, if we insist on writing the solution \eqref{supertranslated schwarzschild} in the standard Bondi gauge with $g_{ra}=0$, the change of coordinates which brings the metric in this form creates an infinite radial expansion. This can also be understood through the residual symmetries. Indeed, when $L_a=0$ the Carroll boost $\Upsilon_a$ becomes related to the supertranslation generator $\xi^u=f$ as $\Upsilon_a=\partial_af$, and one finds from \eqref{AKVs} that the supertranslation percolates to every subleading order in the residual Killing vector. If we then compute a (infinitesimal or finite) supertranslation of \eqref{Schwarzschild} away from $\I^+$, the supertranslation creates an infinite amount of subleading terms in the metric. While this observation is of course not conceptually new, it illustrates the simplifications brought by the twist potential.

Another advantage of working with the twist potential is that it simplifies the study of the charges. Since the supertranslated Schwarzschild solution \eqref{supertranslated schwarzschild} is algebraically special, one can use the results of the previous section to compute the associated charges. Evaluating the $(ur)$ component of the Iyer--Wald charge \eqref{IW charge} with the vector field \eqref{AS AKV}, and assuming that $(f,Y^a,C)$ are field-independent, the result is integrable and given by
\be
Q(\xi)=\sqrt{q}\,M\big(2f-3Y^a\partial_aC\big).
\ee
As expected this result is consistent with \cite{Hawking:2016msc,Hawking:2016sgy,Compere:2016hzt,Compere:2016jwb}, and shows that the supertranslated solution is characterized by the presence of a non-vanishing superrotation charge. Interestingly however, this result is obtained here in finite form and is manifestly independent of the radius $r$. Once again, this is due to the fact that we are working with a non-vanishing twist potential and with the resummed form of the algebraically special solution.

\subsubsection{Supertranslated Kerr--Taub--NUT}

The example which has partly motivated this work is the Kerr--Taub--NUT solution, and the possibility of writing it in finite form in Bondi gauge by allowing for $g_{ra}\neq0$. In outgoing Eddington--Finkelstein coordinates, or equivalently in Bondi gauge with non-vanishing twist potential, it can be obtained from a tetrad of the form \eqref{NP tetrad} and \eqref{AS tetrad} which explicitly reads
\bsub\label{KTN tetrad}
\be
\ell&=\partial_r,\\
n&=\partial_u+\f{a^2\sin^2\theta-\Delta}{2(r^2+\Sigma^2)}\partial_r,\\
m&=\f{1}{r-i\Sigma}\left(L\partial_u+\f{ia\sin\theta}{\sqrt{2}}\partial_r+\f{1}{\sqrt{2}}\big(\delta^a_\theta-i\csc\theta\,\delta^a_\phi\big)\partial_a\right),
\ee
\esub
and where
\be
L=-\f{i\csc\theta}{\sqrt{2}}\chi,
\q
\chi=a\sin^2\theta-2n(\cos\theta+s),
\q
\Sigma=a\cos\theta+n,
\q
\Delta=r^2-2Mr+a^2-n^2.
\ee
Here $(M,a,n)$ are respectively the mass, spin, and NUT parameters, and $s$ is the so-called Manko--Ruiz parameter which encodes the position of the Misner string \cite{Manko:2005nm}. The corresponding line element is
\be\label{KTN}
\de s^2
&=\f{a^2\sin^2\theta-\Delta}{r^2+\Sigma^2}\de u^2-2\de u\,\de r+2\chi\,\de r\,\de\phi-\f{2}{r^2+\Sigma^2}\Big(a\big(r^2+\Sigma^2+a\chi\big)\sin^2\theta-\chi\Delta\Big)\de u\,\de\phi\cr
&\pe+(r^2+\Sigma^2)\de\theta^2+\f{1}{r^2+\Sigma^2}\Big(\big(r^2+\Sigma^2+a\chi\big)^2\sin^2\theta-\chi^2\Delta\Big)\de\phi^2,
\ee
which sits nicely in the algebraically general solution space with non-vanishing twist potential which we have constructed, and more specifically in the algebraically special sector.

As a simple application, one can take advantage of the form of this metric and of the algebraically special asymptotic symmetries \eqref{AS AKV} to compute the charges for Kerr--Taub--NUT. This simplifies the calculations done for Kerr in \cite{Barnich:2011mi} and \cite{Geiller:2024amx} using respectively the Bondi--Sachs gauge and the partial Bondi gauge. For the exact Killing vector $\xi=T\partial_u+Y\partial_\phi$ with constant parameters $T$ and $Y$, we find that the charge is of the form
\be
\slashed{\delta}Q(\xi)=\sqrt{q}\,\delta\Big\{\!-2rn(\cos\theta+s)Y+M\big(2T-3\chi Y\big)\Big\}+\O(r^{-1}),
\ee
where $\sqrt{q}=\sin\theta$. The first orders are integrable, and non-integrable contributions due to products of the parameters $(M,a,n)$ and their variations start to appear at order $\O(r^{-2})$. One can also see that the result contains a divergent piece when $r\to\infty$ due to the NUT parameter $n$. However, for $s=0$ and when the charge aspect is smeared over the asymptotic two-sphere, the integral over $\theta\in[0,\pi]$ of $\sqrt{q}\,\cos\theta$ vanishes. Of course, since the computation of the charge relies on a finite form of the metric and of the vector field, the final result can be written in explicitly closed form without requiring a radial expansion. The result is however lengthy and irrelevant if we are interested in the asymptotic value of the charge\footnote{In the case of Kerr for example, i.e. when $n=0$, one can compute the full charge for the exact Killing vector $\xi=T\partial_u+Y\partial_\phi$ to find
\be
\slashed{\delta}Q(\xi)
&=\sqrt{q}\,\delta\Bigg\{Tr\sec^2\theta\left(\f{r^2+\Sigma^2\cos^2\theta}{r^2+\Sigma^2}+\f{1}{2}(1+\cos^2\theta)\ln(r^2+\Sigma^2)\right)+TM\f{16r^4+8r^2a^2(1+\cos^2\theta)+a^4\big(\cos(4\theta)-1\big)}{8(r^2+\Sigma^2)^2}\nn\\
&\pe\q\quad-YMa\sin^2\theta\f{3r^4+r^2(a^2+\Sigma^2)-a^2\Sigma^2}{(r^2+\Sigma^2)^2}\Bigg\}+\sqrt{q}\,TM\delta(a^2)\f{r^2(1-3\cos^2\theta)-\Sigma^2(1+\cos^2\theta)}{2(r^2+\Sigma^2)^2}.
\ee
As expected, the non-integrability comes from the time translation which probes directions which are not tangential to the codimension-2 boundary (in other words, from the $\iota_\xi\theta$ term in the Iyer--Wald charge \cite{Iyer:1994ys,Compere:2018aar}). This non-integrability is typical of quasi-local charges.}.

Let us now consider a supertranslation of the Kerr--Taub--NUT solution \eqref{KTN}. As in the Schwarzschild case, this is obtained by performing a shift $u\mapsto u-C(x^a)$. The resulting line element is then
\be\label{supertranslated Kerr}
\de s^2=\de\bar{s}^2+\Big(\bar{g}_{uu}\big(\partial_bC\de x^b-2\de u\big)+2\de r-2\bar{g}_{u\phi}\de\phi\Big)\big(\partial_aC\de x^a\big),
\ee
where the bar refers to the initial metric \eqref{KTN} without the supertranslation. We can then compute the charges associated with this supertranslated solution. Since the leading angular metric $q_{ab}$ is fixed and time-independent, we get from \eqref{delta root q} that the Weyl parameter is fixed to $2W=-2\partial_uf=-D_aY^a$. Taking this into account, we find that the charge is given by
\be
\slashed{\delta}Q(\xi)
&=r\sqrt{q}\,\delta\left\{\f{1}{4}\chi\csc^2\theta\partial_\phi\big(D_aY^a\big)-2n(\cos\theta+s)Y^\phi\right\}\cr
&\pe+\sqrt{q}\,\delta\left\{M\big(2f-3Y^a\partial_aC-3\chi Y^\phi\big)+\f{1}{2}a\big(Y^a\partial_a\partial_\phi C-\partial_\phi f\big)-\left(\Sigma^2+\f{1}{4}a^2\sin^2\theta\right)D_aY^a\right\}\q\cr
&\pe+\sqrt{q}\,(a\delta n)(\cos\theta+s)D_aY^a+\O(r^{-1}).
\ee
Just like in the Schwarzschild case, one finds that the supertranslation field enters the charge through the contribution of a superrotation only. It would be interesting to use the finite form \eqref{supertranslated Kerr} of the metric with twist to extend the results of \cite{Hawking:2016msc,Hawking:2016sgy,Compere:2016hzt,Compere:2016jwb} to Kerr, in the spirit of \cite{Bhat:2024cyq}.

\subsubsection{Kinnersley tetrad}
\label{sec:Kinnersley}

We now discuss in more detail how type D spacetimes sit within the algebraically special solution space constructed above with the tetrad \eqref{AS tetrad}. If a spacetime has a repeated principal null direction such that $\Psi_0=\Psi_1=0$, then it is of Petrov type D if and only if the remaining non-vanishing Weyl scalars satisfy
\be\label{type D condition}
3\Psi_2\Psi_4-2(\Psi_3)^2=0.
\ee
This is a Lorentz-invariant condition in the same way as \eqref{S index}. Type D spacetimes are typically characterized by the fact that $n$ is the second repeated null direction (in addition to $\ell$), and therefore by the supplementary vanishing conditions $\Psi_3=\Psi_4=0$. However, these conditions should \textit{not} be imposed in the solution space which we have built above with the tetrad \eqref{AS tetrad}. The reason for this is that we have built this solution space and the tetrad with the gauge condition $\pi=0$, while the description of general type D solutions with $\Psi_3=\Psi_4=0$ requires to have $\pi\neq0$ \cite{Kinnersley:1968zz,Kinnersley:1969zza}. For example, the tetrad \eqref{KTN tetrad} for Kerr--Taub--NUT has the opposite property. It leads to $\Psi_3\neq0$ and $\Psi_4\neq0$ and satisfies $\pi=0$ by construction (since we have chosen this tetrad as a member of our solution space built in the gauge $\pi=0$). With the tetrad \eqref{KTN tetrad}, the type D nature of the solution is therefore not fully manifest, although one can still check that \eqref{type D condition} is satisfied.

A tetrad which is better adapted to the description of solutions of Petrov type D is the so-called Kinnersley tetrad \cite{Kinnersley:1968zz,Kinnersley:1969zza}. It can be obtained from \eqref{NP tetrad} and \eqref{AS tetrad} by performing a Lorentz transformation of class I (or a null rotation about $\ell$) parametrized by
\be\label{parameter p}
\bar{p}=-\f{\Psi_3}{3\Psi_2},
\ee
when $\Psi_2\neq0$. We then obtain a new tetrad given by
\be
\ell\to\ell_\text{K}=\ell,
\q\q
n\to n_\text{K}=n+\bar{p}m+p\bar{m}+p\bar{p}\ell,
\q\q
m\to m_\text{K}=m+p\ell.
\ee
Since algebraically special solutions have $\Psi_0=\Psi_1=0$, the effect of this change of tetrad on the remaining Weyl scalars is
\be
\Psi_2\to\Psi_2,
\q\q
\Psi_3\to0,
\q\q
\Psi_4\to\Psi_4+4\bar{p}\Psi_3+6\bar{p}^2\Psi_2=\Psi_4-\f{2}{3}\f{(\Psi_3)^2}{\Psi_2}.
\ee
Its effect on the spin coefficients $(\kappa,\epsilon,\pi)$, which are initially vanishing with the choice of tetrad \eqref{AS tetrad}, is to preserve $\kappa=\epsilon=0$ while changing
\be
\pi\to\pi+\ell^\mu\partial_\mu\bar{p}=\partial_r\bar{p}.
\ee
From the above transformation law of the Weyl scalars, one can see as announced that solutions which satisfy the Lorentz-invariant type D condition \eqref{type D condition} actually have manifestly vanishing $\Psi_3=\Psi_4=0$ when using the Kinnersley tetrad. In summary, since type D solutions satisfy \eqref{type D condition}, their $\Psi_4$ (for any choice of tetrad) is actually determined by $\Psi_{3,2}$, and using the Kinnersley tetrad then amounts to moving the data contained in $\Psi_3$ into $\pi$.

Explicitly, for the Kerr--Taub--NUT solution \eqref{KTN} we find that the parameter \eqref{parameter p} is actually given by $p=-\Omega$, so that performing the type I transformation on \eqref{KTN tetrad} leads to the Kinnersley tetrad
\bsub\label{KTN Kinnersley tetrad}
\be
\ell_\text{K}&=\partial_r,\\
n_\text{K}&=\f{1}{r^2+\Sigma^2}\left(\big(r^2+\Sigma^2+a\chi\big)\partial_u-\f{\Delta}{2}\partial_r+a\partial_\phi\right)=-g^{r\mu}\partial_\mu+\f{1}{2}g^{rr}\partial_r,\\
m_\text{K}&=\f{1}{r-i\Sigma}\left(L\partial_u+\f{1}{\sqrt{2}}\big(\delta^a_\theta-i\csc\theta\,\delta^a_\phi\big)\partial_a\right).
\ee
\esub
In the literature, the Kinnersley tetrad often appears in Boyer--Lindquist coordinates $(t,r,\theta,\psi)$. These can be obtained from the Eddington--Finkelstein coordinates $(u,r,\theta,\phi)$ by setting
\be\label{BL to EF}
\de u=\de t-\f{1}{\Delta}(r^2+a^2+n^2-2ans)\de r,
\q\q
\de\phi=\de\psi-\f{a}{\Delta}\de r,
\ee
in the metric or in the cotetrad.

\subsubsection{Killing--Yano and Killing--St\"ackel tensors}

One of the remarkable properties of type D solutions is the existence of so-called hidden symmetries arising from Killing--Yano and Killing--St\"ackel\footnote{Symmetric Killing--St\"ackel tensors are often just called Killing tensors.} tensors \cite{Carter1968,Kashiwada1968,WalkerPenrose1970,Debever1971,Floyd1973} (see also \cite{Krtous:2006qy,Frolov:2008jr,Frolov:2009aq,Brink:2009mq,Frolov:2017kze} more recently). The later enable in particular to build conserved quantities of higher order in the momenta. This is illustrated by the famous example of the Carter constant for the Kerr black hole, which arises from a rank-2 Killing--St\"ackel tensor. The advantage of the Kinnersley tetrad is that it enables us to immediately write down rank-2 Killing--St\"ackel and Killing--Yano tensors \cite{Stephani:2003tm}. Dropping the subscript K and denoting the Kinnersley tetrad by $(\ell,n,m,\bar{m})$, one can define
\be\label{K and Y defs}
K_{\mu\nu}\coloneqq\Sigma^2\ell_{(\mu}n_{\nu)}+r^2m_{(\mu}\bar{m}_{\nu)},
\q\q
Y_{\mu\nu}\coloneqq\Sigma\ell_{[\mu}n_{\nu]}+irm_{[\mu}\bar{m}_{\nu]}.
\ee
One can then check that these symmetric and antisymmetric tensors satisfy
\be
\nabla_{(\rho}K_{\mu\nu)}=0,
\q\q
\nabla_{(\rho}Y_{\mu)\nu}=0\quad\Leftrightarrow\quad\nabla_\rho Y_{\mu\nu}=\nabla_{[\rho}Y_{\mu\nu]},
\ee
meaning that $K$ is a Killing--St\"ackel tensor while $Y$ is a Killing--Yano tensor. In addition one has
\be
2{Y_\mu}^\rho Y_{\rho\nu}=-K_{\mu\nu},
\ee
illustrating a particular case of the general property that the symmetrized square of two Killing--Yano tensors is a Killing--St\"ackel tensor. Finally, an additional property is that the Killing--Yano tensor can be obtained from a potential 1-form $y$ as $Y=*\de y$, or in components $Y_{\mu\nu}=\sqrt{-g}\,{\epsilon_{\mu\nu}}^{\rho\sigma}\partial_\rho y_\sigma$. In the case of Kerr--Taub--NUT with Manko--Ruiz parameter this 1-form is given by
\be
4y=\big(\Sigma^2-r^2-2n^2\big)\de t+\Big(a^3\sin^2\theta+2na^2\cos\theta(s\cos\theta-1)+an^2(4s\cos\theta-1)+\chi(r^2+n^2)\Big)\de\psi,\!
\ee
but a compact expression which depends only on the tetrad vectors is not available in general (to the best of our knowledge). Note that here we have given this 1-form in Boyer--Lindquist coordinates, but that the result in Eddington--Finkelstein or Bondi coordinates with twist is obtained immediately using \eqref{BL to EF}.

Although these are standard results when working e.g. with Boyer--Lindquist coordinates for Kerr, we mention them here to show that they naturally extend to the Bondi gauge with non-vanishing twist potential. Given a type D solution whose metric is of the form \eqref{metric} with $g_{ur}=-1$ and $\partial_rg_{ra}=0$, there is an algorithmic way of reconstructing a corresponding tetrad \eqref{AS tetrad} and then a Kinnersley tetrad, from which one can finally obtain the Killing--Yano and Killing--St\"ackel tensors. The later can in turn be used to construct the conserved quantity $Q=K^{\mu\nu}p_\mu p_\nu$ where $p_\mu=g_{\mu\nu}\dot{x}^\nu$ is the momentum along affine geodesics.

As a straightforward application, one can compute the Carter constant of the supertranslated Schwarzschild black hole \eqref{supertranslated schwarzschild} in the case of an axisymmetric supertranslation field $C=C(\theta)$. We find
\be
Q=\f{1}{2}\Big(p_\theta^2+p_\phi^2\csc^2\theta+2p_up_\theta\partial_\theta C+p_u^2(\partial_\theta C)^2\Big),
\ee
which commutes as it should with the Hamiltonian $H=g^{\mu\nu}p_\mu p_\nu$. Using the conserved quantities $E=p_u$ and $L=p_\phi$ arising from the Killing vectors, together with the vanishing of the Lagrangian $g_{\mu\nu}\dot{x}^\mu\dot{x}^\nu$, we then obtain that affine null geodesics satisfy
\be
\dot{u}=\f{r(\dot{r}+E)}{2M-r}+\dot{\theta}\partial_\theta C,\quad\ \ \,
\dot{r}=\f{\sqrt{r^3E^2-2rQ+4QM}}{r^{3/2}},\quad\ \ \,
\dot{\theta}=\f{\sqrt{2Q-L^2\csc^2\theta}}{r^2},\quad\ \ \,
\dot{\phi}=\f{L\csc^2\theta}{r^2}.
\ee
As expected, this result is the same as if we had computed the finite supertranslation $u\mapsto u-C(\theta)$ of the $\dot{x}^\mu$ of the initial Schwarzschild spacetime. Interestingly, one can see that since this supertranslation and the supertranslated metric are \eqref{supertranslated schwarzschild} written in the Bondi gauge with twist potential, the photon orbits inferred from $\de r/\de\theta=\dot{r}/\dot{\theta}$ are not affected by the supertranslation field. This is in stark contrast with the results claimed in \cite{Zhu:2022shb} (see also \cite{Sarkar:2021djs} for supertranslated Vaidya), where in the spirit of \cite{Hawking:2016msc,Hawking:2016sgy} an infinitesimal supertranslation is implemented in the standard Bondi gauge, and where the resulting photon orbits \textit{do} depend on the supertranslation field. This illustrates the subtle role of coordinates transformations when discussing e.g. supertranslated solutions and their possible physical imprints. In a sense, the Bondi gauge with twist potential is adapted to supertranslations since they are very simply accommodated by a change of the retarded time coordinate, without having to change the other coordinates in order to ensure that $g_{ra}=0$. As a consequence of this ``simplicity'' of the action of supertranslations, the photon orbits remain unchanged. The supertranslation field could however affect observables such as the Shapiro delay inferred from $\de r/\de u$, although this should also involve the conversion to the time measured by an asymptotic inertial observer.

\section{Three-dimensional case}
\label{sec:3d}

In this final section we use the NP formalism adapted to three-dimensional spacetimes \cite{Hall:1987vw,Milson:2012ry,Barnich:2016lyg} in order to construct an asymptotically-locally-(A)dS$_3$ solution space in Bondi gauge with $g_{r\phi}\neq0$. For the sake of generality we perform this construction with a non-vanishing cosmological constant, and we also keep all the functions of the coordinates $(u,\phi)$ which arise as radial integration constants in addition to the mass and angular momentum aspects. The upshot is a solution space with 8 such functions. The mass and the angular momentum satisfy evolution equations in $u$, while the remaining 6 functions have an unspecified time dependency. This result encompasses all the known vacuum asymptotically-locally-(A)dS$_3$ solution spaces constructed in Bondi gauge \cite{Barnich:2006av,Barnich:2012aw,Ruzziconi:2020wrb,Alessio:2020ioh,Geiller:2021vpg}, and in particular the derivative expansion gauge \cite{Campoleoni:2018ltl,Ciambelli:2020eba,Ciambelli:2020ftk,Campoleoni:2022wmf} which has $g_{r\phi}\neq0$. As an advantage of working in Bondi gauge, the flat limit $\Lambda\to0$ can always be taken in order to recover results for asymptotically-locally-flat spacetimes.

\subsection{Triad and metric with ``twist''}

For the three-dimensional NP formalism we will follow the notations of \cite{Barnich:2016lyg}, but change the signature to be mostly plus $(-,+,+)$. We pick a set of coordinates $x^\mu=(u,r,\phi)$, consider the internal metric
\be
\eta_{ij}=
\begin{pmatrix}
0&-1&0\\
-1&0&0\\
0&0&1/2
\end{pmatrix},
\ee
and the triad $e_i=(\ell,n,m)$ formed by the real vectors
\be
\ell=\partial_r,
\q\q
n=W\partial_u+U\partial_r+X\partial_\phi,
\q\q
m=Z\partial_u+\Omega\partial_r+F\partial_\phi.
\ee
The inverse spacetime metric is then given by
\be\label{3d inverse metric}
g^{\mu\nu}=e^\mu_ie^\nu_j\eta^{ij}=-2\ell^{(\mu}n^{\nu)}+2m^\mu m^\nu
=
\begin{pmatrix}
2Z^2&-W+2Z\Omega&2ZF\\
-W+2Z\Omega&-2U+2\Omega^2&-X+2F\Omega\\
2ZF&-X+2F\Omega&2F^2
\end{pmatrix},
\ee
while the direct metric is
\bsub\label{3d direct metric}
\be
g_{\mu\nu}&=\f{1}{2G^2}
\begin{pmatrix}
4UF^2+X(X-4\Omega F)&-2GF&4F(W\Omega-UZ)-2G\Omega-WX\\
-2GF&0&2GZ\\
4F(W\Omega-UZ)-2G\Omega-WX&2GZ&W^2+4UZ^2-4WZ\Omega
\end{pmatrix},\\
G&=\f{1}{\sqrt{-2\det(g)}}=WF-ZX.\label{3d NP G}
\ee
\esub
The triad vectors satisfy $\ell^\mu n_\mu=-1$ and $m^\mu m_\mu=1/2$, with all the other contractions vanishing. One should note in particular that $m$ is real and non-null as a consequence of working in three spacetime dimensions.

The three-dimensional NP formalism only features 9 spin coefficients, whose definitions are given in \eqref{3d spin coefficients}. As in the four-dimensional case, $(\kappa,\epsilon,\pi)$ appear in the parallel transport of the triad along $\ell$, and we have
\bsub
\be
\ell^\mu\nabla_\mu\ell&=\epsilon\ell-2\kappa m,\\
\ell^\mu\nabla_\mu n&=-\epsilon n+2\pi m,\\
\ell^\mu\nabla_\mu m&=\pi\ell-\kappa n.
\ee
\esub
Using the metric \eqref{3d direct metric} one can compute
\be
\kappa=\f{Z\partial_rF-\partial_rZF}{WF-ZX},
\q\q
\epsilon=\f{Z\partial_rX-\partial_rWF}{WF-ZX}.
\ee
We do not provide the explicit expressions for the other off-shell spin coefficients since they are not needed. Once the choice $\kappa=0$ is made, implying that $\ell$ describes a congruence of null geodesics, one can reach the gauge $\pi=\epsilon=0$ with internal Lorentz transformations. As in the four-dimensional case, $\epsilon=0$ implies that the congruence is affinely parametrized. One can then easily check that when $\kappa=\epsilon=0$ the Sachs optical scalars defined in \eqref{optical scalars} reduce to
\be\label{3d optical scalars}
\theta=-\partial_r\ln\sqrt{F},
\q\q
|\sigma_\text{Sachs}|^2=0,
\q\q
\omega^2=0.
\ee
This means that affinely parametrized congruences of null geodesics in three-dimensional spacetimes have vanishing shear and twist. It should be noted however that here $|\sigma_\text{Sachs}|^2$ only stands for the RHS of \eqref{optical shear}, and is \textit{not} the square of the three-dimensional NP spin coefficient $\sigma$ defined in \eqref{3d NP sigma}. Instead, one finds (somehow confusingly) that the latter is the expansion since $\sigma=-\theta$. This discussion shows that three-dimensional spacetimes do not admit twist in the sense of four-dimensional spacetimes. In fact, the spin coefficient $\rho$ does not even exist in the three-dimensional NP formalism, and one can see that the role of the four-dimensional $\text{Re}(\rho)$, i.e. the expansion, is played here by $\sigma$. Even if this interpretation of $Z$ as sourcing a twist does not hold in three-dimensional spacetimes, one can still see in \eqref{3d direct metric} that $Z$ sources $g_{r\phi}\neq0$. As in the four-dimensional case, this additional degree of freedom in the metric enables us to restore Carroll covariance on the boundary and to unfreeze the associated Carroll boost symmetry. This is explained in detail in \cite{Campoleoni:2018ltl,Ciambelli:2020eba,Ciambelli:2020ftk,Campoleoni:2022wmf} and \cite{Hartong:2025jpp,Fiorucci:2025twa}, where the Bondi gauge with $g_{r\phi}\neq0$ is called respectively the derivative expansion gauge and the Carroll covariant Bondi gauge. Let us now turn to the NP construction of the solution space.

\subsection{Solution space}

In order to build the solution space we solve the NP equations in the gauge $\kappa=\epsilon=\pi=0$. These equations, gathered in appendix \ref{app:3d NP}, are equivalent in metric form to $G_{\mu\nu}-\Lambda g_{\mu\nu}=0$. The resolution is done by splitting the equations in two sets. First, the 12 (6 scalar and $3+3$ vectorial) equations \eqref{3d NP radial} determine the radial expansion of the spin coefficients and of the triad. Then, the 6 (3 scalar and 3 vectorial) equations \eqref{3d NP temporal} impose certain constraints between some of the coefficients as well as evolution equations.

We start by solving the radial equations \eqref{3d NP radial}. Being first order in $\partial_r$, each equation brings an integration constant which is a function of $(u,\phi)$. We denote these integration constants with a subscript 0, apart from the mass and angular momentum aspects which we denote by $U_0=M$ and $\Omega_0=N$. Explicitly, in addition to $\kappa=\epsilon=\pi=0$ the solutions for the spin coefficients are
\bsub
\be
\sigma&=-\f{1}{2r+\sigma_0},\label{3d NP sigma}\\
\tau&=\sigma\tau_0,\\
\beta&=\sigma\beta_0,\\
\gamma&=\Lambda r+\sigma\tau_0\beta_0+\gamma_0,\\
\mu&=\f{1}{2}\sigma(\Lambda r^2+\Lambda r\sigma_0-2\mu_0),\\
\nu&=\left(\f{\Lambda}{2} r+\mu\right)\tau_0+\nu_0,
\ee
\esub
while for the components of the triad we find
\bsub
\be
Z&=\sigma Z_0,\\
\Omega&=\sigma(N-r\beta_0),\\
F&=\sigma F_0,\\
W&=\sigma\tau_0Z_0+W_0,\\
U&=-\f{\Lambda}{8\sigma^2}+\f{1}{4\sigma}(2\gamma_0-\Lambda\sigma_0)+\f{1}{2}\sigma\tau_0(\sigma_0\beta_0+2N)+M,\\
X&=\sigma\tau_0F_0+X_0.
\ee
\esub
All the radial equations are now solved. As announced, at this stage there are 12 integration constants $(\sigma_0,\tau_0,\beta_0,\gamma_0,\mu_0,\nu_0,Z_0,F_0,W_0,X_0,M,N)$ which are functions of $(u,\phi)$.

The remaining NP equations \eqref{3d NP temporal} are now going to enforce constraints on these free functions, in the form of either algebraic relations or temporal evolution equations. First, equation \eqref{3d NP Delta sigma} is solved by imposing the algebraic relation
\be
\mu_0=M-\f{\Lambda}{8}\sigma_0^2+\f{1}{2}\big(\tau_0^2-2\tau_0\beta_0+X_0\sigma_0'-F_0\tau_0'+W_0\dot{\sigma}_0-Z_0\dot{\tau}_0\big),
\ee
where from now on a prime $'$ denotes $\partial_\phi$ and a dot $\cdot$ denotes $\partial_u$. Since this eliminates $\mu_0$ in favor of $M$ (and other integration constants), equation \eqref{3d NP Delta mu}, which is initially an evolution equation for $\mu_0$, becomes an evolution equation for $M$ which reads
\be\label{3d M EOM}
\dot{M}
&=\f{1}{4W_0}\Big(M\big(4 \Lambda  \sigma_0-8 \gamma_0\big)-4 M'X_0+2\dot{F}_0\tau_0'W_0+2 F_0'\tau_0'X_0+4 F_0\gamma_0\tau_0'-\Lambda  F_0\sigma_0\tau_0'-4 F_0\nu_0'\cr
&\pe\q\quad+2 F_0\dot{\tau}_0'W_0+2 F_0\tau_0''X_0-2 \Lambda  N\tau_0-4 \gamma_0\tau_0^2+\Lambda  \sigma_0\tau_0^2+4 \nu_0\tau_0-2\dot{\sigma}_0\dot{W}_0W_0-2\dot{\sigma}_0W_0'X_0\cr
&\pe\q\quad+4\dot{\beta}_0\tau_0W_0-4 \gamma_0\dot{\sigma}_0W_0+2 \Lambda  \sigma_0\dot{\sigma}_0W_0-2\ddot{\sigma}_0W_0^2-4 \tau_0\dot{\tau}_0W_0-2 \sigma_0'W_0\dot{X}_0\cr
&\pe\q\quad+\beta_0\big(8 \gamma_0\tau_0-3 \Lambda  \sigma_0\tau_0-8 \nu_0+4\dot{\tau}_0W_0+4 \tau_0'X_0\big)-4\dot{\sigma}_0'W_0X_0+2\dot{\tau}_0W_0\dot{Z}_0+2\ddot{\tau}_0W_0Z_0\cr
&\pe\q\quad-2 \sigma_0'X_0X_0'+4 \beta_0'\tau_0X_0-4 \gamma_0\sigma_0'X_0+2 \Lambda  \sigma_0\sigma_0'X_0-2 \sigma_0''X_0^2-4 \tau_0\tau_0'X_0+2\dot{\tau}_0X_0Z_0'\cr
&\pe\q\quad+2\dot{\tau}_0'X_0Z_0+4 \gamma_0\dot{\tau}_0Z_0-\Lambda  \sigma_0\dot{\tau}_0Z_0-4\dot{\nu}_0Z_0\Big).
\ee
Then, \eqref{3d NP Delta beta} is solved by imposing the algebraic relation
\be\label{3d nu0 constraint}
\nu_0=-\f{\Lambda}{2}N-\f{\Lambda}{4}\sigma_0\beta_0+\f{1}{2}\big(\tau_0\gamma_0+X_0\beta_0'-F_0\gamma_0'+W_0\dot{\beta}_0-Z_0\dot{\gamma}_0\big).
\ee
We are then left with the three equations contained in \eqref{3d NP Delta delta}. The $u$ and $\phi$ components give two coupled equations which can be solved to determine $\beta_0$ and $\gamma_0$ algebraically as
\bsub
\be
\beta_0
&=\f{1}{G_0}\Big(Z_0\big(Z_0\dot{X}_0-W_0\dot{F}_0-X_0(\tau_0+F_0')+F_0(X_0'-\dot{W}_0)\big)+F_0\big(X_0Z_0'+W_0(\tau_0+\dot{Z}_0)\big)-F_0^2W_0'\Big),\!\\
\gamma_0&=\f{\Lambda}{2}\sigma_0+\f{1}{G_0}\Big(Z_0\big(W_0\dot{X}_0-X_0\dot{W}_0\big)+F_0\big(W_0X_0'-X_0W_0'\big)+X_0^2Z_0'-W_0^2\dot{F}_0+W_0X_0\big(\dot{Z}_0-F_0'\big)\Big),
\ee
\esub
where we have defined $G_0\coloneqq W_0F_0-Z_0X_0$ by analogy with \eqref{3d NP G}. Finally, the radial component of \eqref{3d NP Delta delta} contains the evolution equation for $N$, which reads
\be\label{3d P EOM}
\dot{N}
&=\f{1}{4W_0}\Big(4 N (\Lambda  \sigma_0-2 \gamma_0)+4 F_0M'+8 M(\beta_0-\tau_0)+4 Z_0\dot{M}-2 \tau_0^3-2 \gamma_0\big(F_0\sigma_0'+Z_0\dot{\sigma}_0\big)+\Lambda  F_0\sigma_0 \sigma_0'\cr
&\pe\q\quad-4 X_0N'-2 \sigma_0 W_0\dot{\beta}_0-2 \tau_0\big(W_0\dot{\sigma}_0+X_0\sigma_0'+Z_0\dot{\beta}_0-F_0\tau_0'\big)-2 \sigma_0 X_0\beta_0'-2 \beta_0 Z_0\dot{\tau}_0\cr
&\pe\q\quad+\Lambda  \sigma_0 Z_0\dot{\sigma}_0+2 \tau_0 Z_0\dot{\tau}_0-4 \beta_0^2 \tau_0-4 \beta_0 \gamma_0 \sigma_0+2 \beta_0 \Lambda  \sigma_0^2+6 \beta_0 \tau_0^2-2 \beta_0 F_0\tau_0'-2 F_0\tau_0 \beta_0'\Big).\q\q\q
\ee
This terminates the resolution of the three-dimensional NP equations, and the resulting metric \eqref{3d direct metric} solves the Einstein equations $G_{\mu\nu}-\Lambda g_{\mu\nu}=0$. One should note that this metric is finite in $r$, as expected from the topological nature of three-dimensional gravity.

At the end of the day, the algebraic relations obtained from the NP equations \eqref{3d NP temporal} have reduced the number of independent functions of $(u,\phi)$ in the solution space from 12 to 8. The remaining 8 functions are $(\sigma_0,\tau_0,Z_0,F_0,W_0,X_0,M,N)$. The mass $M$ and the angular momentum $N$ satisfy the evolution equations \eqref{3d M EOM} and \eqref{3d P EOM}, while the other 6 functions have an unconstrained dependency on $u$ (and in the terminology of \cite{Ciambelli:2024vhy} are therefore kinematical degrees of freedom). In order to understand the role of some of the free functions, it is interesting to compute the induced boundary metric. We find
\be
\de s^2\big|_{\I^+}=\lim_{r\to\infty}\left(\f{\de s^2}{r^2}\right)=\f{2\Lambda}{\Lambda Z_0^2-2W_0^2}\de u^2+\f{2W_0^2-\Lambda Z_0^2}{(W_0F_0-Z_0X_0)^2}\left(\f{2W_0X_0-\Lambda F_0Z_0}{2W_0^2-\Lambda Z_0^2}\de u-\de\phi\right)^2,
\ee
which shows that $(Z_0,F_0,W_0,X_0)$ (over-)parametrize the boundary metric. The role of $\sigma_0$, as can be seen from \eqref{3d NP sigma}, is to parametrize the choice of origin for the coordinate $r$ which serves as the affine parameter for $\ell$. It appears in the solution space because we are working with the NU gauge condition $\epsilon=0$. It has also appeared as $H$ in the solution space constructed in \cite{Geiller:2021vpg}. Moreover, it is the three-dimensional analogue of the ``trace of the shear'' $C$ allowed by the NU gauge in four-dimensional spacetimes \cite{Geiller:2022vto,Geiller:2024amx}. Indeed, as explained in section \ref{sec:gauge conditions} this trace $C$ appears in the four-dimensional NP formalism as $\text{Re}(\rho_2)$, and as explained below \eqref{3d optical scalars} the three-dimensional NP analogue of the expansion $\text{Re}(\rho)$ is precisely $\sigma$.

As far as we are aware, the solution space constructed above is the largest in terms of integration constants. It generalizes the results of \cite{Geiller:2021vpg} by including $\tau_0$ and $Z_0$, and also generalizes the so-called derivative expansion of \cite{Campoleoni:2018ltl,Ciambelli:2020eba,Ciambelli:2020ftk,Campoleoni:2022wmf} by including $\sigma_0$ and $\tau_0$. We summarize in table \ref{3d solution space table} below the various functions of $(u,\phi)$ which appear in these references and in the present work. Of course, since these functions appear as integration constants one can redefine them at will. The table is therefore just indicating the correspondance across the various references, and not the actual map between the functions. One can write the actual maps by studying the metrics in detail. For example when $Z_0=\tau_0=0$ the matching with \cite{Geiller:2021vpg} is
\be
\sigma_0=-2H,
\q\q
F_0=\sqrt{2}e^{-\varphi},
\q\q
W_0=e^{-2\beta},
\q\q
X_0=2Ue^{-2(\varphi+\beta)},
\ee
where the notations on the RHS are from \cite{Geiller:2021vpg} while the LHS are the quantities defined here. It would be interesting to study further the structure of the 8-dimensional solution space which has been constructed here, in particular with a detailed analysis of the charges and the corner terms which can be used to obtain their integrability\footnote{Even if the charges are not conserved because they contain arbitrary functions of $(u,\phi)$, they are expected to be integrable since three-dimensional gravity has no propagating degrees of freedom, and therefore no flux can source the non-integrability. This is indeed what was shown for the solution space constructed in \cite{Geiller:2021vpg}.}. Alternatively, one can wonder whether relaxing further conditions on the initial choice of triad vectors can lead to an even larger set of solutions. We postpone these investigations to future work.

\begin{table}[h]\centering
\quad\begin{tabular}{|c|c|c|c|c|c|c|c|c|c|}
\hline
here&$\sigma_0$&$\tau_0$&$Z_0$&$F_0$&$W_0$&$X_0$&$M$&$N$\\ \hline
\cite{Geiller:2021vpg}&$H$&--&--&$\varphi$&$\beta$&$U$&$M$&$N$\\ \hline
\cite{Ciambelli:2020eba,Ciambelli:2020ftk} in Bondi gauge&--&--&--&$\varphi$&$\beta_0$&$U_0$&$M$&$N$\\ \hline
\cite{Ciambelli:2020eba,Ciambelli:2020ftk} in derivative expansion&--&--&$u_\phi$&$\tilde{g}_{\phi\phi}$&$u_u$&$\tilde{g}_{u\phi}$&$\varepsilon$&$\chi$\\ \hline
\end{tabular}
\caption{List of the functions of $(u,\phi)$ which parametrize various solution spaces available in the literature.}
\label{3d solution space table}
\end{table}

\section{Perspectives}

In this work we have presented a framework for describing algebraically general and asymptotically-flat solutions of general relativity in the presence of twist. We did so using both the Newman--Penrose and metric formulations, where the twist is sourced by a potential appearing in the tetrad component $m^u$ and in the metric component $g_{ra}\neq0$. This was motivated in part by the desire to properly embed the algebraically special solutions within the Bondi formalism for asymptotically-flat spacetimes, in the sense of preserving the property that such solutions are resummed and have a finite radial expansion. In particular, the Kerr--Taub--NUT solution and its finitely supertranslated version take a finite form in this extended Bondi gauge with twist, which reduces simply to their expression in Eddington--Finkelstein coordinates. Another motivation came from the recent works \cite{Campoleoni:2023fug} and \cite{Hartong:2025jpp,Fiorucci:2025twa}, where a Carroll covariant Bondi gauge (respectively in Newman--Unti and Bondi--Sachs form) was constructed by introducing $g_{ra}\neq0$, however without explicit reference to the twist potential. In the present work we have worked out the most salient properties of this Carroll covariant Bondi gauge, by providing in particular $i)$ a resolution of the field equations in the NP and metric formulations; $ii)$ a dictionary between the two formulations; $iii)$ an extension of the Bondi hierarchy organizing the solution space; $iv)$ a study of the asymptotic symmetries; $v)$ a study of the algebraically special solutions and especially of type D examples; $vi)$ a study of the Carroll covariant Bondi gauge in three-dimensional spacetimes with non-vanishing cosmological constant.

There are many interesting directions in which this work could be pushed, and which we plan on exploring in the future: 
\begin{itemize}
\item \textbf{Turning on the twist in various other contexts.} First, it should be possible to extend the present work to various contexts which are slightly more general, namely: in the presence of a cosmological constant, in the presence of matter, in the presence of logarithmic terms and violations of peeling, and in the presence of a free boundary metric parametrized by the extra integration constants $W_0(u,x^a)$ and $X^a_0(u,x^b)$.
\item \textbf{Finite distance null hypersurfaces.} It would be interesting to relax the hypersurface-orthogonality condition and to study the role of the twist for finite null surfaces. There is indeed a large literature on symmetries and charges at null boundaries \cite{Donnay:2016ejv,Akhmedov:2017ftb,Chandrasekaran:2021hxc,Odak:2023pga}, but these considerations are studied with vanishing twist. The twist could be in particular relevant in order to understand the mapping between finite null surfaces and null infinity in the case of algebraically special solutions. In this context, the recent work \cite{Agrawal:2025fsv,Ruzziconi:2025fct,Ruzziconi:2025fuy} could be extended to include the twist, and one could hope to map null infinity to the horizon by exploiting the finite form of the metric and charges for algebraically special solutions. One can also wonder whether the results of \cite{Ashtekar:2024bpi,Ashtekar:2024stm,Ashtekar:2024mme} on the link between isolated horizons and null infinity can be extended in the presence of twist.
\item \textbf{Black hole perturbation theory.} An interesting question is whether the Carroll covariant Bondi gauge with non-vanishing twist potential could simplify the study of black hole perturbation theory. Black hole perturbation theory is typically performed in the NP formalism, especially for rotating black holes, but one can wonder if the Bondi hierarchy of Einstein equations in the presence of twist potential, presented in section \ref{sec:hierarchy}, could be used in order to study perturbation theory in the metric formalism \cite{Loutrel:2020wbw,BenAchour:2025hns}, or if it could simplify the study of the algebraically special perturbations \cite{Chandrasekhar:1984mgh,Chandrasekhar:1985kt}.
\item \textbf{Numerical relativity.} Since black hole perturbation theory and gravitational radiation are most commonly studied with numerical methods in the NP formalism, it would also be interesting to see if the generalized tetrad \eqref{NP tetrad} with twist potential and the resulting solution space can bring useful technical and/or conceptual simplifications \cite{Nerozzi:2005hz,Moxon:2020gha,Iozzo:2020jcu}.
\item \textbf{Map from harmonic to Bondi gauge with twist.} Another question relevant to the description of gravitational radiation at null infinity is whether the Bondi gauge with twist can be easily reached from the harmonic gauge. The subtle task of transforming from harmonic coordinates to Bondi coordinates (more precisely to the Bondi--Sachs gauge) was tackled in \cite{Blanchet:2020ngx,Blanchet:2023pce} (see also \cite{Mao:2024urq,Mao:2025lwk}), but it would be interesting to see if the diffeomorphism from harmonic coordinates is easier to implement if allowing for non-vanishing twist at null infinity.
\item \textbf{Link with twistor space and $\boldsymbol{w_{1+\infty}}$ symmetries.} Finally, an intriguing question is whether the twist could play a role when mapping structures from null infinity to twistor space. This is an important question in the context of the $w_{1+\infty}$ symmetries which were revealed in celestial holography \cite{Ball:2021tmb,Strominger:2021lvk}, studied in the gravitational phase space in \cite{Freidel:2021ytz,Compere:2022zdz,Geiller:2024bgf,Cresto:2024mne,Cresto:2024fhd}, and studied in twistor space in \cite{Adamo:2021lrv,Kmec:2024nmu,Donnay:2024qwq}. Indeed, it was proposed in \cite{Cresto:2024mne} (see section 8) that one can relate $\I$ to twistor space by introducing a twist potential in the null transverse vectors $(m,\bar{m})$. This is done by performing a null boost whose parameter is the stereographic angle describing the null direction of the geodesics intersecting $\I$ \cite{Adamo:2009vu,Adamo:2010ey} (see equation (198)). In light of this proposal, a natural question is whether the map between $\I$ and twistor space could be better understood within asymptotically-flat spacetimes with non-vanishing twist.
\end{itemize}

\section*{Acknowledgements}

We thank Glenn Barnich, Luca Ciambelli, Marios Petropoulos, and Simone Speziale for discussions and comments. The work of PM is supported in part by the National Natural Science Foundation of China under Grants No. 12475059 and No. 11935009, and by Tianjin University Self-Innovation Fund Extreme Basic Research Project Grant No. 2025XJ21-0007.

\newpage

\appendix

\section{Newman--Penrose formalism}
\label{app:NP}

\subsection{Notations and conventions}
\label{app:notations}

Our choice of signature is $(-,+,+,+)$. Spacetime indices are denoted by $\mu,\nu,\dots$. Two-dimensional indices in the angular direction of the metric are denoted by $a,b,\dots$. Internal Lorentzian indices are lowered and raised with $\eta_{ij}$ and $\eta^{ij}$, and are denoted by $i,j,\dots$.

In order for the NP equations listed in appendix \ref{appendix:Ricci} and \ref{app:4d NP equations} to take the same form as in \cite{Chandrasekhar:1985kt}, we define the Weyl scalars and the spin coefficients with the opposite sign, following the conventions of \cite{Ashtekar:2000hw}. The spin coefficients are given in appendix \ref{app:spin coeffs} below, and the Weyl scalars are defined as
\be\label{Weyl scalars}
\Psi_0\coloneqq W_{\ell m\ell m},\quad\q
\Psi_1\coloneqq W_{\ell n\ell m},\quad\q
\Psi_2\coloneqq W_{\ell m\bar{m}n},\quad\q
\Psi_3\coloneqq W_{n\bar{m}n\ell},\quad\q
\Psi_4\coloneqq W_{n\bar{m}n\bar{m}}.
\ee
Our definitions of the Riemann and Weyl tensors are
\bsub
\be
{R^\lambda}_{\sigma\mu\nu}&=\partial_\mu\Gamma^\lambda_{\sigma\nu}-\partial_\nu\Gamma^\lambda_{\mu\sigma}+\Gamma^\lambda_{\mu\rho}\Gamma^\rho_{\sigma\nu}-\Gamma^\lambda_{\rho\nu}\Gamma^\rho_{\mu\sigma},\\
W_{\mu\nu\rho\sigma}&=R_{\mu\nu\rho\sigma}+\f{1}{2}\big(g_{\mu\sigma}R_{\nu\rho}+g_{\nu\rho}R_{\mu\sigma}-g_{\mu\rho}R_{\nu\sigma}-g_{\nu\sigma}R_{\mu\rho}\big)+\f{R}{6}\big(g_{\mu\rho}g_{\nu\sigma}-g_{\nu\rho}g_{\mu\sigma}\big).
\ee
\esub
The spin-weighed derivative operator $\eth$ and its generalization $\Eth$ in the presence of twist and time dependency in $\gamma_0\neq0$ are defined as
\bsub\label{generalized eth}
\be
\Eth&\coloneqq\eth-2s\gamma_0L+L\partial_u,\\
\Ethb&\coloneqq\ethb+2s\gamma_0\bar{L}+\bar{L}\partial_u,\\
\eth&\coloneqq m_1^a\partial_a+2s(\bar{\alpha}_1+\gamma_0L),\\
\ethb&\coloneqq\bar{m}_1^a\partial_a-2s(\alpha_1+\gamma_0\bar{L}),
\ee
\esub
where $m^a_1$ and $\alpha_1$ are given explicitly in \eqref{frame m1} and \eqref{NP alpha1}. We have the generalized commutation relation
\be
\big[\Ethb,\Eth\big]=2(s\mu_R-i\Sigma\partial_u),
\ee
where $\mu_R$ is given by \eqref{NP mu R} in NP form, and by \eqref{metric mu R} in metric form. In the main text we use many formulas extracted from appendix B of \cite{Geiller:2024bgf}, as well as the identities
\bsub\label{Eth formulas}
\be
\Eth F&=\big(\partial_aF+L_a\partial_uF\big)m^a_1,\\
\Eth\big(F_a\bar{m}^a_1\big)&=\big(D_aF_b+L_a\partial_uF_b\big)m^a_1\bar{m}^b_1,\\
\Eth\big(F_{ab}\bar{m}^a_1\bar{m}^b_1\big)&=\big(D^aF_{\la ab\ra}+L^a\partial_uF_{\la ab\ra}\big)\bar{m}^b_1,\\
\Eth\big(F_a\bar{m}^a_1\big)-\Ethb\big(F_am^a_1\big)&=i\big(D^a+L^a\partial_u\big)\widetilde{F}_a,
\ee
\esub
where $\widetilde{F}_a=\eps^{ab}_{(q)}F_b=\f{1}{\sqrt{q}}\epsilon^{ab}F_b$ with $\epsilon^{ab}$ the Levi--Civita symbol. Here $F$, $F_a$, and $F_{ab}$ represent any scalar, vector or rank-2 tensor.

\newpage

\subsection{Spin coefficients}
\label{app:spin coeffs}

Here we give the explicit expressions of the spin coefficients corresponding to the tetrad \eqref{NP tetrad}. In order to make these expressions compact, we use the notations introduced in section \ref{sec:metric with twist} in order to trade some of the complex scalars for real two-dimensional vectors. Denoting by $\D_a$ the covariant derivative with respect to $\gamma_{ab}$, we have
\bsub\label{spin coefficients}
\be
\kappa
&=-m^\mu\ell^\nu\nabla_\nu\ell_\mu
=g_{ur}\partial_rZ_am^a,\label{NP kappa}\\
\sigma
&=-m^\mu m^\nu\nabla_\nu\ell_\mu
=-\f{1}{2}\partial_r\gamma_{ab}m^am^b+\kappa X,\\
\rho
&=-m^\mu\bar{m}^\nu\nabla_\nu\ell_\mu
=-\f{1}{2}\partial_r\ln\sqrt{-g}+\text{Re}(\epsilon)+i\text{Im}(\kappa\bar{\Omega})+\f{i}{2}g_{ur}\big(\partial_uZ_a-\D_a\big)\widetilde{Z}^a,\label{full rho}\\
\pi
&=-n^\mu\ell^\nu\nabla_\nu\bar{m}_\mu
=\f{1}{2}\bar{m}^\mu\partial_\mu\ln g_{ur}+\f{1}{2}\bar{m}^a\big(\partial_uZ_a+\partial_r\Omega_a+\gamma_{ab}\partial_rX^b\big)+\bar{X}\big(\text{Re}(\epsilon)+i\text{Im}(\rho)\big),\\
\lambda
&=\bar{m}^\mu\bar{m}^\nu\nabla_\nu n_\mu\\
&=\partial_u\big(g_{\mu a}\bar{m}^\mu\big)\big(W\bar{m}^a-\bar{Z}X^a\big)+\bar{m}_a\big(\bar{\Omega}\partial_rX^a-U\partial_r\bar{m}^a\big)+\bar{X}\big(2\bar{\Omega}\,\text{Re}(\epsilon)-\bar{\kappa}U\big)\cr
&\pe+\bar{m}^a\big(\partial_a\bar{X}+\bar{X}\partial_a\ln g_{ur}\big)+2\big(\partial_a\bar{m}_b-g_{ur}\bar{X}\partial_aZ_b\big)X^{[a}\bar{m}^{b]},\cr
\tau
&=-m^\mu n^\nu\nabla_\nu\ell_\mu\\
&=X\big(2\text{Re}(\epsilon)-i\text{Im}(\rho)\big)-\bar{\pi}+\f{\kappa}{2}\Omega_aX^a+\f{1}{2}g_{ur}X^a\big(Z_a\partial_uZ_bm^b+\partial_aZ_bm^b-m^\mu\partial_\mu Z_a\big)+m_a\partial_rX^a,\cr
\mu
&=-n^\mu m^\nu\nabla_\nu\bar{m}_\mu\\
&=2\gamma+\partial_rU+2\big(U-\Omega_aX^a\big)\text{Re}(\epsilon)-\Omega_a\partial_rX^a+\big(m^\mu\partial_\mu X^a-n^\mu\partial_\mu m^a\big)\bar{m}_a\cr
&\pe+\bar{X}\Big(g_{ur}\partial_uZ_a\big(ZX^a-Wm^a\big)+g_{ur}\Omega X^a\partial_rZ_a+m^\mu\partial_\mu\ln g_{ur}-\kappa U-2g_{ur}\partial_aZ_bX^{[a}m^{b]}\Big),\cr
\nu
&=-n^\mu n^\nu\nabla_\nu\bar{m}_\mu\\
&=\Big(\bar{\kappa}U-2\bar{\Omega}\,\text{Re}(\epsilon)+\partial_u\big(g_{ur}Z_a\big)\big(W\bar{m}^a-\bar{Z}X^a\big)+2g_{ur}\partial_aZ_bX^{[a}\bar{m}^{b]}-\bar{m}^a\partial_a\ln g_{ur}-\bar{m}^\mu\partial_\mu\Big)\big(U-\Omega_aX^a\big)\cr
&\pe+\big(n^\mu\bar{m}^a-\bar{m}^\mu X^a\big)\partial_\mu\Omega_a,\cr
\epsilon
&=\f{1}{2}\big(\bar{m}^\mu\ell^\nu\nabla_\nu m_\mu-n^\mu\ell^\nu\nabla_\nu\ell_\mu\big)
=\f{1}{2}g_{ur}\big(\partial_rW-Z_a\partial_rX^a\big)+\f{i}{2}\text{Im}\big(\bar{m}_a\partial_rm^a+\rho+\kappa\bar{X}\big),\label{NP epsilon}\\
\gamma
&=\f{1}{2}\big(\bar{m}^\mu n^\nu\nabla_\nu m_\mu-n^\mu n^\nu\nabla_\nu\ell_\mu\big)\\
&=\left(\text{Re}(\epsilon)+\f{1}{2}\partial_r\right)\big(\Omega_aX^a-U\big)-\f{1}{2}X^a\partial_r\Omega_a+i\text{Re}(\epsilon)\text{Im}(X\bar{\Omega})+\f{i}{2}U\,\text{Im}(\kappa\bar{X})+\f{i}{2}\text{Im}(X\partial_u\bar{Z})\cr
&\pe+\f{i}{2}\text{Im}\Big(n^\mu\partial_\mu m_a\bar{m}^a-m^\mu\partial_\mu X^a\bar{m}_a+m^\mu\partial_\mu\Omega_a\bar{m}^a+g_{ur}\big(U-\Omega_bX^b\big)m^\mu\partial_\mu Z_a\bar{m}^a+X\bar{m}^a\partial_a\ln g_{ur}\Big)\cr
&\pe+\f{i}{2}g_{ur}\,\text{Im}\Big(X\partial_u\big(W\bar{m}^a-\bar{Z}X^a\big)Z_a-2X\partial_aZ_bX^{[a}\bar{m}^{b]}\Big),\cr
\alpha
&=\f{1}{2}\big(\bar{m}^\mu\bar{m}^\nu\nabla_\nu m_\mu-n^\mu\bar{m}^\nu\nabla_\nu\ell_\mu\big)
=\f{1}{2}\Big(\pi-\overline{\gamma_{433}}+\bar{\kappa}\big(\Omega_aX^a-U\big)-\bar{m}^a\partial_r\Omega_a\Big),\\
\beta
&=\f{1}{2}\big(\bar{m}^\mu m^\nu\nabla_\nu m_\mu-n^\mu m^\nu\nabla_\nu\ell_\mu\big)
=\bar{\alpha}+\gamma_{433},\\
\gamma_{433}&=2iX\,\text{Im}(\kappa\bar{\Omega})+i\widetilde{Z}^a\big(g_{ur}X\partial_uZ_a-\partial_um_a\big)-i\widetilde{\Omega}^a\partial_rm_a+i\D_a\widetilde{m}^a-ig_{ur}X\D_a\widetilde{Z}^a.
\ee
\esub
We recall that $\D_a$ appearing in these expressions is the covariant derivative with respect to $\gamma_{ab}$.

\newpage

\subsection{Decomposition of the Ricci tensor}
\label{appendix:Ricci}

In this appendix we explain how the cotetrad and the expressions \eqref{spin coefficients} for the spin coefficients can be used to rewrite explicitly the components of the Ricci tensor. Using the notations introduced in section \ref{sec:metric with twist} and recalling that $g_{ur}=(Z_aX^a-W)^{-1}$, we find that the cotetrad can be written as
\bsub\label{cotetrad}
\be
\ell&=g_{ur}\big(\de u-Z_a\de x^a\big),\\
n&=g_{ur}\Big(\big(\Omega_aX^a-U\big)\de u-g_{ur}^{-1}\de r+\big(UZ_a-W\Omega_a-(Z_b\widetilde{\Omega}^b)\widetilde{X}_a\big)\de x^a\Big),\\
m&=g_{ur}\Big(X\de u+\big(iZ\widetilde{X}_a-Wm_a\big)\de x^a\Big).
\ee
\esub
As explained in section \ref{sec:gauge conditions}, with the gauge conditions $\kappa=\text{Re}(\epsilon)=0$ we actually have that $g_{ur}=-1$. For the sake of generality however, we give here expressions which are valid regardless of the choice of gauge and which follow uniquely from the use of the tetrad \eqref{NP tetrad}. Denoting $e_i=(\ell,n,m,\bar{m})$ and $R_{ij}=R_{\mu\nu}e^\mu_ie^\nu_j$, we have the decomposition
\be\label{Ricci decomposition}
R_{\mu\nu}
&=R_{11}n_\mu n_\nu+R_{22}\ell_\mu\ell_\nu+R_{33}\bar{m}_\mu\bar{m}_\nu+R_{44}m_\mu m_\nu\cr
&\pe+2R_{12}n_{(\mu}\ell_{\nu)}-2R_{13}n_{(\mu}m_{\nu)}-2R_{14}n_{(\mu}\bar{m}_{\nu)}-2R_{23}\ell_{(\mu}\bar{m}_{\nu)}-2R_{24}\ell_{(\mu}m_{\nu)}+2R_{34}m_{(\mu}\bar{m}_{\nu)},\q
\ee
and the Ricci scalar is given by
\be
R=2(R_{34}-R_{12}).
\ee
This implies in particular that the radial components of the Ricci tensor (those involved in the hypersurface equations) are given by
\be
R_{r\mu}=R_{14}m_\mu+R_{13}\bar{m}_\mu-R_{11}n_\mu-R_{12}\ell_\mu.
\ee
We have in particular that $R_{rr}=R_{11}$, which leads to the Einstein equation \eqref{rr EFE} in the main text.

Similarly, all the components of the Ricci tensor can be reconstructed from \eqref{Ricci decomposition}, the cotetrad \eqref{cotetrad}, and the projections $R_{ij}$. These can in turn be expressed in terms of the spin coefficients \eqref{spin coefficients} using the so-called non-vacuum, or ``off-shell'' NP equations (i.e. where the Ricci tensor and scalar have not been set to zero using the field equations). Only a subset of these NP equations is required in order to express the Ricci tensor, and one can chose
\bsub
\be
D\rho&=\rho^2+\sigma\bar{\sigma}+\f{1}{2}R_{11},\\
D\alpha&=\rho\alpha+\bar{\sigma}\beta+\f{1}{2}R_{14},\\
D\beta&=\bar{\rho}\beta+\sigma\alpha+\Psi_{1},\\
D\tau&=\rho\tau+\sigma\bar{\tau}+\Psi_1+\f{1}{2}R_{13},\\
D\lambda&=\rho\lambda+\bar{\sigma}\mu+\f{1}{2}R_{44},\\
D\mu&=\bar{\rho}\mu+\sigma\lambda+\Psi_{2}+\f{1}{12}R,\\
D\gamma&=\tau\alpha+\bar{\tau}\beta+\Psi_2+\f{1}{4}(R_{12}+R_{34})-\f{1}{24}R,\\
D\nu&=\tau\lambda+\bar{\tau}\mu+\Psi_3+\f{1}{2}R_{24},\\
\Delta\mu&=\delta\nu-\mu^2-\lambda\bar{\lambda}-\mu(\gamma+\bar{\gamma})+\nu(\bar{\alpha}+3\beta-\tau)-\f{1}{2}R_{22},\\
\Delta\beta&=\delta\gamma-\mu\tau+\sigma\nu+\beta(\gamma-\bar{\gamma}-\mu)-\alpha\bar{\lambda}+\gamma(\bar{\alpha}+\beta-\tau)-\f{1}{2}R_{23},\\
\Delta\sigma&=\delta\tau-\sigma\mu-\rho\bar{\lambda}-\tau(\tau-\bar{\alpha}+\beta)+\sigma(3\gamma-\bar{\gamma})-\f{1}{2}R_{33},\\
\delta\alpha&=\bar{\delta}\beta+\mu\rho-\lambda\sigma+\alpha\bar{\alpha}+\beta\bar{\beta}-2\alpha\beta+\gamma(\rho-\bar{\rho})-\Psi_2+\f{1}{4}(R_{12}+R_{34})+\f{1}{24}R,\\
\delta\lambda&=\bar{\delta}\mu+\nu(\rho-\bar{\rho})-\mu(\alpha+\bar{\beta})+\lambda(\bar{\alpha}-3\beta)-\Psi_3+\f{1}{2}R_{24}.
\ee
\esub
Note that some of these equations have contributions from the Weyl scalars as well, but that they can be combined to reduce to 10 expressions involving only the components of $R_{ij}$ and the spin coefficients \eqref{spin coefficients}.

\subsection{4d field equations}
\label{app:4d NP equations}

Here we list the four-dimensional NP equations in the case of the gauge choice $\kappa=\epsilon=\pi=0$. They can be split into two subsets, involving respectively $D$ and $(\Delta,\delta)$. First, we have the radial equations
\bsub\label{4d NP radial}
\be
D\rho&=\rho^2+\sigma\bar{\sigma},\label{NP1a}\\
D\sigma&=\sigma(\rho+\bar{\rho})+\Psi_{0},\label{NP1b}\\
D\alpha&=\rho\alpha+\bar{\sigma}\beta,\label{NP1c}\\
D\beta&=\bar{\rho}\beta+\sigma\alpha+\Psi_{1},\label{NP1d}\\
D\tau&=\rho\tau+\sigma\bar{\tau}+\Psi_1,\label{NP1e}\\
D\lambda&=\rho\lambda+\bar{\sigma}\mu,\label{NP1f}\\
D\mu&=\bar{\rho}\mu+\sigma\lambda+\Psi_{2},\label{NP1g}\\
D\gamma&=\tau\alpha+\bar{\tau}\beta+\Psi_2,\label{NP1h}\\
D\nu&=\tau\lambda+\bar{\tau}\mu+\Psi_3,\label{NP1i}\\
D\Delta&=\Delta D-(\gamma+\bar{\gamma})D+\bar{\tau}\delta+\tau\bar{\delta},\label{NP1j}\\
D\delta&=\delta D-(\bar{\alpha}+\beta)D+\bar{\rho}\delta+\sigma\bar{\delta},\label{NP1k}\\
D\Psi_1&=\bar{\delta}\Psi_0+4\rho\Psi_1-4\alpha\Psi_0,\label{NP1l}\\
D\Psi_2&=\bar{\delta}\Psi_1+3\rho\Psi_2-2\alpha\Psi_1-\lambda\Psi_0,\label{NP1m}\\
D\Psi_3&=\bar{\delta}\Psi_2+2\rho\Psi_3-2\lambda\Psi_1,\label{NP1n}\\
D\Psi_4&=\bar{\delta}\Psi_3+\rho\Psi_4+2\alpha\Psi_3-3\lambda\Psi_2.\label{NP1o}
\ee
\esub
Then, we have
\bsub\label{4d NP supplementary}
\be
\Delta\lambda&=\bar{\delta}\nu-\lambda(\mu+\bar{\mu})-\lambda(3\gamma-\bar{\gamma})+\nu(3\alpha+\bar{\beta}-\bar{\tau})-\Psi_4,\label{NP2a}\\
\Delta\rho&=\bar{\delta}\tau-\rho\bar{\mu}-\sigma\lambda-\tau(\bar{\tau}+\alpha-\bar{\beta})+\rho(\gamma+\bar{\gamma})-\Psi_2,\label{NP2b}\\
\Delta\alpha&=\bar{\delta}\gamma+\rho\nu-\lambda(\tau+\beta)+\alpha(\bar{\gamma}-\bar{\mu})+\gamma(\bar{\beta}-\bar{\tau})-\Psi_3,\label{NP2c}\\
\Delta\mu&=\delta\nu-\mu^2-\lambda\bar{\lambda}-\mu(\gamma+\bar{\gamma})+\nu(\bar{\alpha}+3\beta-\tau),\label{NP2d}\\
\Delta\beta&=\delta\gamma-\mu\tau+\sigma\nu+\beta(\gamma-\bar{\gamma}-\mu)-\alpha\bar{\lambda}+\gamma(\bar{\alpha}+\beta-\tau),\label{NP2e}\\
\Delta\sigma&=\delta\tau-\sigma\mu-\rho\bar{\lambda}-\tau(\tau-\bar{\alpha}+\beta)+\sigma(3\gamma-\bar{\gamma}),\label{NP2f}\\
\Delta\delta&=\delta\Delta+\bar{\nu}D+(\bar{\alpha}+\beta-\tau)\Delta+(\gamma-\bar{\gamma}-\mu)\delta-\bar{\lambda}\bar{\delta},\label{NP2g}\\
\delta\rho&=\bar{\delta}\sigma+\rho(\bar{\alpha}+\beta)-\sigma(3\alpha-\bar{\beta})+\tau(\rho-\bar{\rho})-\Psi_1,\label{NP2h}\\
\delta\alpha&=\bar{\delta}\beta+\mu\rho-\lambda\sigma+\alpha\bar{\alpha}+\beta\bar{\beta}-2\alpha\beta+\gamma(\rho-\bar{\rho})-\Psi_2,\label{NP2i}\\
\delta\lambda&=\bar{\delta}\mu+\nu(\rho-\bar{\rho})-\mu(\alpha+\bar{\beta})+\lambda(\bar{\alpha}-3\beta)-\Psi_3,\label{NP2j}\\
\bar{\delta}\delta&=\delta\bar{\delta}+(\bar{\mu}-\mu)D+(\bar{\rho}-\rho)\Delta+(\alpha-\bar{\beta})\delta+(\beta-\bar{\alpha})\bar{\delta},\label{NP2k}\\
\Delta\Psi_0&=\delta\Psi_1+(4\gamma-\mu)\Psi_0-(4\tau+2\beta)\Psi_1+3\sigma\Psi_2,\label{NP2l}\\
\Delta\Psi_1&=\delta\Psi_2+\nu\Psi_0+(2\gamma-2\mu)\Psi_1-3\tau\Psi_2+2\sigma\Psi_3,\label{NP2m}\\
\Delta\Psi_2&=\delta\Psi_3+2\nu\Psi_1-3\mu\Psi_2+(2\beta-2\tau)\Psi_3+\sigma\Psi_4,\label{NP2n}\\
\Delta\Psi_3&=\delta\Psi_4+3\nu\Psi_2-(2\gamma+4\mu)\Psi_3+(4\beta-\tau)\Psi_4.\label{NP2o}
\ee
\esub

\subsection[3d field equations with $\Lambda\neq0$]{3d field equations with $\boldsymbol{\Lambda\neq0}$}
\label{app:3d NP}

With the conventions and notations introduced in section \ref{sec:3d}, the three-dimensional NP spin coefficients are given by
\bsub\label{3d spin coefficients}
\be
\kappa&=-m^\mu\ell^\nu\nabla_\nu\ell_\mu,\\
\epsilon&=-n^\mu\ell^\nu\nabla_\nu\ell_\mu,\\
\pi&=m^\mu\ell^\nu\nabla_\nu n_\mu,\\
\tau&=-m^\mu n^\nu\nabla_\nu\ell_\mu,\\
\gamma&=-n^\mu n^\nu\nabla_\nu\ell_\mu,\\
\nu&=m^\mu n^\nu\nabla_\nu n_\mu,\\
\sigma&=-m^\mu m^\nu\nabla_\nu\ell_\mu,\label{3d NP sigma}\\
\beta&=-n^\mu m^\nu\nabla_\nu\ell_\mu,\\
\mu&=m^\mu m^\nu\nabla_\nu n_\mu.
\ee
\esub
When $\kappa=\epsilon=\pi=0$, the three-dimensional NP equations with non-vanishing cosmological constant $\Lambda$ can be split into a set corresponding to the radial equations
\bsub\label{3d NP radial}
\be
D\sigma&=2\sigma^2,\\
D\tau&=2\tau\sigma,\\
D\beta&=2\beta\sigma,\\
D\gamma&=2\tau\beta+\Lambda,\\
D\mu&=2\mu\sigma-\f{\Lambda}{2},\\
D\nu&=2\tau\mu,\\
D\delta&=\delta D-\beta D+2\sigma\delta,\\
D\Delta&=\Delta D-\gamma D+2\tau\delta,
\ee
\esub
and a set corresponding to the temporal equations
\bsub\label{3d NP temporal}
\be
\Delta\sigma&=\delta\tau+(\gamma-2\mu)\sigma-\tau^2+\f{\Lambda}{2},\label{3d NP Delta sigma}\\
\Delta\mu&=\delta\nu-(2\mu+\gamma)\mu+(2\beta-\tau)\nu,\label{3d NP Delta mu}\\
\Delta\beta&=\delta\gamma+(\beta-\tau)\gamma-2(\beta+\tau)\mu+2\sigma\nu,\label{3d NP Delta beta}\\
\Delta\delta&=\delta\Delta+\nu D+(\beta-\tau)\Delta-2\mu\delta.\label{3d NP Delta delta}
\ee
\esub

\newpage

\subsection{Resolution of the 4d NP equations}
\label{app:NP resolution}

In this appendix we explain briefly how the NP equations are solved in order to obtain the solution space described in section \ref{sec:4d NP solution space}.
\begin{itemize}
\item The radial expansions \eqref{radial spin coefficients} of the spin coefficients are found by solving the radial NP equations \eqref{NP1a} to \eqref{NP1i} exactly in that order.
\item The radial expansions \eqref{NP tetrad solution} for the 8 components of the tetrad are found by expanding radially and solving the $4+4$ commutator relations \eqref{NP1i} and \eqref{NP1j}.
\item The radial expansions \eqref{Psi} for the Weyl scalars are found by expanding the radial equations from \eqref{NP1l} to \eqref{NP1o}, with the exception of $\Psi_0$ whose radial expansion is an input data.
\item We now explain which information is obtained by expanding at leading order the equations from \eqref{NP2a} to \eqref{NP2k}. We obtain
\bsub
\be
\eqref{NP2a}&&\to&&&\Psi_4^0\\
\eqref{NP2b}&&\to&&&U_0\text{ in \eqref{U expansion}}\\
\eqref{NP2c}&&\to&&&\nu_0\\
\eqref{NP2d}&&\to&&&\text{equation \eqref{mu1 evolution}}\\
\eqref{NP2e}&&\to&&&\text{equation \eqref{alpha1 evolution}}\\
\eqref{NP2f}&&\to&&&\lambda_1\\
\hspace{-1cm}\text{4 components of }\eqref{NP2g}&&\to&&&\tau_1,\; \Psi_3^0,\; \text{equation \eqref{ma and gamma0 constraints}}\\
\eqref{NP2h}&&\to&&&\omega_1\\
\eqref{NP2i}&&\to&&&\mu_1\\
\eqref{NP2j}&&\to&&&\Psi_3^0\\
\hspace{-1cm}\text{4 components of }\eqref{NP2k}&&\to&&&\Sigma,\; \alpha_1,\; \text{Im}(\Psi_2^0)
\ee
\esub
\item Finally, the remaining equations from \eqref{NP2l} to \eqref{NP2o} provide at leading order the evolution equations \eqref{EOM Psi00}, \eqref{EOM Psi10}, \eqref{EOM Psi20}, and \eqref{Psi30 evolution}.
\end{itemize}

\newpage

\section{C-metric}
\label{app:C-metric}

The line element for the C-metric in Bondi gauge is \cite{Bonnor:1990tw,Bini:2005qyt,Griffiths:2006tk}
\be
\de s^2=-H\de u^2-2\de u\,\de r+2Ar^2\sin\theta\,\de u\,\de\theta+r^2q_{ab}\de x^a\de x^b,
\ee
where
\bsub
\be
G&=(1+2AM\cos\theta)\sin^2\theta,\\
H&=1-\f{2M}{r}-A^2Gr^2-Ar\big(2\cos\theta+AM(1+3\cos2\theta)\big)+6AM\cos\theta,\\
q_{ab}&=\text{diag}\big(\sin^2\theta\,G^{-1},G\big).\label{C-metric 2d q}
\ee
\esub
The parameters $A$ and $M$ denote respectively the acceleration and the mass of the source. The tetrad giving rise to this line element is of the form \eqref{NP tetrad} with
\be
W=1,
\q\q
U=-\f{1}{2}(H+A^2Gr^2),
\q\q
X^a=(-AG\csc\theta,0),
\q\q
Z=\Omega=0,
\ee
and with $rm^a=m^a_1$ obtained from \eqref{frame m1} with \eqref{C-metric 2d q}.

Although this tetrad has vanishing twist potential, the fact that $X^a=X^a_0=\O(1)$ makes it lie a priori outside of the algebraically special solution space which we have constructed in section \ref{sec: algebraically special}. This is because we have set the integration constant to $X^a_0(u,x^b)=0$ in \eqref{AS Xa}. In principle, one can of course study the resolution of the NP equations once this integration constant is reintroduced. This was done in the metric formulation for the algebraically general solution space in \cite{Geiller:2022vto,MG-AdS,McNees:2025acf}. In particular, it was shown that the constraint which arises from the Einstein equations $G^\tf_{ab}|_{\O(r)}=0$ is $(\partial_u-4\gamma_0)q_{ab}+2D_{\la a}X^0_{b\ra}=0$. Here, with the time-independent metric \eqref{C-metric 2d q} one can indeed verify that this constraint is consistently satisfied. The construction of the algebraically special solution space with twist and $X^a_0\neq0$ requires to understand (among other things) how this constraint propagates in the solution space. This will be the focus of future work.

It is however possible, at least in principle, to change coordinates and to bring the C-metric in a form which has $X^a=0$ while preserving $W=1$ and $Z=\Omega=0$. For this, we first go to stereographic coordinates $(z,\bar{z})$ by redefining the angular coordinates as
\be\label{stereographic}
\theta=2\,\text{arccot}\sqrt{z\bar{z}},
\q\q
\phi=-\f{i}{2}\ln\f{z}{\bar{z}}.
\ee
Then, one must further redefine the angular coordinates as
\be
z=\zeta e^f,
\q\q
\bar{z}=\bar{\zeta}e^f,
\ee
for some function $f(u,\zeta,\bar{\zeta})$. When this function satisfies
\be
\partial_uf=A\left(1-2AM\f{1-\zeta\bar{\zeta}e^{2f}}{1+\zeta\bar{\zeta}e^{2f}}\right),
\ee
we obtain the C-metric with $X^a=Z=\Omega=0$ and $W=1$, and with a more complicated form of the angular metric $q_{ab}$. Unfortunately, this function cannot be computed analytically. This motivates further the enlargement of the solution space by introducing $X^a_0\neq0$, as it enables us to easily include certain solutions such as the C-metric without cumbersome changes of coordinates.

\newpage

\bibliography{Biblio.bib}

\providecommand{\href}[2]{#2}\begingroup\raggedright\begin{thebibliography}{100}

\bibitem{Bondi:1960jsa}
H.~Bondi, ``{Gravitational Waves in General Relativity}'',
  \href{http://dx.doi.org/10.1038/186535a0}{\emph{Nature} {\bfseries 186}
  (1960) 535--535}.

\bibitem{Bondi:1962px}
H.~Bondi, M.~G.~J. van~der Burg and A.~W.~K. Metzner, ``{Gravitational waves in
  general relativity. 7. Waves from axisymmetric isolated systems}'',
  \href{http://dx.doi.org/10.1098/rspa.1962.0161}{\emph{Proc. Roy. Soc. Lond.}
  {\bfseries A269} (1962) 21--52}.

\bibitem{Bondi:1962rkt}
H.~Bondi, ``{Radiation from an isolated system}'',  in \emph{{International
  Conference on Relativistic Theories of Gravitation}}, pp.~115--122, 1964.

\bibitem{Sachs:1962zza}
R.~Sachs, ``{Asymptotic symmetries in gravitational theory}'',
  \href{http://dx.doi.org/10.1103/PhysRev.128.2851}{\emph{Phys. Rev.}
  {\bfseries 128} (1962) 2851--2864}.

\bibitem{Sachs:1962wk}
R.~K. Sachs, ``{Gravitational waves in general relativity. 8. Waves in
  asymptotically flat space-times}'',
  \href{http://dx.doi.org/10.1098/rspa.1962.0206}{\emph{Proc. Roy. Soc. Lond.}
  {\bfseries A270} (1962) 103--126}.

\bibitem{Penrose:1962ij}
R.~Penrose, ``{Asymptotic properties of fields and space-times}'',
  \href{http://dx.doi.org/10.1103/PhysRevLett.10.66}{\emph{Phys. Rev. Lett.}
  {\bfseries 10} (1963) 66--68}.

\bibitem{Penrose:1964ge}
R.~Penrose, ``Conformal treatment of infinity'',
  \href{http://dx.doi.org/10.1007/s10714-010-1110-5}{\emph{General Relativity
  and Gravitation} {\bfseries 43} (1964) 901--922}.

\bibitem{Penrose:1965am}
R.~Penrose, ``{Zero rest mass fields including gravitation: Asymptotic
  behavior}'', \href{http://dx.doi.org/10.1098/rspa.1965.0058}{\emph{Proc. Roy.
  Soc. Lond. A} {\bfseries 284} (1965) 159}.

\bibitem{Geroch1977}
R.~Geroch, \emph{Asymptotic Structure of Space-Time}, pp.~1--105.
\newblock Springer US, Boston, MA, 1977.

\bibitem{Braginsky:1986ia}
V.~B. Braginsky and L.~P. Grishchuk, ``{Kinematic Resonance and Memory Effect
  in Free Mass Gravitational Antennas}'', {\emph{Sov. Phys. JETP} {\bfseries
  62} (1985) 427--430}.

\bibitem{1987Natur.327..123B}
V.~B. {Braginskii} and K.~S. {Thorne}, ``{Gravitational-wave bursts with memory
  and experimental prospects}'',
  \href{http://dx.doi.org/10.1038/327123a0}{\emph{Nature} {\bfseries 327} (May,
  1987) 123--125}.

\bibitem{Christodoulou:1991cr}
D.~Christodoulou, ``{Nonlinear nature of gravitation and gravitational wave
  experiments}'',
  \href{http://dx.doi.org/10.1103/PhysRevLett.67.1486}{\emph{Phys. Rev. Lett.}
  {\bfseries 67} (1991) 1486--1489}.

\bibitem{Blanchet:1992br}
L.~Blanchet and T.~Damour, ``{Hereditary effects in gravitational radiation}'',
  \href{http://dx.doi.org/10.1103/PhysRevD.46.4304}{\emph{Phys. Rev.}
  {\bfseries D46} (1992) 4304--4319}.

\bibitem{Thorne:1992sdb}
K.~S. Thorne, ``{Gravitational-wave bursts with memory: The Christodoulou
  effect}'', \href{http://dx.doi.org/10.1103/PhysRevD.45.520}{\emph{Phys. Rev.}
  {\bfseries D45} (1992) 520--524}.

\bibitem{Low:1954kd}
F.~E. Low, ``{Scattering of light of very low frequency by systems of spin
  1/2}'', \href{http://dx.doi.org/10.1103/PhysRev.96.1428}{\emph{Phys. Rev.}
  {\bfseries 96} (1954) 1428--1432}.

\bibitem{Weinberg:1965nx}
S.~Weinberg, ``{Infrared photons and gravitons}'',
  \href{http://dx.doi.org/10.1103/PhysRev.140.B516}{\emph{Phys. Rev.}
  {\bfseries 140} (1965) B516--B524}.

\bibitem{Strominger:2013jfa}
A.~Strominger, ``{On BMS Invariance of Gravitational Scattering}'',
  \href{http://dx.doi.org/10.1007/JHEP07(2014)152}{\emph{JHEP} {\bfseries 07}
  (2014) 152}, [\href{https://arxiv.org/abs/1312.2229}{{\ttfamily 1312.2229}}].

\bibitem{He:2014laa}
T.~He, V.~Lysov, P.~Mitra and A.~Strominger, ``{BMS supertranslations and
  Weinberg's soft graviton theorem}'',
  \href{http://dx.doi.org/10.1007/JHEP05(2015)151}{\emph{JHEP} {\bfseries 05}
  (2015) 151}, [\href{https://arxiv.org/abs/1401.7026}{{\ttfamily 1401.7026}}].

\bibitem{Strominger:2014pwa}
A.~Strominger and A.~Zhiboedov, ``{Gravitational Memory, BMS Supertranslations
  and Soft Theorems}'',
  \href{http://dx.doi.org/10.1007/JHEP01(2016)086}{\emph{JHEP} {\bfseries 01}
  (2016) 086}, [\href{https://arxiv.org/abs/1411.5745}{{\ttfamily 1411.5745}}].

\bibitem{Pasterski:2015tva}
S.~Pasterski, A.~Strominger and A.~Zhiboedov, ``{New Gravitational Memories}'',
  \href{http://dx.doi.org/10.1007/JHEP12(2016)053}{\emph{JHEP} {\bfseries 12}
  (2016) 053}, [\href{https://arxiv.org/abs/1502.06120}{{\ttfamily
  1502.06120}}].

\bibitem{Compere:2016gwf}
G.~Comp\`ere, ``{Bulk supertranslation memories: a concept reshaping the vacua
  and black holes of general relativity}'',
  \href{http://dx.doi.org/10.1142/S0218271816440065}{\emph{Int. J. Mod. Phys.
  D} {\bfseries 25} (2016) 1644006},
  [\href{https://arxiv.org/abs/1606.00377}{{\ttfamily 1606.00377}}].

\bibitem{Compere:2019odm}
G.~Comp\`ere, ``{Infinite towers of supertranslation and superrotation
  memories}'',
  \href{http://dx.doi.org/10.1103/PhysRevLett.123.021101}{\emph{Phys. Rev.
  Lett.} {\bfseries 123} (2019) 021101},
  [\href{https://arxiv.org/abs/1904.00280}{{\ttfamily 1904.00280}}].

\bibitem{Nichols:2017rqr}
D.~A. Nichols, ``{Spin memory effect for compact binaries in the post-Newtonian
  approximation}'',
  \href{http://dx.doi.org/10.1103/PhysRevD.95.084048}{\emph{Phys. Rev. D}
  {\bfseries 95} (2017) 084048},
  [\href{https://arxiv.org/abs/1702.03300}{{\ttfamily 1702.03300}}].

\bibitem{Nichols:2018qac}
D.~A. Nichols, ``{Center-of-mass angular momentum and memory effect in
  asymptotically flat spacetimes}'',
  \href{http://dx.doi.org/10.1103/PhysRevD.98.064032}{\emph{Phys. Rev. D}
  {\bfseries 98} (2018) 064032},
  [\href{https://arxiv.org/abs/1807.08767}{{\ttfamily 1807.08767}}].

\bibitem{Flanagan:2018yzh}
E.~E. Flanagan, A.~M. Grant, A.~I. Harte and D.~A. Nichols, ``{Persistent
  gravitational wave observables: general framework}'',
  \href{http://dx.doi.org/10.1103/PhysRevD.99.084044}{\emph{Phys. Rev. D}
  {\bfseries 99} (2019) 084044},
  [\href{https://arxiv.org/abs/1901.00021}{{\ttfamily 1901.00021}}].

\bibitem{Flanagan:2019ezo}
E.~E. Flanagan, A.~M. Grant, A.~I. Harte and D.~A. Nichols, ``{Persistent
  gravitational wave observables: Nonlinear plane wave spacetimes}'',
  \href{http://dx.doi.org/10.1103/PhysRevD.101.104033}{\emph{Phys. Rev. D}
  {\bfseries 101} (2020) 104033},
  [\href{https://arxiv.org/abs/1912.13449}{{\ttfamily 1912.13449}}].

\bibitem{Grant:2021hga}
A.~M. Grant and D.~A. Nichols, ``{Persistent gravitational wave observables:
  Curve deviation in asymptotically flat spacetimes}'',
  \href{http://dx.doi.org/10.1103/PhysRevD.105.024056}{\emph{Phys. Rev. D}
  {\bfseries 105} (2022) 024056},
  [\href{https://arxiv.org/abs/2109.03832}{{\ttfamily 2109.03832}}].

\bibitem{Grant:2023ged}
A.~M. Grant, ``{Persistent gravitational wave observables: Nonlinearities in
  (non-)geodesic deviation}'',
  [\href{https://arxiv.org/abs/2401.00047}{{\ttfamily 2401.00047}}].

\bibitem{Compere:2018ylh}
G.~Comp\`ere, A.~Fiorucci and R.~Ruzziconi, ``{Superboost transitions,
  refraction memory and super-Lorentz charge algebra}'',
  \href{http://dx.doi.org/10.1007/JHEP11(2018)200}{\emph{JHEP} {\bfseries 11}
  (2018) 200}, [\href{https://arxiv.org/abs/1810.00377}{{\ttfamily
  1810.00377}}].

\bibitem{Seraj:2021rxd}
A.~Seraj and B.~Oblak, ``{Gyroscopic gravitational memory}'',
  \href{http://dx.doi.org/10.1007/JHEP11(2023)057}{\emph{JHEP} {\bfseries 11}
  (2023) 057}, [\href{https://arxiv.org/abs/2112.04535}{{\ttfamily
  2112.04535}}].

\bibitem{Seraj:2022qyt}
A.~Seraj and B.~Oblak, ``{Precession Caused by Gravitational Waves}'',
  \href{http://dx.doi.org/10.1103/PhysRevLett.129.061101}{\emph{Phys. Rev.
  Lett.} {\bfseries 129} (2022) 061101},
  [\href{https://arxiv.org/abs/2203.16216}{{\ttfamily 2203.16216}}].

\bibitem{White:2014qia}
C.~D. White, ``{Diagrammatic insights into next-to-soft corrections}'',
  \href{http://dx.doi.org/10.1016/j.physletb.2014.08.041}{\emph{Phys. Lett. B}
  {\bfseries 737} (2014) 216--222},
  [\href{https://arxiv.org/abs/1406.7184}{{\ttfamily 1406.7184}}].

\bibitem{Cachazo:2014fwa}
F.~Cachazo and A.~Strominger, ``{Evidence for a New Soft Graviton Theorem}'',
  [\href{https://arxiv.org/abs/1404.4091}{{\ttfamily 1404.4091}}].

\bibitem{Zlotnikov:2014sva}
M.~Zlotnikov, ``{Sub-sub-leading soft-graviton theorem in arbitrary
  dimension}'', \href{http://dx.doi.org/10.1007/JHEP10(2014)148}{\emph{JHEP}
  {\bfseries 10} (2014) 148},
  [\href{https://arxiv.org/abs/1407.5936}{{\ttfamily 1407.5936}}].

\bibitem{Kalousios:2014uva}
C.~Kalousios and F.~Rojas, ``{Next to subleading soft-graviton theorem in
  arbitrary dimensions}'',
  \href{http://dx.doi.org/10.1007/JHEP01(2015)107}{\emph{JHEP} {\bfseries 01}
  (2015) 107}, [\href{https://arxiv.org/abs/1407.5982}{{\ttfamily 1407.5982}}].

\bibitem{Conde:2016rom}
E.~Conde and P.~Mao, ``{BMS Supertranslations and Not So Soft Gravitons}'',
  \href{http://dx.doi.org/10.1007/JHEP05(2017)060}{\emph{JHEP} {\bfseries 05}
  (2017) 060}, [\href{https://arxiv.org/abs/1612.08294}{{\ttfamily
  1612.08294}}].

\bibitem{Campiglia:2016efb}
M.~Campiglia and A.~Laddha, ``{Sub-subleading soft gravitons and large
  diffeomorphisms}'',
  \href{http://dx.doi.org/10.1007/JHEP01(2017)036}{\emph{JHEP} {\bfseries 01}
  (2017) 036}, [\href{https://arxiv.org/abs/1608.00685}{{\ttfamily
  1608.00685}}].

\bibitem{Campiglia:2016jdj}
M.~Campiglia and A.~Laddha, ``{Sub-subleading soft gravitons: New symmetries of
  quantum gravity?}'',
  \href{http://dx.doi.org/10.1016/j.physletb.2016.11.046}{\emph{Phys. Lett. B}
  {\bfseries 764} (2017) 218--221},
  [\href{https://arxiv.org/abs/1605.09094}{{\ttfamily 1605.09094}}].

\bibitem{Banerjee:2021cly}
S.~Banerjee, S.~Ghosh and S.~S. Samal, ``{Subsubleading soft graviton symmetry
  and MHV graviton scattering amplitudes}'',
  \href{http://dx.doi.org/10.1007/JHEP08(2021)067}{\emph{JHEP} {\bfseries 08}
  (2021) 067}, [\href{https://arxiv.org/abs/2104.02546}{{\ttfamily
  2104.02546}}].

\bibitem{Freidel:2021dfs}
L.~Freidel, D.~Pranzetti and A.-M. Raclariu, ``{Sub-subleading Soft Graviton
  Theorem from Asymptotic Einstein's Equations}'',
  [\href{https://arxiv.org/abs/2111.15607}{{\ttfamily 2111.15607}}].

\bibitem{Laddha:2018myi}
A.~Laddha and A.~Sen, ``{Logarithmic Terms in the Soft Expansion in Four
  Dimensions}'', \href{http://dx.doi.org/10.1007/JHEP10(2018)056}{\emph{JHEP}
  {\bfseries 10} (2018) 056},
  [\href{https://arxiv.org/abs/1804.09193}{{\ttfamily 1804.09193}}].

\bibitem{Sahoo:2018lxl}
B.~Sahoo and A.~Sen, ``{Classical and Quantum Results on Logarithmic Terms in
  the Soft Theorem in Four Dimensions}'',
  \href{http://dx.doi.org/10.1007/JHEP02(2019)086}{\emph{JHEP} {\bfseries 02}
  (2019) 086}, [\href{https://arxiv.org/abs/1808.03288}{{\ttfamily
  1808.03288}}].

\bibitem{Saha:2019tub}
A.~P. Saha, B.~Sahoo and A.~Sen, ``{Proof of the classical soft graviton
  theorem in $D$ = 4}'',
  \href{http://dx.doi.org/10.1007/JHEP06(2020)153}{\emph{JHEP} {\bfseries 06}
  (2020) 153}, [\href{https://arxiv.org/abs/1912.06413}{{\ttfamily
  1912.06413}}].

\bibitem{Barnich:2009se}
G.~Barnich and C.~Troessaert, ``{Symmetries of asymptotically flat 4
  dimensional spacetimes at null infinity revisited}'',
  \href{http://dx.doi.org/10.1103/PhysRevLett.105.111103}{\emph{Phys. Rev.
  Lett.} {\bfseries 105} (2010) 111103},
  [\href{https://arxiv.org/abs/0909.2617}{{\ttfamily 0909.2617}}].

\bibitem{Barnich:2010eb}
G.~Barnich and C.~Troessaert, ``{Aspects of the BMS/CFT correspondence}'',
  \href{http://dx.doi.org/10.1007/JHEP05(2010)062}{\emph{JHEP} {\bfseries 05}
  (2010) 062}, [\href{https://arxiv.org/abs/1001.1541}{{\ttfamily 1001.1541}}].

\bibitem{Barnich:2011mi}
G.~Barnich and C.~Troessaert, ``{BMS charge algebra}'',
  \href{http://dx.doi.org/10.1007/JHEP12(2011)105}{\emph{JHEP} {\bfseries 12}
  (2011) 105}, [\href{https://arxiv.org/abs/1106.0213}{{\ttfamily 1106.0213}}].

\bibitem{Barnich:2011ct}
G.~Barnich and C.~Troessaert, ``{Supertranslations call for superrotations}'',
  \href{http://dx.doi.org/10.22323/1.127.0010}{\emph{PoS} {\bfseries CNCFG2010}
  (2010) 010}, [\href{https://arxiv.org/abs/1102.4632}{{\ttfamily 1102.4632}}].

\bibitem{Barnich:2013axa}
G.~Barnich and C.~Troessaert, ``{Comments on holographic current algebras and
  asymptotically flat four dimensional spacetimes at null infinity}'',
  \href{http://dx.doi.org/10.1007/JHEP11(2013)003}{\emph{JHEP} {\bfseries 11}
  (2013) 003}, [\href{https://arxiv.org/abs/1309.0794}{{\ttfamily 1309.0794}}].

\bibitem{Campiglia:2020qvc}
M.~Campiglia and J.~Peraza, ``{Generalized BMS charge algebra}'',
  \href{http://dx.doi.org/10.1103/PhysRevD.101.104039}{\emph{Phys. Rev. D}
  {\bfseries 101} (2020) 104039},
  [\href{https://arxiv.org/abs/2002.06691}{{\ttfamily 2002.06691}}].

\bibitem{Campiglia:2014yka}
M.~Campiglia and A.~Laddha, ``{Asymptotic symmetries and subleading soft
  graviton theorem}'',
  \href{http://dx.doi.org/10.1103/PhysRevD.90.124028}{\emph{Phys. Rev. D}
  {\bfseries 90} (2014) 124028},
  [\href{https://arxiv.org/abs/1408.2228}{{\ttfamily 1408.2228}}].

\bibitem{Campiglia:2015yka}
M.~Campiglia and A.~Laddha, ``{New symmetries for the Gravitational
  S-matrix}'', \href{http://dx.doi.org/10.1007/JHEP04(2015)076}{\emph{JHEP}
  {\bfseries 04} (2015) 076},
  [\href{https://arxiv.org/abs/1502.02318}{{\ttfamily 1502.02318}}].

\bibitem{Flanagan:2015pxa}
E.~E. Flanagan and D.~A. Nichols, ``{Conserved charges of the extended
  Bondi-Metzner-Sachs algebra}'',
  \href{http://dx.doi.org/10.1103/PhysRevD.95.044002}{\emph{Phys. Rev. D}
  {\bfseries 95} (2017) 044002},
  [\href{https://arxiv.org/abs/1510.03386}{{\ttfamily 1510.03386}}].

\bibitem{Freidel:2021fxf}
L.~Freidel, R.~Oliveri, D.~Pranzetti and S.~Speziale, ``{The Weyl BMS group and
  Einstein\textquoteright{}s equations}'',
  \href{http://dx.doi.org/10.1007/JHEP07(2021)170}{\emph{JHEP} {\bfseries 07}
  (2021) 170}, [\href{https://arxiv.org/abs/2104.05793}{{\ttfamily
  2104.05793}}].

\bibitem{Geiller:2022vto}
M.~Geiller and C.~Zwikel, ``{The partial Bondi gauge: Further enlarging the
  asymptotic structure of gravity}'',
  \href{http://dx.doi.org/10.21468/SciPostPhys.13.5.108}{\emph{SciPost Phys.}
  {\bfseries 13} (2022) 108},
  [\href{https://arxiv.org/abs/2205.11401}{{\ttfamily 2205.11401}}].

\bibitem{Geiller:2024amx}
M.~Geiller and C.~Zwikel, ``{The partial Bondi gauge: Gauge fixings and
  asymptotic charges}'',
  \href{http://dx.doi.org/10.21468/SciPostPhys.16.3.076}{\emph{SciPost Phys.}
  {\bfseries 16} (2024) 076},
  [\href{https://arxiv.org/abs/2401.09540}{{\ttfamily 2401.09540}}].

\bibitem{Pasterski:2016qvg}
S.~Pasterski, S.-H. Shao and A.~Strominger, ``{Flat Space Amplitudes and
  Conformal Symmetry of the Celestial Sphere}'',
  \href{http://dx.doi.org/10.1103/PhysRevD.96.065026}{\emph{Phys. Rev. D}
  {\bfseries 96} (2017) 065026},
  [\href{https://arxiv.org/abs/1701.00049}{{\ttfamily 1701.00049}}].

\bibitem{Pasterski:2017kqt}
S.~Pasterski and S.-H. Shao, ``{Conformal basis for flat space amplitudes}'',
  \href{http://dx.doi.org/10.1103/PhysRevD.96.065022}{\emph{Phys. Rev. D}
  {\bfseries 96} (2017) 065022},
  [\href{https://arxiv.org/abs/1705.01027}{{\ttfamily 1705.01027}}].

\bibitem{Donnay:2020guq}
L.~Donnay, S.~Pasterski and A.~Puhm, ``{Asymptotic Symmetries and Celestial
  CFT}'', \href{http://dx.doi.org/10.1007/JHEP09(2020)176}{\emph{JHEP}
  {\bfseries 09} (2020) 176},
  [\href{https://arxiv.org/abs/2005.08990}{{\ttfamily 2005.08990}}].

\bibitem{Pasterski:2021raf}
S.~Pasterski, M.~Pate and A.-M. Raclariu, ``{Celestial Holography}'',  in
  \emph{{2022 Snowmass Summer Study}}, 11, 2021.
\newblock \href{https://arxiv.org/abs/2111.11392}{{\ttfamily 2111.11392}}.

\bibitem{Pasterski:2021rjz}
S.~Pasterski, ``{Lectures on celestial amplitudes}'',
  \href{http://dx.doi.org/10.1140/epjc/s10052-021-09846-7}{\emph{Eur. Phys. J.
  C} {\bfseries 81} (2021) 1062},
  [\href{https://arxiv.org/abs/2108.04801}{{\ttfamily 2108.04801}}].

\bibitem{Raclariu:2021zjz}
A.-M. Raclariu, ``{Lectures on Celestial Holography}'',
  [\href{https://arxiv.org/abs/2107.02075}{{\ttfamily 2107.02075}}].

\bibitem{Donnay:2022aba}
L.~Donnay, A.~Fiorucci, Y.~Herfray and R.~Ruzziconi, ``{Carrollian Perspective
  on Celestial Holography}'',
  \href{http://dx.doi.org/10.1103/PhysRevLett.129.071602}{\emph{Phys. Rev.
  Lett.} {\bfseries 129} (2022) 071602},
  [\href{https://arxiv.org/abs/2202.04702}{{\ttfamily 2202.04702}}].

\bibitem{Donnay:2022wvx}
L.~Donnay, A.~Fiorucci, Y.~Herfray and R.~Ruzziconi, ``{Bridging Carrollian and
  celestial holography}'',
  \href{http://dx.doi.org/10.1103/PhysRevD.107.126027}{\emph{Phys. Rev. D}
  {\bfseries 107} (2023) 126027},
  [\href{https://arxiv.org/abs/2212.12553}{{\ttfamily 2212.12553}}].

\bibitem{Mason:2023mti}
L.~Mason, R.~Ruzziconi and A.~Yelleshpur~Srikant, ``{Carrollian Amplitudes and
  Celestial Symmetries}'',  [\href{https://arxiv.org/abs/2312.10138}{{\ttfamily
  2312.10138}}].

\bibitem{Madler:2016xju}
T.~M\"adler and J.~Winicour, ``{Bondi-Sachs Formalism}'',
  \href{http://dx.doi.org/10.4249/scholarpedia.33528}{\emph{Scholarpedia}
  {\bfseries 11} (2016) 33528},
  [\href{https://arxiv.org/abs/1609.01731}{{\ttfamily 1609.01731}}].

\bibitem{Newman:1962cia}
E.~T. Newman and T.~W.~J. Unti, ``{Behavior of Asymptotically Flat Empty
  Spaces}'', \href{http://dx.doi.org/10.1063/1.1724303}{\emph{J. Math. Phys.}
  {\bfseries 3} (1962) 891}.

\bibitem{PhysRevD.110.024018}
C.~Gundlach, D.~Hilditch and T.~W. Baumgarte, ``Simulations of gravitational
  collapse in null coordinates. i. formulation and weak-field tests in
  generalized bondi gauges'',
  \href{http://dx.doi.org/10.1103/PhysRevD.110.024018}{\emph{Phys. Rev. D}
  {\bfseries 110} (Jul, 2024) 024018}.

\bibitem{PhysRevD.110.024020}
C.~Gundlach, ``Simulations of gravitational collapse in null coordinates. iii.
  hyperbolicity'',
  \href{http://dx.doi.org/10.1103/PhysRevD.110.024020}{\emph{Phys. Rev. D}
  {\bfseries 110} (Jul, 2024) 024020}.

\bibitem{Tamburino:1966zz}
L.~A. Tamburino and J.~H. Winicour, ``{Gravitational Fields in Finite and
  Conformal Bondi Frames}'',
  \href{http://dx.doi.org/10.1103/PhysRev.150.1039}{\emph{Phys. Rev.}
  {\bfseries 150} (1966) 1039--1053}.

\bibitem{Geroch:1981ut}
R.~P. Geroch and J.~Winicour, ``{Linkages in general relativity}'',
  \href{http://dx.doi.org/10.1063/1.524987}{\emph{J. Math. Phys.} {\bfseries
  22} (1981) 803--812}.

\bibitem{Ashtekar:1981bq}
A.~Ashtekar and M.~Streubel, ``{Symplectic Geometry of Radiative Modes and
  Conserved Quantities at Null Infinity}'',
  \href{http://dx.doi.org/10.1098/rspa.1981.0109}{\emph{Proc. Roy. Soc. Lond.
  A} {\bfseries 376} (1981) 585--607}.

\bibitem{Dray:1984rfa}
T.~Dray and M.~Streubel, ``{Angular momentum at null infinity}'',
  \href{http://dx.doi.org/10.1088/0264-9381/1/1/005}{\emph{Class. Quant. Grav.}
  {\bfseries 1} (1984) 15--26}.

\bibitem{Wald:1999wa}
R.~M. Wald and A.~Zoupas, ``{A General definition of 'conserved quantities' in
  general relativity and other theories of gravity}'',
  \href{http://dx.doi.org/10.1103/PhysRevD.61.084027}{\emph{Phys. Rev. D}
  {\bfseries 61} (2000) 084027},
  [\href{https://arxiv.org/abs/gr-qc/9911095}{{\ttfamily gr-qc/9911095}}].

\bibitem{Ciambelli:2017wou}
L.~Ciambelli, A.~C. Petkou, P.~M. Petropoulos and K.~Siampos, ``{The
  Robinson-Trautman spacetime and its holographic fluid}'',
  \href{http://dx.doi.org/10.22323/1.292.0076}{\emph{PoS} {\bfseries CORFU2016}
  (2017) 076}, [\href{https://arxiv.org/abs/1707.02995}{{\ttfamily
  1707.02995}}].

\bibitem{Adami:2024mtu}
H.~Adami, A.~Parvizi, M.~M. Sheikh-Jabbari and V.~Taghiloo, ``{Heisenberg soft
  hair on Robinson-Trautman spacetimes}'',
  \href{http://dx.doi.org/10.1007/JHEP05(2024)191}{\emph{JHEP} {\bfseries 05}
  (2024) 191}, [\href{https://arxiv.org/abs/2402.17658}{{\ttfamily
  2402.17658}}].

\bibitem{barnich-seraj-RT}
G.~Barnich and A.~Seraj{\emph{~\!\!\!} (to appear)}.

\bibitem{Poole:2018koa}
A.~Poole, K.~Skenderis and M.~Taylor, ``{(A)dS$\mathbf{_4}$ in Bondi gauge}'',
  \href{http://dx.doi.org/10.1088/1361-6382/ab117c}{\emph{Class. Quant. Grav.}
  {\bfseries 36} (2019) 095005},
  [\href{https://arxiv.org/abs/1812.05369}{{\ttfamily 1812.05369}}].

\bibitem{Compere:2019bua}
G.~Comp\`ere, A.~Fiorucci and R.~Ruzziconi, ``{The $\Lambda$-BMS$_4$ group of
  dS$_4$ and new boundary conditions for AdS$_4$}'',
  \href{http://dx.doi.org/10.1088/1361-6382/ab3d4b}{\emph{Class. Quant. Grav.}
  {\bfseries 36} (2019) 195017},
  [\href{https://arxiv.org/abs/1905.00971}{{\ttfamily 1905.00971}}].

\bibitem{MG-AdS}
M.~Geiller, ``{On radiation and the Weyl tensor in asymptotically-(A)dS$_4$
  spacetimes}'', {\emph{~\!\!\!} (to appear)}.

\bibitem{McNees:2025acf}
R.~McNees and C.~Zwikel, ``{(Anti)-de Sitter with leaky boundaries and
  corners}'',  [\href{https://arxiv.org/abs/2512.03170}{{\ttfamily
  2512.03170}}].

\bibitem{Compere:2023ktn}
G.~Comp\`ere, S.~J. Hoque and E.~c. Kutluk, ``{Quadrupolar radiation in de
  Sitter: Displacement memory and Bondi metric}'',
  [\href{https://arxiv.org/abs/2309.02081}{{\ttfamily 2309.02081}}].

\bibitem{Compere:2024ekl}
G.~Comp\`ere, S.~J. Hoque and E.~c. Kutluk, ``{The $SO(1,4)$ flux-balance laws
  of de Sitter at quadrupolar order}'',
  [\href{https://arxiv.org/abs/2411.16215}{{\ttfamily 2411.16215}}].

\bibitem{Arenas-Henriquez:2025rpt}
G.~Arenas-Henriquez, L.~Ciambelli, F.~Diaz, W.~Jia and D.~Rivera-Betancour,
  ``{Radiation in Fluid/Gravity and the Flat Limit}'',
  [\href{https://arxiv.org/abs/2508.01446}{{\ttfamily 2508.01446}}].

\bibitem{1985FoPh...15..605W}
J.~{Winicour}, ``{Logarithmic asymptotic flatness}'',
  \href{http://dx.doi.org/10.1007/BF01882485}{\emph{Foundations of Physics}
  {\bfseries 15} (May, 1985) 605--616}.

\bibitem{Chrusciel:1993hx}
P.~T. Chrusciel, M.~A.~H. MacCallum and D.~B. Singleton, ``{Gravitational waves
  in general relativity: 14. Bondi expansions and the polyhomogeneity of
  Scri}'',  [\href{https://arxiv.org/abs/gr-qc/9305021}{{\ttfamily
  gr-qc/9305021}}].

\bibitem{Kroon:1998tu}
J.~A.~V. Kroon, ``{Conserved quantities for polyhomogeneous space-times}'',
  \href{http://dx.doi.org/10.1088/0264-9381/15/8/023}{\emph{Class. Quant.
  Grav.} {\bfseries 15} (1998) 2479--2491},
  [\href{https://arxiv.org/abs/gr-qc/9805094}{{\ttfamily gr-qc/9805094}}].

\bibitem{ValienteKroon:2002gb}
J.~A. Valiente~Kroon, ``{Polyhomogeneous expansions close to null and spatial
  infinity}'', {\emph{Lect. Notes Phys.} {\bfseries 604} (2002) 135--160},
  [\href{https://arxiv.org/abs/gr-qc/0202001}{{\ttfamily gr-qc/0202001}}].

\bibitem{Godazgar:2020peu}
M.~Godazgar and G.~Long, ``{BMS charges in polyhomogeneous spacetimes}'',
  \href{http://dx.doi.org/10.1103/PhysRevD.102.064036}{\emph{Phys. Rev. D}
  {\bfseries 102} (2020) 064036},
  [\href{https://arxiv.org/abs/2007.15672}{{\ttfamily 2007.15672}}].

\bibitem{Freidel:2024tpl}
L.~Freidel and A.~Riello, ``{Renormalization of conformal infinity as a
  stretched horizon}'',  [\href{https://arxiv.org/abs/2402.03097}{{\ttfamily
  2402.03097}}].

\bibitem{Geiller:2024ryw}
M.~Geiller, A.~Laddha and C.~Zwikel, ``{Symmetries of the gravitational
  scattering in the absence of peeling}'',
  \href{http://dx.doi.org/10.1007/JHEP12(2024)081}{\emph{JHEP} {\bfseries 12}
  (2024) 081}, [\href{https://arxiv.org/abs/2407.07978}{{\ttfamily
  2407.07978}}].

\bibitem{1986mgm..conf..365D}
T.~{Damour}, ``{Analytical calculations of gravitational radiation.}'',  in
  \emph{Fourth Marcel Grossmann Meeting on General Relativity}, pp.~365--392,
  Jan., 1986.

\bibitem{2002nmgm.meet...44C}
D.~{Christodoulou}, ``{The Global Initial Value Problem in General
  Relativity}'',  in \emph{The Ninth Marcel Grossmann Meeting} (V.~G.
  {Gurzadyan}, R.~T. {Jantzen} and R.~{Ruffini}, eds.), pp.~44--54, Dec., 2002.
\newblock \href{http://dx.doi.org/10.1142/9789812777386_0004}{DOI}.

\bibitem{Friedrich:1983vx}
H.~Friedrich, ``Cauchy problems for the conformal vacuum field equations in
  general relativity'',
  \href{http://dx.doi.org/10.1007/BF01206015}{\emph{Communications in
  Mathematical Physics} {\bfseries 91} (1983) 445--472}.

\bibitem{PhysRevD.19.3483}
M.~Walker and C.~M. Will, ``{Relativistic Kepler problem. I. Behavior in the
  distant past of orbits with gravitational radiation damping}'',
  \href{http://dx.doi.org/10.1103/PhysRevD.19.3483}{\emph{Phys. Rev. D}
  {\bfseries 19} (Jun, 1979) 3483--3494}.

\bibitem{PhysRevD.19.3495}
M.~Walker and C.~M. Will, ``{Relativistic Kepler problem. II. Asymptotic
  behavior of the field in the infinite past}'',
  \href{http://dx.doi.org/10.1103/PhysRevD.19.3495}{\emph{Phys. Rev. D}
  {\bfseries 19} (Jun, 1979) 3495--3508}.

\bibitem{doi:10.1098/rspa.1981.0101}
J.~Porrill, J.~M. Stewart and R.~Penrose, ``Electromagnetic and gravitational
  fields in a schwarzschild space-time'',
  \href{http://dx.doi.org/10.1098/rspa.1981.0101}{\emph{Proceedings of the
  Royal Society of London. A. Mathematical and Physical Sciences} {\bfseries
  376} (1981) 451--463},
  [\href{https://arxiv.org/abs/https://royalsocietypublishing.org/doi/pdf/10.1098/rspa.1981.0101}{{\ttfamily
  https://royalsocietypublishing.org/doi/pdf/10.1098/rspa.1981.0101}}].

\bibitem{Andersson:1993we}
L.~Andersson and P.~T. Chrusciel, ``{On 'hyperboloidal' Cauchy data for vacuum
  Einstein equations and obstructions to smoothness of 'null infinity'}'',
  \href{http://dx.doi.org/10.1103/PhysRevLett.70.2829}{\emph{Phys. Rev. Lett.}
  {\bfseries 70} (1993) 2829--2832},
  [\href{https://arxiv.org/abs/gr-qc/9304019}{{\ttfamily gr-qc/9304019}}].

\bibitem{Valiente-Kroon:2002xys}
J.~A. Valiente-Kroon, ``{A New class of obstructions to the smoothness of null
  infinity}'', \href{http://dx.doi.org/10.1007/s00220-003-0967-5}{\emph{Commun.
  Math. Phys.} {\bfseries 244} (2004) 133--156},
  [\href{https://arxiv.org/abs/gr-qc/0211024}{{\ttfamily gr-qc/0211024}}].

\bibitem{Kroon:2004me}
J.~A.~V. Kroon, ``{Time asymmetric spacetimes near null and spatial infinity.
  II. Expansions of developments of initial data sets with non-smooth conformal
  metrics}'', \href{http://dx.doi.org/10.1088/0264-9381/22/9/015}{\emph{Class.
  Quant. Grav.} {\bfseries 22} (2005) 1683--1707},
  [\href{https://arxiv.org/abs/gr-qc/0412045}{{\ttfamily gr-qc/0412045}}].

\bibitem{Kehrberger:2021uvf}
L.~M.~A. Kehrberger, ``{The Case Against Smooth Null Infinity I: Heuristics and
  Counter-Examples}'',
  \href{http://dx.doi.org/10.1007/s00023-021-01108-2}{\emph{Annales Henri
  Poincare} {\bfseries 23} (2022) 829--921},
  [\href{https://arxiv.org/abs/2105.08079}{{\ttfamily 2105.08079}}].

\bibitem{Kehrberger:2021vhp}
L.~M.~A. Kehrberger, ``{The Case Against Smooth Null Infinity II: A
  Logarithmically Modified Price's Law}'',
  [\href{https://arxiv.org/abs/2105.08084}{{\ttfamily 2105.08084}}].

\bibitem{Kehrberger:2021azo}
L.~M.~A. Kehrberger, ``{The Case Against Smooth Null Infinity III: Early-Time
  Asymptotics for Higher $\ell $-Modes of Linear Waves on a Schwarzschild
  Background}'', \href{http://dx.doi.org/10.1007/s40818-022-00129-2}{\emph{Ann.
  PDE} {\bfseries 8} (2022) 12},
  [\href{https://arxiv.org/abs/2106.00035}{{\ttfamily 2106.00035}}].

\bibitem{Kehrberger:2024clh}
L.~Kehrberger, ``{The case against smooth null infinity IV: Linearized gravity
  around Schwarzschild -- an overview}'',
  \href{http://dx.doi.org/10.1098/rsta.2023.0039}{\emph{Phil. Trans. Roy. Soc.
  Lond. A} {\bfseries 382} (2024) 20230039},
  [\href{https://arxiv.org/abs/2401.04170}{{\ttfamily 2401.04170}}].

\bibitem{Kehrberger:2024aak}
L.~Kehrberger and H.~Masaood, ``{The Case Against Smooth Null Infinity V:
  Early-Time Asymptotics of Linearised Gravity Around Schwarzschild for Fixed
  Spherical Harmonic Modes}'',
  [\href{https://arxiv.org/abs/2401.04179}{{\ttfamily 2401.04179}}].

\bibitem{Gajic:2022pst}
D.~Gajic and L.~M.~A. Kehrberger, ``{On the relation between asymptotic
  charges, the failure of peeling and late-time tails}'',
  \href{http://dx.doi.org/10.1088/1361-6382/ac8863}{\emph{Class. Quant. Grav.}
  {\bfseries 39} (2022) 195006},
  [\href{https://arxiv.org/abs/2202.04093}{{\ttfamily 2202.04093}}].

\bibitem{Kehrberger:2023btg}
L.~Kehrberger, \emph{{Mathematical Studies on the Asymptotic Behaviour of
  Gravitational Radiation in General Relativity}}.
\newblock PhD thesis, Department of Applied Mathematics And Theoretical
  Physics, Cambridge U., 2023.
\newblock 10.17863/CAM.99689.

\bibitem{Bieri:2023cyn}
L.~Bieri, ``{Radiation and Asymptotics for Spacetimes with Non-Isotropic
  Mass}'',  [\href{https://arxiv.org/abs/2304.00611}{{\ttfamily 2304.00611}}].

\bibitem{Laddha:2018vbn}
A.~Laddha and A.~Sen, ``{Observational Signature of the Logarithmic Terms in
  the Soft Graviton Theorem}'',
  \href{http://dx.doi.org/10.1103/PhysRevD.100.024009}{\emph{Phys. Rev. D}
  {\bfseries 100} (2019) 024009},
  [\href{https://arxiv.org/abs/1806.01872}{{\ttfamily 1806.01872}}].

\bibitem{Agrawal:2023zea}
S.~Agrawal, L.~Donnay, K.~Nguyen and R.~Ruzziconi, ``{Logarithmic soft graviton
  theorems from superrotation Ward identities}'',
  \href{http://dx.doi.org/10.1007/JHEP02(2024)120}{\emph{JHEP} {\bfseries 02}
  (2024) 120}, [\href{https://arxiv.org/abs/2309.11220}{{\ttfamily
  2309.11220}}].

\bibitem{Choi:2024ygx}
S.~Choi, A.~Laddha and A.~Puhm, ``{Asymptotic Symmetries for Logarithmic Soft
  Theorems in Gauge Theory and Gravity}'',
  [\href{https://arxiv.org/abs/2403.13053}{{\ttfamily 2403.13053}}].

\bibitem{Choi:2024ajz}
S.~Choi, A.~Laddha and A.~Puhm, ``{The Classical Super-Rotation Infrared
  Triangle}'',  [\href{https://arxiv.org/abs/2412.16142}{{\ttfamily
  2412.16142}}].

\bibitem{DeAngelis:2025vlf}
S.~De~Angelis, A.~Herderschee, R.~Roiban and F.~Teng, ``{Asymptotic Simplicity
  and Scattering in General Relativity from Quantum Field Theory}'',
  [\href{https://arxiv.org/abs/2511.10637}{{\ttfamily 2511.10637}}].

\bibitem{McNees:2024iyu}
R.~McNees and C.~Zwikel, ``{The symplectic potential for leaky boundaries}'',
  [\href{https://arxiv.org/abs/2408.13203}{{\ttfamily 2408.13203}}].

\bibitem{Bonga:2023eml}
B.~Bonga, C.~Bunster and A.~P\'erez, ``{Gravitational radiation with
  \ensuremath{\Lambda}\ensuremath{>}0}'',
  \href{http://dx.doi.org/10.1103/PhysRevD.108.064039}{\emph{Phys. Rev. D}
  {\bfseries 108} (2023) 064039},
  [\href{https://arxiv.org/abs/2306.08029}{{\ttfamily 2306.08029}}].

\bibitem{He:2023qha}
T.~He, A.-M. Raclariu and K.~M. Zurek, ``{From shockwaves to the gravitational
  memory effect}'',
  \href{http://dx.doi.org/10.1007/JHEP01(2024)006}{\emph{JHEP} {\bfseries 01}
  (2024) 006}, [\href{https://arxiv.org/abs/2305.14411}{{\ttfamily
  2305.14411}}].

\bibitem{Mao:2019ahc}
P.~Mao, ``{Asymptotics with a cosmological constant: The solution space}'',
  \href{http://dx.doi.org/10.1103/PhysRevD.99.104024}{\emph{Phys. Rev. D}
  {\bfseries 99} (2019) 104024},
  [\href{https://arxiv.org/abs/1901.04010}{{\ttfamily 1901.04010}}].

\bibitem{Compere:2020lrt}
G.~Comp\`ere, A.~Fiorucci and R.~Ruzziconi, ``{The $\Lambda$-BMS$_4$ charge
  algebra}'', \href{http://dx.doi.org/10.1007/JHEP10(2020)205}{\emph{JHEP}
  {\bfseries 10} (2020) 205},
  [\href{https://arxiv.org/abs/2004.10769}{{\ttfamily 2004.10769}}].

\bibitem{Bonga:2020fhx}
B.~Bonga and K.~Prabhu, ``{BMS-like symmetries in cosmology}'',
  [\href{https://arxiv.org/abs/2009.01243}{{\ttfamily 2009.01243}}].

\bibitem{Enriquez-Rojo:2020miw}
M.~Enriquez-Rojo and T.~Heckelbacher, ``{Asymptotic symmetries in spatially
  flat FRW spacetimes}'',
  \href{http://dx.doi.org/10.1103/PhysRevD.103.064009}{\emph{Phys. Rev. D}
  {\bfseries 103} (2021) 064009},
  [\href{https://arxiv.org/abs/2011.01960}{{\ttfamily 2011.01960}}].

\bibitem{Enriquez-Rojo:2021blc}
M.~Enriquez-Rojo and T.~Heckelbacher, ``{Holography and black holes in
  asymptotically flat FLRW spacetimes}'',
  \href{http://dx.doi.org/10.1103/PhysRevD.103.104035}{\emph{Phys. Rev. D}
  {\bfseries 103} (2021) 104035},
  [\href{https://arxiv.org/abs/2102.02234}{{\ttfamily 2102.02234}}].

\bibitem{Enriquez-Rojo:2022onp}
M.~Enriquez-Rojo, T.~Heckelbacher and R.~Oliveri, ``{Asymptotic dynamics and
  charges for FLRW spacetimes}'',
  \href{http://dx.doi.org/10.1103/PhysRevD.107.024039}{\emph{Phys. Rev. D}
  {\bfseries 107} (2023) 024039},
  [\href{https://arxiv.org/abs/2201.07600}{{\ttfamily 2201.07600}}].

\bibitem{Ashtekar:1996cd}
A.~Ashtekar, J.~Bicak and B.~G. Schmidt, ``{Asymptotic structure of symmetry
  reduced general relativity}'',
  \href{http://dx.doi.org/10.1103/PhysRevD.55.669}{\emph{Phys. Rev. D}
  {\bfseries 55} (1997) 669--686},
  [\href{https://arxiv.org/abs/gr-qc/9608042}{{\ttfamily gr-qc/9608042}}].

\bibitem{Barnich:2006av}
G.~Barnich and G.~Compere, ``{Classical central extension for asymptotic
  symmetries at null infinity in three spacetime dimensions}'',
  \href{http://dx.doi.org/10.1088/0264-9381/24/5/F01}{\emph{Class. Quant.
  Grav.} {\bfseries 24} (2007) F15--F23},
  [\href{https://arxiv.org/abs/gr-qc/0610130}{{\ttfamily gr-qc/0610130}}].

\bibitem{Brown:1986nw}
J.~D. Brown and M.~Henneaux, ``{Central Charges in the Canonical Realization of
  Asymptotic Symmetries: An Example from Three-Dimensional Gravity}'',
  \href{http://dx.doi.org/10.1007/BF01211590}{\emph{Commun. Math. Phys.}
  {\bfseries 104} (1986) 207--226}.

\bibitem{Barnich:2012aw}
G.~Barnich, A.~Gomberoff and H.~A. Gonz\'alez, ``{The Flat limit of three
  dimensional asymptotically anti-de Sitter spacetimes}'',
  \href{http://dx.doi.org/10.1103/PhysRevD.86.024020}{\emph{Phys. Rev. D}
  {\bfseries 86} (2012) 024020},
  [\href{https://arxiv.org/abs/1204.3288}{{\ttfamily 1204.3288}}].

\bibitem{Ciambelli:2020eba}
L.~Ciambelli, C.~Marteau, P.~M. Petropoulos and R.~Ruzziconi, ``{Gauges in
  Three-Dimensional Gravity and Holographic Fluids}'',
  \href{http://dx.doi.org/10.1007/JHEP11(2020)092}{\emph{JHEP} {\bfseries 11}
  (2020) 092}, [\href{https://arxiv.org/abs/2006.10082}{{\ttfamily
  2006.10082}}].

\bibitem{Ciambelli:2020ftk}
L.~Ciambelli, C.~Marteau, P.~M. Petropoulos and R.~Ruzziconi,
  ``{Fefferman-Graham and Bondi Gauges in the Fluid/Gravity Correspondence}'',
  \href{http://dx.doi.org/10.22323/1.376.0154}{\emph{PoS} {\bfseries CORFU2019}
  (2020) 154}, [\href{https://arxiv.org/abs/2006.10083}{{\ttfamily
  2006.10083}}].

\bibitem{Campoleoni:2022wmf}
A.~Campoleoni, L.~Ciambelli, A.~Delfante, C.~Marteau, P.~M. Petropoulos and
  R.~Ruzziconi, ``{Holographic Lorentz and Carroll frames}'',
  \href{http://dx.doi.org/10.1007/JHEP12(2022)007}{\emph{JHEP} {\bfseries 12}
  (2022) 007}, [\href{https://arxiv.org/abs/2208.07575}{{\ttfamily
  2208.07575}}].

\bibitem{Ruzziconi:2020wrb}
R.~Ruzziconi and C.~Zwikel, ``{Conservation and Integrability in
  Lower-Dimensional Gravity}'',
  \href{http://dx.doi.org/10.1007/JHEP04(2021)034}{\emph{JHEP} {\bfseries 04}
  (2021) 034}, [\href{https://arxiv.org/abs/2012.03961}{{\ttfamily
  2012.03961}}].

\bibitem{Geiller:2021vpg}
M.~Geiller, C.~Goeller and C.~Zwikel, ``{3d gravity in Bondi-Weyl gauge:
  charges, corners, and integrability}'',
  \href{http://dx.doi.org/10.1007/JHEP09(2021)029}{\emph{JHEP} {\bfseries 09}
  (2021) 029}, [\href{https://arxiv.org/abs/2107.01073}{{\ttfamily
  2107.01073}}].

\bibitem{Kerr:1963ud}
R.~P. Kerr, ``{Gravitational field of a spinning mass as an example of
  algebraically special metrics}'',
  \href{http://dx.doi.org/10.1103/PhysRevLett.11.237}{\emph{Phys. Rev. Lett.}
  {\bfseries 11} (1963) 237--238}.

\bibitem{Visser:2007fj}
M.~Visser, ``{The Kerr spacetime: A Brief introduction}'',  in \emph{{Kerr
  Fest: Black Holes in Astrophysics, General Relativity and Quantum Gravity}},
  6, 2007.
\newblock \href{https://arxiv.org/abs/0706.0622}{{\ttfamily 0706.0622}}.

\bibitem{FletcherLun1996}
S.~J. Fletcher and A.~W.~C. Lun, ``{Bondi-Sachs metrics and exact solutions}'',
   in \emph{Proceedings of the 7th Marcel Grossmann Meeting on General
  Relativity} (R.~T. Jantzen and G.~M. Keiser, eds.), (Singapore),
  pp.~296--298, World Scientific, 1996.

\bibitem{2003CQGra..20.4153F}
S.~J. {Fletcher} and A.~W.~C. {Lun}, ``{The Kerr spacetime in generalized Bondi
  Sachs coordinates}'',
  \href{http://dx.doi.org/10.1088/0264-9381/20/19/302}{\emph{Classical and
  Quantum Gravity} {\bfseries 20} (Oct., 2003) 4153--4167}.

\bibitem{Hoque:2021nti}
S.~J. Hoque and A.~Virmani, ``{The Kerr\textendash{}de Sitter spacetime in
  Bondi coordinates}'',
  \href{http://dx.doi.org/10.1088/1361-6382/ac2c1f}{\emph{Class. Quant. Grav.}
  {\bfseries 38} (2021) 225002},
  [\href{https://arxiv.org/abs/2108.01098}{{\ttfamily 2108.01098}}].

\bibitem{Hawking:2016msc}
S.~W. Hawking, M.~J. Perry and A.~Strominger, ``{Soft Hair on Black Holes}'',
  \href{http://dx.doi.org/10.1103/PhysRevLett.116.231301}{\emph{Phys. Rev.
  Lett.} {\bfseries 116} (2016) 231301},
  [\href{https://arxiv.org/abs/1601.00921}{{\ttfamily 1601.00921}}].

\bibitem{Hawking:2016sgy}
S.~W. Hawking, M.~J. Perry and A.~Strominger, ``{Superrotation Charge and
  Supertranslation Hair on Black Holes}'',
  \href{http://dx.doi.org/10.1007/JHEP05(2017)161}{\emph{JHEP} {\bfseries 05}
  (2017) 161}, [\href{https://arxiv.org/abs/1611.09175}{{\ttfamily
  1611.09175}}].

\bibitem{Compere:2016hzt}
G.~Comp\`ere and J.~Long, ``{Classical static final state of collapse with
  supertranslation memory}'',
  \href{http://dx.doi.org/10.1088/0264-9381/33/19/195001}{\emph{Class. Quant.
  Grav.} {\bfseries 33} (2016) 195001},
  [\href{https://arxiv.org/abs/1602.05197}{{\ttfamily 1602.05197}}].

\bibitem{Compere:2016jwb}
G.~Comp\`ere and J.~Long, ``{Vacua of the gravitational field}'',
  \href{http://dx.doi.org/10.1007/JHEP07(2016)137}{\emph{JHEP} {\bfseries 07}
  (2016) 137}, [\href{https://arxiv.org/abs/1601.04958}{{\ttfamily
  1601.04958}}].

\bibitem{Stephani:2003tm}
H.~Stephani, D.~Kramer, M.~A.~H. MacCallum, C.~Hoenselaers and E.~Herlt,
  \emph{{Exact solutions of Einstein's field equations}}.
\newblock Cambridge Monographs on Mathematical Physics. Cambridge Univ. Press,
  Cambridge, 2003,
  \href{http://dx.doi.org/10.1017/CBO9780511535185}{10.1017/CBO9780511535185}.

\bibitem{Mao:2024jpt}
P.~Mao and W.~Zhao, ``{Twisting asymptotic symmetries and algebraically special
  vacuum solutions}'',  [\href{https://arxiv.org/abs/2401.12054}{{\ttfamily
  2401.12054}}].

\bibitem{Mittal:2022ywl}
N.~Mittal, P.~M. Petropoulos, D.~Rivera-Betancour and M.~Vilatte, ``{Ehlers,
  Carroll, charges and dual charges}'',
  \href{http://dx.doi.org/10.1007/JHEP07(2023)065}{\emph{JHEP} {\bfseries 07}
  (2023) 065}, [\href{https://arxiv.org/abs/2212.14062}{{\ttfamily
  2212.14062}}].

\bibitem{Campoleoni:2023fug}
A.~Campoleoni, A.~Delfante, S.~Pekar, P.~M. Petropoulos, D.~Rivera-Betancour
  and M.~Vilatte, ``{Flat from anti de Sitter}'',
  \href{http://dx.doi.org/10.1007/JHEP12(2023)078}{\emph{JHEP} {\bfseries 12}
  (2023) 078}, [\href{https://arxiv.org/abs/2309.15182}{{\ttfamily
  2309.15182}}].

\bibitem{Fiorucci:2025twa}
A.~Fiorucci, S.~Pekar, P.~Marios~Petropoulos and M.~Vilatte,
  ``{Carrollian-Holographic Derivation of Gravitational Flux-Balance Laws}'',
  \href{http://dx.doi.org/10.1103/qv17-ks32}{\emph{Phys. Rev. Lett.} {\bfseries
  135} (2025) 261602}, [\href{https://arxiv.org/abs/2505.00077}{{\ttfamily
  2505.00077}}].

\bibitem{Hartong:2025jpp}
J.~Hartong, E.~Have, V.~Nenmeli and G.~Oling, ``{Boundary Energy-Momentum
  Tensors for Asymptotically Flat Spacetimes}'',
  [\href{https://arxiv.org/abs/2505.05432}{{\ttfamily 2505.05432}}].

\bibitem{Adami:2024rkr}
H.~Adami, M.~M. Sheikh-Jabbari and V.~Taghiloo, ``{Gravitational stress tensor
  and current at null infinity in three dimensions}'',
  \href{http://dx.doi.org/10.1016/j.physletb.2024.138835}{\emph{Phys. Lett. B}
  {\bfseries 855} (2024) 138835},
  [\href{https://arxiv.org/abs/2405.00149}{{\ttfamily 2405.00149}}].

\bibitem{Barnich:2019vzx}
G.~Barnich, P.~Mao and R.~Ruzziconi, ``{BMS current algebra in the context of
  the Newman\textendash{}Penrose formalism}'',
  \href{http://dx.doi.org/10.1088/1361-6382/ab7c01}{\emph{Class. Quant. Grav.}
  {\bfseries 37} (2020) 095010},
  [\href{https://arxiv.org/abs/1910.14588}{{\ttfamily 1910.14588}}].

\bibitem{Chandrasekhar:1985kt}
S.~Chandrasekhar, \emph{{The mathematical theory of black holes}}.
\newblock Clarendon Press, 1985.

\bibitem{Barnich:2016lyg}
G.~Barnich and C.~Troessaert, ``{Finite BMS transformations}'',
  \href{http://dx.doi.org/10.1007/JHEP03(2016)167}{\emph{JHEP} {\bfseries 03}
  (2016) 167}, [\href{https://arxiv.org/abs/1601.04090}{{\ttfamily
  1601.04090}}].

\bibitem{Barnich:2011ty}
G.~Barnich and P.-H. Lambert, ``{A Note on the Newman-Unti group and the BMS
  charge algebra in terms of Newman-Penrose coefficients}'',
  \href{http://dx.doi.org/10.1155/2012/197385}{\emph{Adv. Math. Phys.}
  {\bfseries 2012} (2012) 197385},
  [\href{https://arxiv.org/abs/1102.0589}{{\ttfamily 1102.0589}}].

\bibitem{Barnich:2012nkq}
G.~Barnich and P.-H. Lambert, ``{Asymptotic symmetries at null infinity and
  local conformal properties of spin coefficients}'', {\emph{TSPU Bulletin}
  {\bfseries 2012} (2012) 28--31},
  [\href{https://arxiv.org/abs/1301.5754}{{\ttfamily 1301.5754}}].

\bibitem{Geiller:2024bgf}
M.~Geiller, ``{Celestial $w_{1+\infty}$ charges and the subleading structure of
  asymptotically-flat spacetimes}'',
  \href{http://dx.doi.org/10.21468/SciPostPhys.18.1.023}{\emph{SciPost Phys.}
  {\bfseries 18} (2025) 023},
  [\href{https://arxiv.org/abs/2403.05195}{{\ttfamily 2403.05195}}].

\bibitem{Compere:2022zdz}
G.~Comp\`ere, R.~Oliveri and A.~Seraj, ``{Metric reconstruction from celestial
  multipoles}'', \href{http://dx.doi.org/10.1007/JHEP11(2022)001}{\emph{JHEP}
  {\bfseries 11} (2022) 001},
  [\href{https://arxiv.org/abs/2206.12597}{{\ttfamily 2206.12597}}].

\bibitem{Newman:1965ik}
E.~T. Newman and R.~Penrose, ``{10 exact gravitationally-conserved
  quantities}'',
  \href{http://dx.doi.org/10.1103/PhysRevLett.15.231}{\emph{Phys. Rev. Lett.}
  {\bfseries 15} (1965) 231--233}.

\bibitem{Newman:1968uj}
E.~T. Newman and R.~Penrose, ``{New conservation laws for zero rest-mass fields
  in asymptotically flat space-time}'',
  \href{http://dx.doi.org/10.1098/rspa.1968.0112}{\emph{Proc. Roy. Soc. Lond.
  A} {\bfseries 305} (1968) 175--204}.

\bibitem{Freidel:2021ytz}
L.~Freidel, D.~Pranzetti and A.-M. Raclariu, ``Higher spin dynamics in gravity
  and ${w}_{1+\ensuremath{\infty}}$ celestial symmetries'',
  \href{http://dx.doi.org/10.1103/PhysRevD.106.086013}{\emph{Phys. Rev. D}
  {\bfseries 106} (Oct, 2022) 086013},
  [\href{https://arxiv.org/abs/2112.15573}{{\ttfamily 2112.15573}}].

\bibitem{Cresto:2024mne}
N.~Cresto and L.~Freidel, ``{Asymptotic Higher Spin Symmetries II: Noether
  Realization in Gravity}'',
  [\href{https://arxiv.org/abs/2410.15219}{{\ttfamily 2410.15219}}].

\bibitem{Cresto:2024fhd}
N.~Cresto and L.~Freidel, ``{Asymptotic higher spin symmetries I: covariant
  wedge algebra in gravity}'',
  \href{http://dx.doi.org/10.1007/s11005-025-01921-4}{\emph{Lett. Math. Phys.}
  {\bfseries 115} (2025) 39},
  [\href{https://arxiv.org/abs/2409.12178}{{\ttfamily 2409.12178}}].

\bibitem{Duval:2014uoa}
C.~Duval, G.~W. Gibbons, P.~A. Horvathy and P.~M. Zhang, ``{Carroll versus
  Newton and Galilei: two dual non-Einsteinian concepts of time}'',
  \href{http://dx.doi.org/10.1088/0264-9381/31/8/085016}{\emph{Class. Quant.
  Grav.} {\bfseries 31} (2014) 085016},
  [\href{https://arxiv.org/abs/1402.0657}{{\ttfamily 1402.0657}}].

\bibitem{Ciambelli:2019lap}
L.~Ciambelli, R.~G. Leigh, C.~Marteau and P.~M. Petropoulos, ``{Carroll
  Structures, Null Geometry and Conformal Isometries}'',
  \href{http://dx.doi.org/10.1103/PhysRevD.100.046010}{\emph{Phys. Rev. D}
  {\bfseries 100} (2019) 046010},
  [\href{https://arxiv.org/abs/1905.02221}{{\ttfamily 1905.02221}}].

\bibitem{Ciambelli:2025unn}
L.~Ciambelli and P.~Jai-akson, ``{Foundations of Carrollian Geometry}'',
  [\href{https://arxiv.org/abs/2510.21651}{{\ttfamily 2510.21651}}].

\bibitem{Ciambelli:2025mex}
L.~Ciambelli, ``{Asymptotic Limit of Null Hypersurfaces}'',
  [\href{https://arxiv.org/abs/2501.17357}{{\ttfamily 2501.17357}}].

\bibitem{Iyer:1994ys}
V.~Iyer and R.~M. Wald, ``{Some properties of Noether charge and a proposal for
  dynamical black hole entropy}'',
  \href{http://dx.doi.org/10.1103/PhysRevD.50.846}{\emph{Phys. Rev.} {\bfseries
  D50} (1994) 846--864}, [\href{https://arxiv.org/abs/gr-qc/9403028}{{\ttfamily
  gr-qc/9403028}}].

\bibitem{Compere:2018aar}
G.~Compere and A.~Fiorucci, ``{Advanced Lectures on General Relativity}'',
  \href{http://dx.doi.org/10.1007/978-3-030-04260-8}{\emph{Lect. Notes Phys.}
  {\bfseries 952} (2019) 150},
  [\href{https://arxiv.org/abs/1801.07064}{{\ttfamily 1801.07064}}].

\bibitem{Freidel:2019ohg}
L.~Freidel, F.~Hopfmueller and A.~Riello, ``{Asymptotic Renormalization in Flat
  Space: Symplectic Potential and Charges of Electromagnetism}'',
  \href{http://dx.doi.org/10.1007/JHEP10(2019)126}{\emph{JHEP} {\bfseries 10}
  (2019) 126}, [\href{https://arxiv.org/abs/1904.04384}{{\ttfamily
  1904.04384}}].

\bibitem{McNees:2023tus}
R.~McNees and C.~Zwikel, ``{Finite charges from the bulk action}'',
  \href{http://dx.doi.org/10.1007/JHEP08(2023)154}{\emph{JHEP} {\bfseries 08}
  (2023) 154}, [\href{https://arxiv.org/abs/2306.16451}{{\ttfamily
  2306.16451}}].

\bibitem{Riello:2024uvs}
A.~Riello and L.~Freidel, ``{Renormalization of conformal infinity as a
  stretched horizon}'',
  \href{http://dx.doi.org/10.1088/1361-6382/ad5cbb}{\emph{Class. Quant. Grav.}
  {\bfseries 41} (2024) 175013},
  [\href{https://arxiv.org/abs/2402.03097}{{\ttfamily 2402.03097}}].

\bibitem{Delfante:2024npo}
A.~Delfante, \emph{{Of asymptotic charges and renormalizations}}.
\newblock PhD thesis, U. Mons, 6, 2024.

\bibitem{Grant:2021sxk}
A.~M. Grant, K.~Prabhu and I.~Shehzad, ``{The Wald\textendash{}Zoupas
  prescription for asymptotic charges at null infinity in general
  relativity}'', \href{http://dx.doi.org/10.1088/1361-6382/ac571a}{\emph{Class.
  Quant. Grav.} {\bfseries 39} (2022) 085002},
  [\href{https://arxiv.org/abs/2105.05919}{{\ttfamily 2105.05919}}].

\bibitem{Odak:2022ndm}
G.~Odak, A.~Rignon-Bret and S.~Speziale, ``{Wald-Zoupas prescription with soft
  anomalies}'',
  \href{http://dx.doi.org/10.1103/PhysRevD.107.084028}{\emph{Phys. Rev. D}
  {\bfseries 107} (2023) 084028},
  [\href{https://arxiv.org/abs/2212.07947}{{\ttfamily 2212.07947}}].

\bibitem{Rignon-Bret:2024gcx}
A.~Rignon-Bret and S.~Speziale, ``{Centerless-BMS charge algebra}'',
  [\href{https://arxiv.org/abs/2405.01526}{{\ttfamily 2405.01526}}].

\bibitem{Oliveri:2019gvm}
R.~Oliveri and S.~Speziale, ``{Boundary effects in General Relativity with
  tetrad variables}'',  [\href{https://arxiv.org/abs/1912.01016}{{\ttfamily
  1912.01016}}].

\bibitem{Oliveri:2020xls}
R.~Oliveri and S.~Speziale, ``{A note on dual gravitational charges}'',
  [\href{https://arxiv.org/abs/2010.01111}{{\ttfamily 2010.01111}}].

\bibitem{Freidel:2020xyx}
L.~Freidel, M.~Geiller and D.~Pranzetti, ``{Edge modes of gravity. Part I.
  Corner potentials and charges}'',
  \href{http://dx.doi.org/10.1007/JHEP11(2020)026}{\emph{JHEP} {\bfseries 11}
  (2020) 026}, [\href{https://arxiv.org/abs/2006.12527}{{\ttfamily
  2006.12527}}].

\bibitem{Freidel:2020svx}
L.~Freidel, M.~Geiller and D.~Pranzetti, ``{Edge modes of gravity. Part II.
  Corner metric and Lorentz charges}'',
  \href{http://dx.doi.org/10.1007/JHEP11(2020)027}{\emph{JHEP} {\bfseries 11}
  (2020) 027}, [\href{https://arxiv.org/abs/2007.03563}{{\ttfamily
  2007.03563}}].

\bibitem{Venter:2005cs}
L.~R. Venter and N.~T. Bishop, ``{Numerical validation of the Kerr metric in
  Bondi-Sachs form}'',
  \href{http://dx.doi.org/10.1103/PhysRevD.73.084023}{\emph{Phys. Rev. D}
  {\bfseries 73} (2006) 084023},
  [\href{https://arxiv.org/abs/gr-qc/0506077}{{\ttfamily gr-qc/0506077}}].

\bibitem{Newman:1961qr}
E.~Newman and R.~Penrose, ``{An Approach to gravitational radiation by a method
  of spin coefficients}'', \href{http://dx.doi.org/10.1063/1.1724257}{\emph{J.
  Math. Phys.} {\bfseries 3} (1962) 566--578}.

\bibitem{Newman:2009}
E.~T. Newman and R.~Penrose, ``{S}pin-coefficient formalism'',
  \href{http://dx.doi.org/10.4249/scholarpedia.7445}{\emph{Scholarpedia}
  {\bfseries 4} (2009) 7445}.

\bibitem{Timofeev:1996vg}
V.~N. Timofeev, ``Algebraically special vacuum gravitational fields with a
  cosmological constant'',
  \href{http://dx.doi.org/10.1007/BF02437026}{\emph{Russian Physics Journal}
  {\bfseries 39} (1996) 585--590}.

\bibitem{Petropoulos:2015fba}
P.~M. Petropoulos and K.~Siampos, ``{Integrability, Einstein spaces and
  holographic fluids}'',  [\href{https://arxiv.org/abs/1510.06456}{{\ttfamily
  1510.06456}}].

\bibitem{Ciambelli:2018wre}
L.~Ciambelli, C.~Marteau, A.~C. Petkou, P.~M. Petropoulos and K.~Siampos,
  ``{Flat holography and Carrollian fluids}'',
  \href{http://dx.doi.org/10.1007/JHEP07(2018)165}{\emph{JHEP} {\bfseries 07}
  (2018) 165}, [\href{https://arxiv.org/abs/1802.06809}{{\ttfamily
  1802.06809}}].

\bibitem{LJMason_1998}
L.~J. Mason, ``The asymptotic structure of algebraically special spacetimes'',
  \href{http://dx.doi.org/10.1088/0264-9381/15/4/022}{\emph{Classical and
  Quantum Gravity} {\bfseries 15} (apr, 1998) 1019}.

\bibitem{1962AcPPS..22...13G}
J.~N. {Goldberg} and R.~K. {Sachs}, ``{A theorem on Petrov types}'',
  {\emph{Acta Physica Polonica B, Proceedings Supplement} {\bfseries 22} (Jan.,
  1962) 13}.

\bibitem{Krasinski:2009vz}
A.~Krasi{\'n}ski and M.~Przanowski, ``{Editorial note to: J. N. Goldberg and R.
  K. Sachs, A theorem on Petrov types}'',
  \href{http://dx.doi.org/10.1007/s10714-008-0721-6}{\emph{General Relativity
  and Gravitation} {\bfseries 41} (2009) 421--432}.

\bibitem{Lu:2025fzm}
H.~Lu and P.~Mao, ``{Four-dimensional Stationary Algebraically Special
  Solutions, Weyl Invariants, and Soft Hairs}'',
  [\href{https://arxiv.org/abs/2503.14586}{{\ttfamily 2503.14586}}].

\bibitem{Manko:2005nm}
V.~S. Manko and E.~Ruiz, ``{Physical interpretation of NUT solution}'',
  \href{http://dx.doi.org/10.1088/0264-9381/22/17/014}{\emph{Class. Quant.
  Grav.} {\bfseries 22} (2005) 3555--3560},
  [\href{https://arxiv.org/abs/gr-qc/0505001}{{\ttfamily gr-qc/0505001}}].

\bibitem{Bhat:2024cyq}
S.~A. Bhat, S.~Bhattacharjee and S.~J. Kapadia, ``{Can the near-horizon black
  hole memory be detected through binary inspirals?}'',
  \href{http://dx.doi.org/10.1103/c344-fm5w}{\emph{Phys. Rev. D} {\bfseries
  112} (2025) 024068}, [\href{https://arxiv.org/abs/2406.15604}{{\ttfamily
  2406.15604}}].

\bibitem{Kinnersley:1968zz}
W.~M. Kinnersley, \emph{{Type D gravitational fields}}.
\newblock PhD thesis, Caltech, 1968.

\bibitem{Kinnersley:1969zza}
W.~Kinnersley, ``{Type D Vacuum Metrics}'',
  \href{http://dx.doi.org/10.1063/1.1664958}{\emph{J. Math. Phys.} {\bfseries
  10} (1969) 1195--1203}.

\bibitem{Carter1968}
B.~Carter, ``{Global Structure of the Kerr Family of Gravitational Fields}'',
  \href{http://dx.doi.org/10.1103/PhysRev.174.1559}{\emph{Physical Review}
  {\bfseries 174} (1968) 1559--1571}.

\bibitem{Kashiwada1968}
T.~Kashiwada, ``On conformal killing tensor'', {\emph{Natural Science Report,
  Ochanomizu University} {\bfseries 19} (1968) 67--74}.

\bibitem{WalkerPenrose1970}
M.~Walker and R.~Penrose, ``On quadratic first integrals of the geodesic
  equations for type \text{D} spacetimes'',
  \href{http://dx.doi.org/10.1007/BF01649445}{\emph{Communications in
  Mathematical Physics} {\bfseries 18} (1970) 265--274}.

\bibitem{Debever1971}
R.~Debever, N.~Kamran and R.~G. McLenaghan, ``{Petrov Type D vacuum metrics and
  Killing tensors}'', \href{http://dx.doi.org/10.1063/1.1665655}{\emph{Journal
  of Mathematical Physics} {\bfseries 12} (1971) 986--990}.

\bibitem{Floyd1973}
R.~Floyd, ``{The dynamics of Kerr fields}'', {\emph{PhD Thesis, University of
  London} (1973) }.

\bibitem{Krtous:2006qy}
P.~Krtous, D.~Kubiznak, D.~N. Page and V.~P. Frolov, ``{Killing-Yano Tensors,
  Rank-2 Killing Tensors, and Conserved Quantities in Higher Dimensions}'',
  \href{http://dx.doi.org/10.1088/1126-6708/2007/02/004}{\emph{JHEP} {\bfseries
  02} (2007) 004}, [\href{https://arxiv.org/abs/hep-th/0612029}{{\ttfamily
  hep-th/0612029}}].

\bibitem{Frolov:2008jr}
V.~P. Frolov and D.~Kubiznak, ``{Higher-Dimensional Black Holes: Hidden
  Symmetries and Separation of Variables}'',
  \href{http://dx.doi.org/10.1088/0264-9381/25/15/154005}{\emph{Class. Quant.
  Grav.} {\bfseries 25} (2008) 154005},
  [\href{https://arxiv.org/abs/0802.0322}{{\ttfamily 0802.0322}}].

\bibitem{Frolov:2009aq}
V.~P. Frolov, ``{Hidden Symmetries and Black Holes}'',
  \href{http://dx.doi.org/10.1088/1742-6596/189/1/012015}{\emph{J. Phys. Conf.
  Ser.} {\bfseries 189} (2009) 012015},
  [\href{https://arxiv.org/abs/0901.1472}{{\ttfamily 0901.1472}}].

\bibitem{Brink:2009mq}
J.~Brink, ``{Spacetime Encodings III - Second Order Killing Tensors}'',
  \href{http://dx.doi.org/10.1103/PhysRevD.81.022001}{\emph{Phys. Rev. D}
  {\bfseries 81} (2010) 022001},
  [\href{https://arxiv.org/abs/0911.1589}{{\ttfamily 0911.1589}}].

\bibitem{Frolov:2017kze}
V.~P. Frolov, P.~Krtous and D.~Kubiznak, ``{Black holes, hidden symmetries, and
  complete integrability}'',
  \href{http://dx.doi.org/10.1007/s41114-017-0009-9}{\emph{Living Rev. Rel.}
  {\bfseries 20} (2017) 6}, [\href{https://arxiv.org/abs/1705.05482}{{\ttfamily
  1705.05482}}].

\bibitem{Zhu:2022shb}
Q.-H. Zhu, Y.-X. Han and Q.-G. Huang, ``{The shadow of supertranslated black
  hole}'', \href{http://dx.doi.org/10.1140/epjc/s10052-023-11232-4}{\emph{Eur.
  Phys. J. C} {\bfseries 83} (2023) 88},
  [\href{https://arxiv.org/abs/2205.14554}{{\ttfamily 2205.14554}}].

\bibitem{Sarkar:2021djs}
S.~Sarkar, S.~Kumar and S.~Bhattacharjee, ``{Can we detect a supertranslated
  black hole?}'',
  \href{http://dx.doi.org/10.1103/PhysRevD.105.084001}{\emph{Phys. Rev. D}
  {\bfseries 105} (2022) 084001},
  [\href{https://arxiv.org/abs/2110.03547}{{\ttfamily 2110.03547}}].

\bibitem{Hall:1987vw}
G.~S. Hall, T.~Morgan and Z.~Perj{\'e}s, ``Three-dimensional space-times'',
  \href{http://dx.doi.org/10.1007/BF00759150}{\emph{General Relativity and
  Gravitation} {\bfseries 19} (1987) 1137--1147}.

\bibitem{Milson:2012ry}
R.~Milson and L.~Wylleman, ``{Three-dimensional spacetimes of maximal order}'',
  \href{http://dx.doi.org/10.1088/0264-9381/30/9/095004}{\emph{Class. Quant.
  Grav.} {\bfseries 30} (2013) 095004},
  [\href{https://arxiv.org/abs/1210.6920}{{\ttfamily 1210.6920}}].

\bibitem{Alessio:2020ioh}
F.~Alessio, G.~Barnich, L.~Ciambelli, P.~Mao and R.~Ruzziconi, ``{Weyl charges
  in asymptotically locally AdS$_3$ spacetimes}'',
  \href{http://dx.doi.org/10.1103/PhysRevD.103.046003}{\emph{Phys. Rev. D}
  {\bfseries 103} (2021) 046003},
  [\href{https://arxiv.org/abs/2010.15452}{{\ttfamily 2010.15452}}].

\bibitem{Campoleoni:2018ltl}
A.~Campoleoni, L.~Ciambelli, C.~Marteau, P.~M. Petropoulos and K.~Siampos,
  ``{Two-dimensional fluids and their holographic duals}'',
  \href{http://dx.doi.org/10.1016/j.nuclphysb.2019.114692}{\emph{Nucl. Phys. B}
  {\bfseries 946} (2019) 114692},
  [\href{https://arxiv.org/abs/1812.04019}{{\ttfamily 1812.04019}}].

\bibitem{Ciambelli:2024vhy}
L.~Ciambelli and M.~Geiller, ``{Field-dependent diffeomorphisms and the
  transformation of surface charges between gauges}'',
  \href{http://dx.doi.org/10.1007/JHEP05(2025)022}{\emph{JHEP} {\bfseries 05}
  (2025) 022}, [\href{https://arxiv.org/abs/2412.14992}{{\ttfamily
  2412.14992}}].

\bibitem{Donnay:2016ejv}
L.~Donnay, G.~Giribet, H.~A. Gonz\'alez and M.~Pino, ``{Extended Symmetries at
  the Black Hole Horizon}'',
  \href{http://dx.doi.org/10.1007/JHEP09(2016)100}{\emph{JHEP} {\bfseries 09}
  (2016) 100}, [\href{https://arxiv.org/abs/1607.05703}{{\ttfamily
  1607.05703}}].

\bibitem{Akhmedov:2017ftb}
E.~T. Akhmedov and M.~Godazgar, ``{Symmetries at the black hole horizon}'',
  \href{http://dx.doi.org/10.1103/PhysRevD.96.104025}{\emph{Phys. Rev. D}
  {\bfseries 96} (2017) 104025},
  [\href{https://arxiv.org/abs/1707.05517}{{\ttfamily 1707.05517}}].

\bibitem{Chandrasekaran:2021hxc}
V.~Chandrasekaran, E.~E. Flanagan, I.~Shehzad and A.~J. Speranza, ``{Brown-York
  charges at null boundaries}'',
  \href{http://dx.doi.org/10.1007/JHEP01(2022)029}{\emph{JHEP} {\bfseries 01}
  (2022) 029}, [\href{https://arxiv.org/abs/2109.11567}{{\ttfamily
  2109.11567}}].

\bibitem{Odak:2023pga}
G.~Odak, A.~Rignon-Bret and S.~Speziale, ``{General gravitational charges on
  null hypersurfaces}'',
  \href{http://dx.doi.org/10.1007/JHEP12(2023)038}{\emph{JHEP} {\bfseries 12}
  (2023) 038}, [\href{https://arxiv.org/abs/2309.03854}{{\ttfamily
  2309.03854}}].

\bibitem{Agrawal:2025fsv}
S.~Agrawal, P.~Charalambous and L.~Donnay, ``{Null infinity as an inverted
  extremal horizon: Matching an infinite set of conserved quantities for
  gravitational perturbations}'',
  [\href{https://arxiv.org/abs/2506.15526}{{\ttfamily 2506.15526}}].

\bibitem{Ruzziconi:2025fct}
R.~Ruzziconi and C.~Zwikel, ``{Celestial Symmetries of Black Hole Horizons}'',
  [\href{https://arxiv.org/abs/2504.08027}{{\ttfamily 2504.08027}}].

\bibitem{Ruzziconi:2025fuy}
R.~Ruzziconi and C.~Zwikel, ``{Celestial $Lw_{1+\infty}$ Symmetries and
  Subleading Phase Space of Null Hypersurfaces}'',
  [\href{https://arxiv.org/abs/2511.07525}{{\ttfamily 2511.07525}}].

\bibitem{Ashtekar:2024bpi}
A.~Ashtekar and S.~Speziale, ``{Null infinity as a weakly isolated horizon}'',
  \href{http://dx.doi.org/10.1103/PhysRevD.110.044048}{\emph{Phys. Rev. D}
  {\bfseries 110} (2024) 044048},
  [\href{https://arxiv.org/abs/2402.17977}{{\ttfamily 2402.17977}}].

\bibitem{Ashtekar:2024stm}
A.~Ashtekar and S.~Speziale, ``{Null infinity and horizons: A new approach to
  fluxes and charges}'',
  \href{http://dx.doi.org/10.1103/PhysRevD.110.044049}{\emph{Phys. Rev. D}
  {\bfseries 110} (2024) 044049},
  [\href{https://arxiv.org/abs/2407.03254}{{\ttfamily 2407.03254}}].

\bibitem{Ashtekar:2024mme}
A.~Ashtekar and S.~Speziale, ``{Horizons and null infinity: A fugue in four
  voices}'', \href{http://dx.doi.org/10.1103/PhysRevD.109.L061501}{\emph{Phys.
  Rev. D} {\bfseries 109} (2024) L061501},
  [\href{https://arxiv.org/abs/2401.15618}{{\ttfamily 2401.15618}}].

\bibitem{Loutrel:2020wbw}
N.~Loutrel, J.~L. Ripley, E.~Giorgi and F.~Pretorius, ``{Second Order
  Perturbations of Kerr Black Holes: Reconstruction of the Metric}'',
  \href{http://dx.doi.org/10.1103/PhysRevD.103.104017}{\emph{Phys. Rev. D}
  {\bfseries 103} (2021) 104017},
  [\href{https://arxiv.org/abs/2008.11770}{{\ttfamily 2008.11770}}].

\bibitem{BenAchour:2025hns}
J.~Ben~Achour, C.~Montagnon and H.~Roussille, ``{Algebraically special
  perturbations of the Kerr black hole: a metric formulation}'',
  \href{http://dx.doi.org/10.1088/1475-7516/2025/11/050}{\emph{JCAP} {\bfseries
  11} (2025) 050}, [\href{https://arxiv.org/abs/2507.10384}{{\ttfamily
  2507.10384}}].

\bibitem{Chandrasekhar:1984mgh}
S.~Chandrasekhar, ``{On algebraically special perturbations of black holes}'',
  \href{http://dx.doi.org/10.1098/rspa.1984.0021}{\emph{Proc. Roy. Soc. Lond.
  A} {\bfseries 392} (1984) 1--13}.

\bibitem{Nerozzi:2005hz}
A.~Nerozzi, M.~Bruni, V.~Re and L.~M. Burko, ``{Towards a wave-extraction
  method for numerical relativity: IV. Testing the quasi-Kinnersley method in
  the Bondi-Sachs framework}'',
  \href{http://dx.doi.org/10.1103/PhysRevD.73.044020}{\emph{Phys. Rev. D}
  {\bfseries 73} (2006) 044020},
  [\href{https://arxiv.org/abs/gr-qc/0507068}{{\ttfamily gr-qc/0507068}}].

\bibitem{Moxon:2020gha}
J.~Moxon, M.~A. Scheel and S.~A. Teukolsky, ``{Improved Cauchy-characteristic
  evolution system for high-precision numerical relativity waveforms}'',
  \href{http://dx.doi.org/10.1103/PhysRevD.102.044052}{\emph{Phys. Rev. D}
  {\bfseries 102} (2020) 044052},
  [\href{https://arxiv.org/abs/2007.01339}{{\ttfamily 2007.01339}}].

\bibitem{Iozzo:2020jcu}
D.~A.~B. Iozzo, M.~Boyle, N.~Deppe, J.~Moxon, M.~A. Scheel, L.~E. Kidder
  et~al., ``{Extending gravitational wave extraction using Weyl characteristic
  fields}'', \href{http://dx.doi.org/10.1103/PhysRevD.103.024039}{\emph{Phys.
  Rev. D} {\bfseries 103} (2021) 024039},
  [\href{https://arxiv.org/abs/2010.15200}{{\ttfamily 2010.15200}}].

\bibitem{Blanchet:2020ngx}
L.~Blanchet, G.~Comp\`ere, G.~Faye, R.~Oliveri and A.~Seraj, ``{Multipole
  expansion of gravitational waves: from harmonic to Bondi coordinates}'',
  \href{http://dx.doi.org/10.1007/JHEP02(2021)029}{\emph{JHEP} {\bfseries 02}
  (2021) 029}, [\href{https://arxiv.org/abs/2011.10000}{{\ttfamily
  2011.10000}}].

\bibitem{Blanchet:2023pce}
L.~Blanchet, G.~Comp\`ere, G.~Faye, R.~Oliveri and A.~Seraj, ``{Multipole
  expansion of gravitational waves: memory effects and Bondi aspects}'',
  \href{http://dx.doi.org/10.1007/JHEP07(2023)123}{\emph{JHEP} {\bfseries 07}
  (2023) 123}, [\href{https://arxiv.org/abs/2303.07732}{{\ttfamily
  2303.07732}}].

\bibitem{Mao:2024urq}
P.~Mao and B.~Zeng, ``{Note on post-Minkowskian expansion and Bondi
  coordinates}'',
  \href{http://dx.doi.org/10.1016/j.nuclphysb.2025.117111}{\emph{Nucl. Phys. B}
  {\bfseries 1019} (2025) 117111},
  [\href{https://arxiv.org/abs/2405.11953}{{\ttfamily 2405.11953}}].

\bibitem{Mao:2025lwk}
P.~Mao and B.~Zeng, ``{From harmonic to Newman-Unti coordinates at the second
  post-Minkowskian order}'',
  [\href{https://arxiv.org/abs/2511.07647}{{\ttfamily 2511.07647}}].

\bibitem{Ball:2021tmb}
A.~Ball, S.~A. Narayanan, J.~Salzer and A.~Strominger, ``{Perturbatively exact
  w$_{1+\infty}$ asymptotic symmetry of quantum self-dual gravity}'',
  \href{http://dx.doi.org/10.1007/JHEP01(2022)114}{\emph{JHEP} {\bfseries 01}
  (2022) 114}, [\href{https://arxiv.org/abs/2111.10392}{{\ttfamily
  2111.10392}}].

\bibitem{Strominger:2021lvk}
A.~Strominger, ``{w(1+infinity) and the Celestial Sphere}'',
  [\href{https://arxiv.org/abs/2105.14346}{{\ttfamily 2105.14346}}].

\bibitem{Adamo:2021lrv}
T.~Adamo, L.~Mason and A.~Sharma, ``{Celestial $w_{1+\infty}$ Symmetries from
  Twistor Space}'',
  \href{http://dx.doi.org/10.3842/SIGMA.2022.016}{\emph{SIGMA} {\bfseries 18}
  (2022) 016}, [\href{https://arxiv.org/abs/2110.06066}{{\ttfamily
  2110.06066}}].

\bibitem{Kmec:2024nmu}
A.~Kmec, L.~Mason, R.~Ruzziconi and A.~Yelleshpur~Srikant, ``{Celestial
  $Lw_{1+\infty}$ charges from a twistor action}'',
  [\href{https://arxiv.org/abs/2407.04028}{{\ttfamily 2407.04028}}].

\bibitem{Donnay:2024qwq}
L.~Donnay, L.~Freidel and Y.~Herfray, ``{Carrollian $Lw_{1+\infty}$
  representation from twistor space}'',
  [\href{https://arxiv.org/abs/2402.00688}{{\ttfamily 2402.00688}}].

\bibitem{Adamo:2009vu}
T.~M. Adamo, C.~N. Kozameh and E.~T. Newman, ``{Null Geodesic Congruences,
  Asymptotically Flat Space-Times and Their Physical Interpretation}'',
  \href{http://dx.doi.org/10.12942/lrr-2009-6}{\emph{Living Rev. Rel.}
  {\bfseries 12} (2009) 6}, [\href{https://arxiv.org/abs/0906.2155}{{\ttfamily
  0906.2155}}].

\bibitem{Adamo:2010ey}
T.~M. Adamo and E.~T. Newman, ``{The Generalized Good Cut Equation}'',
  \href{http://dx.doi.org/10.1088/0264-9381/27/24/245004}{\emph{Class. Quant.
  Grav.} {\bfseries 27} (2010) 245004},
  [\href{https://arxiv.org/abs/1007.4215}{{\ttfamily 1007.4215}}].

\bibitem{Ashtekar:2000hw}
A.~Ashtekar, S.~Fairhurst and B.~Krishnan, ``{Isolated horizons: Hamiltonian
  evolution and the first law}'',
  \href{http://dx.doi.org/10.1103/PhysRevD.62.104025}{\emph{Phys. Rev. D}
  {\bfseries 62} (2000) 104025},
  [\href{https://arxiv.org/abs/gr-qc/0005083}{{\ttfamily gr-qc/0005083}}].

\bibitem{Bonnor:1990tw}
W.~B. Bonnor, ``{The C-metric in Bondi's coordinates}'',
  \href{http://dx.doi.org/10.1088/0264-9381/7/10/002}{\emph{Classical and
  Quantum Gravity} {\bfseries 7} (1990) L229}.

\bibitem{Bini:2005qyt}
D.~Bini, C.~Cherubini, S.~Filippi and A.~Geralico, ``{C metric: The equatorial
  plane and Fermi coordinates}'',
  \href{http://dx.doi.org/10.1088/0264-9381/22/23/015}{\emph{Class. Quant.
  Grav.} {\bfseries 22} (2005) 5157--5168},
  [\href{https://arxiv.org/abs/1408.4277}{{\ttfamily 1408.4277}}].

\bibitem{Griffiths:2006tk}
J.~B. Griffiths, P.~Krtous and J.~Podolsky, ``{Interpreting the C-metric}'',
  \href{http://dx.doi.org/10.1088/0264-9381/23/23/008}{\emph{Class. Quant.
  Grav.} {\bfseries 23} (2006) 6745--6766},
  [\href{https://arxiv.org/abs/gr-qc/0609056}{{\ttfamily gr-qc/0609056}}].

\end{thebibliography}\endgroup
\bibliographystyle{Biblio}

\end{document}